\documentclass[11pt]{article}

\usepackage{tikz}
\usetikzlibrary{decorations.pathreplacing, arrows.meta}
\tikzset{vert/.style = {circle, fill, inner sep = 0, minimum size = 5}}
\newcommand{\circled}[2][inner sep=1pt]{\ifmmode
\tikz[baseline=(X.base),outer sep=0pt]{\node[circle,draw,#1](X){\ensuremath{#2}};}
\else
\tikz[baseline=(X.base),outer sep=0pt]{\node[circle,draw,#1](X){#2};}
\fi
}
\newcommand{\tboxed}[2][inner sep=-0.5pt]{\ifmmode\tikz[baseline=(X.base),outer
sep=0pt]{\node[draw,#1](X){\ensuremath{#2}};}
\else
\tikz[baseline=(X.base),outer sep=0pt]{\node[draw,#1](X){#2};}
\fi
}
\newcommand{\sboxed}[2][inner sep=3.5pt]{\ifmmode\tikz[baseline=(X.base),outer
sep=0pt]{\node[draw,#1](X){\ensuremath{#2}};}
\else
\tikz[baseline=(X.base),outer sep=0pt]{\node[square,draw,#1](X){#2};}
\fi
}

\usepackage{amssymb}
\usepackage{amsmath}
\usepackage{amsthm}
\usepackage{amssymb}
\usepackage{amssymb}
\usepackage[multiple]{footmisc}
\usepackage{enumerate}
\usepackage[textwidth=2cm]{todonotes}
\usepackage{epic,xcolor}
\usepackage{graphicx}
\usepackage{natbib}
\usepackage{tabularx}
\usepackage{booktabs}
\usepackage{rotating}
\usepackage{bm,upgreek}
\usepackage{caption}
\usepackage{subcaption}
\usepackage{multirow}
\usepackage{setspace}
\usepackage{comment}
\usepackage{multirow}
\usepackage{xr-hyper}

\definecolor{chicago-maroon}{RGB}{128,0,0}
\definecolor{northwestern-purple}{RGB}{82,0,99}

\usepackage[unicode=true, pdfusetitle,
bookmarks=true,bookmarksnumbered=false,bookmarksopen=false,
breaklinks=false,pdfborder={0 0 1},backref=false,colorlinks=false]
{hyperref}
\hypersetup{
	colorlinks = true,
	urlcolor = chicago-maroon,
	linkcolor = chicago-maroon,
	citecolor = northwestern-purple,
}

\newtheorem{prop}{\sc{Proposition}}

\newtheorem{example}{\sc{Example}}
\newtheorem{thm}{\sc{Theorem}}
\newtheorem{lem}{\sc{Lemma}}

\newtheorem{cor}{\sc{Corollary}}

\newtheorem{defn}{\sc{Definition}}
\newtheorem*{definition*}{\sc{Definition}}

\newcommand{\tabextralines}[1]{}

\newcommand{\ul}{\ensuremath{\underline{u}}}
\newcommand{\uh}{\ensuremath{\overline{u}}}
\def\I{\mathbb{I}}
\def\R{\mathcal{R}}
\def\D{\Delta}
\def\t{\theta}
\def\m{\mu}
\def\T{\Theta}
\def\e{\epsilon}
\def\bsigma{\bm{\sigma}}
\def\ca{c_1}
\def\cb{c_2}
\def\cg{c_3}
\def\-{\text{-}}
\newcommand{\apa}[1]{i^1_{#1}}
\newcommand{\apb}[1]{i^2_{#1}}
\newcommand{\apc}[1]{i^3_{#1}}

\def\q{\bm{q}}
\def\p{\bm{p}}
\def\P{\bm{P}}
\def\Q{\bm{Q}}
\def\D{\bm{D}}
\def\bkappa{\bm{\kappa}}
\def\S{\bm{S}}
\def\s{\bm{s}}
\def\C{\bm{C}}
\def\bu{\bm{u}}
\def\x{\bm{x}}

\newenvironment{tabnotes}[2][1]{\begin{minipage}[t]{#1\textwidth}\vspace{0.1cm}\scriptsize{\emph{Notes:} #2}}{\end{minipage}}

\newcommand{\graphique}[2][1]{\begin{minipage}{\linewidth}\begin{center}\includegraphics[width=#1\linewidth,clip]{#2}\end{center}\end{minipage}}

\newenvironment{mfignotesin}[4][1]{
\begin{figure}[!ht]\begin{center}
\begin{minipage}{#1\textwidth}\begin{center}#3\caption{#2}\end{center}\end{minipage}
\if!#4!\empty \else \\
\begin{footnotesize}\begin{minipage}{\textwidth}\scriptsize{\medskip\par
\emph{Notes:} #4}\end{minipage}\end{footnotesize} \fi }
{\end{center}\end{figure}}

\usepackage{geometry}[margin=1in]
\usepackage{longtable}
\usepackage[normalem]{ulem}
\DeclareMathOperator*{\argmax}{arg\,max}

\newcommand\listappendixname{List of Appendices}
\newcommand\appcaption[1]{%
  \addcontentsline{app}{section}{#1}}
\makeatletter
\newcommand\listofappendices{%
  \section*{\listappendixname}\@starttoc{app}}
\makeatother

 \definecolor{cornell-red}{RGB}{179,27,27}

\newcommand{\GAo}[1]{{}}

\usepackage{pagecolor}
\begin{document}
	\setstretch{1}
	
\allowdisplaybreaks
	\title{Stable Matching with Mistaken Agents\thanks{This paper largely supersedes another paper of ours, entitled ``Strategic Mistakes: Implications for Market Design Research.'' We thank Xingye Wu, who has provided excellent research assistance for the theory part, and Julien Grenet for his generous help with the Monte Carlo simulations.  We are grateful to the seminar/conference participants at ANU, Asia-Pacific IO Conference, ASSA Meeting, Barcelona GSE Summer Forum, Boston College, Deakin, Duke, Columbia, Conference on Economic Design, ``Econometrics Meets Theory'' Conference at NYU, European Meeting on Game Theory, ``Dynamic Models in Economics'' Workshop at NUS, Hitotsubashi, Higher School of Economics, Game Theory and Management Conference, MIT, NBER Market Design Group Meeting, PET Conference, Workshop ``Matching in Practice'', Paris School of Economics, Stony Brook Conference on  Game Theory, UC Irvine, University of Queensland, and Waseda for their comments.	
	Authors acknowledge support from the Australian Research Council (DP160101350) and the University of Melbourne (Artemov); National Research Foundation of Korea (NRF-2020S1A5A2A03043516, Che); National Science Foundation (SES-1851821, Che; SES-1730636, He).}}
	\author{Georgy Artemov\thanks{Department of Economics, University of Melbourne,  Australia.  Email: {\texttt{georgy@gmail.com}}}	\,\and \, Yeon-Koo Che\thanks{Department of Economics, Columbia University,  USA.  Email: {\texttt{yeonkooche@gmail.com}}.}
		\,\and \, YingHua He\thanks{Department of Economics, Rice University, USA. \ Email:
			{\texttt{yinghua.he@rice.edu}}.}}
	
	\date{\today \endgraf}

\maketitle
\begin{abstract}
Motivated by growing evidence of agents' mistakes in strategically simple environments, we propose a solution concept---robust equilibrium---that requires only an asymptotically optimal behavior. We use it to study large random matching markets operated by the applicant-proposing Deferred Acceptance (DA). Although truth-telling is a dominant strategy, almost all applicants may be non-truthful in robust equilibrium; however, the outcome must be arbitrarily close to the stable matching. 
Our results imply that one can assume truthful agents to study DA outcomes, theoretically or counterfactually. However, to estimate the preferences of mistaken agents, one should assume stable matching but not truth-telling.

\end{abstract}

\newpage	
\setstretch{1.2}

\section{Introduction} 

Strategy-proofness and stability are two important desiderata in market design for two-sided matching \citep{abdulkadiroglu/sonmez:03}.
One describes agents' optimal behavior, and the other is a property of a matching outcome. Strategy-proofness---making it a weakly dominant strategy to truthfully reveal one's own preferences---minimizes the scope for mistakes and thus levels the playing field. It also aids empirical research by making agents' choices easy to interpret. 
Stability---ensuring that each agent is matched with her favorite partner among those who are willing to match with her---is crucial for the long-term sustainability of a mechanism \citep[see, e.g.,][]{roth:91} and the fairness of matching, particularly in the context of centralized college admissions and school choice \citep{abdulkadiroglu/sonmez:03}.

These two desiderata are satisfied under one of the most popular mechanisms in practice, the Deferred Acceptance (DA): it is strategy-proof for agents on one side of the market, and the matching outcome is always stable when agents report truthfully. Alarmingly, there is growing evidence that non-truthful behaviors, which are called strategic mistakes in the literature, are common in DA.\footnote{The literature uses the term  ``mistake'' to refer to the play of a {\it weakly dominated strategy}, regardless of whether it entails an actual payoff loss (which depends on other individuals' actions). We call a mistake \emph{payoff-relevant} when it leads to a payoff loss for an agent, i.e., her truthful reporting of preferences would have matched her with a more preferred partner given other agents' actions.} Laboratory experiments \citep[see, e.g.,][]{chen/sonmez:02} and studies of high-stake real-life matching markets \citep{rees-jones:16,shorrer/sovag:2017,chen_li:2015,ACH:Proj0WP,HRR:16} show that a significant fraction of participants misreport their preferences. When agents make mistakes, DA no longer guarantees stability.

Mistaken agents may pose a broad challenge to market-design research, both theoretical and empirical. Most theoretical studies on strategy-proof mechanisms assume that agents play their unique dominant strategy, truth-telling (TT), which guarantees a stable matching under DA. A natural question is: Do the documented mistakes, i.e., non-truthful behaviors, imply that the theoretical predictions about DA outcomes in the literature are incorrect? Moreover, much of the empirical literature relies on assumptions that ignore mistakes. Does it mean that the estimates from that literature are biased? These are the questions that our paper aims to answer.

We examine agent behavior and outcome in many-to-one matching economies operated by DA. Participants on one side, which are labeled ``colleges,'' use priority scores to strictly rank those on the other side, whom we call ``applicants.'' Each applicant knows her own score when applying to colleges. This setting captures many practices in the field. For example, the markets with mistaken agents mentioned above fit this description and cover admissions to secondary schools, universities, or post-graduate programs in four different countries. In these settings, priority scores may be a measure of an applicant's academic performance such as scores from entrance exams. 
We define a college as {\it feasible} to an applicant if her score is above the college's {\it cutoff}, which is the lowest score among the college's accepted applicants. When each applicant is matched with her favorite feasible college, the matching is stable. Therefore, a stable matching can be characterized by a set of ``market-clearing'' cutoffs \citep{azevedo/leshno:16}.

An important empirical finding from the literature is that, although mistakes are frequent, only a small fraction of them have payoff consequences; that is, unilaterally correcting a mistake changes the assignment for only few applicants. In general, mistakes are difficult to identify in the field because applicant preferences are unknown to researchers. Some recent studies  \citep{shorrer/sovag:2017,ACH:Proj0WP,HRR:16} focus on pairs of education programs that differ only in a financial component (e.g., scholarship vs.\ no scholarship) and hence an applicant's preference order between the two in a pair can be unambiguously determined. 
As summarized in Table~\ref{tab:otherStudies}, 17--35\% of applicants in these studies make an identifiable mistake: when reporting their ordinal preferences to the mechanism, they rank a program \textit{without} a scholarship \textit{above} the identical program with a scholarship (column~3). 
However, among the applicants with an identified mistake, only 1--20\% would have a different assignment if each applicant's mistake is corrected unilaterally (columns~5 and 7). 

\begin{table}[h!]
	\centering \footnotesize
	\caption{Mistakes and Payoff-relevance across Different Studies}\label{tab:otherStudies} 
	\resizebox{1\textwidth}{!}{%
    \begin{tabular}{rcccccccccc}
		\toprule
          & Size of the & &\multicolumn{2}{c}{Identified mistakes}  &    & \multicolumn{5}{c}{Payoff-relevant mistakes}  \\
          \cline{4-5} \cline{7-11}
          & relevant  & & \multirow{2}{*}{Freq.} & \multirow{2}{*}{share: $\frac{(2)}{(1)}$}  & & \multicolumn{2}{c}{upper bound}  & & \multicolumn{2}{c}{lower bound}  \\
          \cline{7-8} \cline{10-11}
          & sample & &  &   & & Freq. & share: $\frac{(4)}{(2)}$  & & Freq. & share: $\frac{(6)}{(2)}$  \\
          & (1)   &  &(2)   & (3)  &  & (4)   & (5)  &  & (6)   & (7) \\
		\midrule
        \cite{shorrer/sovag:2017} & \multirow{2}{*}{92,777} &  & \multirow{2}{*}{15,653} & \multirow{2}{*}{17\%}  & & \multirow{2}{*}{1,479}  & \multirow{2}{*}{9\%}   &  & \multirow{2}{*}{669}   & \multirow{2}{*}{4\%} \\
        College admissions in Hungary \\
        [0.5em]
        \cite{ACH:Proj0WP} & \multirow{2}{*}{2,915}  &  &\multirow{2}{*}{1,009}  & \multirow{2}{*}{35\%}  & & \multirow{2}{*}{201}   & \multirow{2}{*}{20\%}   & & \multirow{2}{*}{14}    & \multirow{2}{*}{1\%} \\
        \multicolumn{1}{r}{College admissions in Australia}\\
        [0.5em]
        \cite{HRR:16} & \multirow{2}{*}{672}   & & \multirow{2}{*}{130}   & \multirow{2}{*}{19\%}   && \multirow{2}{*}{10}    & \multirow{2}{*}{8\%}  &  & \multirow{2}{*}{3}     & \multirow{2}{*}{2\%}   \\
        \multicolumn{1}{r}{Graduate admissions in Israel} \\

    	\bottomrule
	\end{tabular}}
	\begin{tabnotes} 
	All studies identify instances where an applicant appears to prefer an education program without a scholarship to the same program with a scholarship (i.e., she ranks the former higher than the latter in her application or applies only to the former). We call these instances \textit{identifiable mistakes.} Column~(1) is the number of applicants that can possibly make an identifiable mistake. Columns~(2) and (3) show the number of applicants who made an identifiable mistake. If unilaterally correcting an applicant's mistake leads to a different outcome for the applicant, the mistake is payoff relevant. As evaluating payoff-relevance requires knowledge of true applicant preferences, these studies find an upper (columns~4 and 5) and a lower (columns~6 and 7) bounds.
	\end{tabnotes} 
\end{table}%

Motivated by such an empirical pattern---a significant presence of mistakes but largely of little payoff consequences---, we employ a new solution concept, which we call \textit{robust equilibrium}, that relaxes Bayesian Nash equilibrium to allow for mistakes with ``small'' payoff consequences.  We operationalize the payoff ``smallness'' by studying a sequence of DA-run matching economies that grow large both in the number of applicants and the number of seats per college, with a fixed number of colleges. Along the sequence, applicant types---i.e., their preferences and their priority scores at colleges---are randomly drawn from a well-behaved distribution (to be made precise later). This random sampling maintains certain tractability while, at the same time, approximating applicants' uncertainty about the types of other applicants in real life. We define a strategy profile as a (possibly asymmetric) function that maps randomly drawn types to the rank-order lists (ROLs) submitted by each applicant. While TT is a weakly dominant strategy under DA, our concept allows for possible mistakes or deviations from TT. Specifically, robust equilibrium is a strategy profile in which, for any $\epsilon>0$, each applicant obtains within $\epsilon$ of their highest possible payoff in a sufficiently large economy.

Recall that one of our research questions is about the empirical literature that assumes no mistakes by agents. In DA, such an assumption implies TT. Our first main result (Theorem~\ref{p1}) says that this assumption is not justified in a robust equilibrium, as all but a vanishing fraction of applicants may submit untruthful ROLs.\GAo{Our first main result says that this assumption is not justified in a robust equilibrium. Specifically, Theorem~\ref{p1} shows that a dramatic departure from TT---all but a vanishing fraction of applicants submitting untruthful ROLs---is supported as a robust equilibrium.} To the extent that robust equilibrium captures applicants' behavior, this result suggests that we should not be surprised by the documented mistakes. 
Furthermore, our theorem does not impose any structure on  mistakes: applicants may omit their more preferred colleges or flip the order of colleges in their ROLs as long as the probability of admission to these colleges is low. Both of these behaviors are consistent with the evidence reported in Table~\ref{tab:otherStudies}.

In contrast, regarding our other research question, we obtain a positive answer: the theoretical predictions about stable matching under DA are generally valid, at least in large economies. Despite the behavioral multiplicity and ambiguity, under mild conditions, {\it all} robust equilibria yield a virtually unique outcome in a sufficiently large economy (Theorem~\ref{p2}). The outcome is asymptotically stable---the fraction of applicants who obtain their favorite feasible college converges to one---and converges to the outcome that would arise from TT. In other words, even if applicants make mistakes, the outcome is well approximated by the outcome that would arise with fully rational applicants (Corollary~\ref{cor2}).

At first glance, asymptotic stability may appear to be an unsurprising consequence of robust equilibrium. One may conjecture that, as payoff losses vanish, fewer applicants suffer a loss, which would imply that most applicants must obtain their favorite feasible college. However, a robust equilibrium allows everyone to have a vanishing loss, so a possibility remains that an arbitrarily large number of applicants do not obtain their favorite feasible college. In other words, robustness does not conceptually imply asymptotic stability. 

Alternatively, one may expect asymptotic stability to result from applicants becoming ``price-takers'' in a large economy; namely, every applicant may simply perceive the colleges' admission cutoffs as fixed and unaffected by her unilateral deviation. While acting optimally against fixed cutoffs might lead to a stable matching, such a price-taking hypothesis cannot be taken for granted even in an arbitrarily large economy. 

    \def\ca{\texttt{a}}
    \def\cb{\texttt{b}}
    \def\cg{\texttt{c}}
    \def\apj{\theta}
    \def\apa{\upalpha}
    \def\apb{\upbeta}
    \def\apc{\upgamma}

The lack of price-taking can be illustrated using a version of an example due to \cite{roth:82}. There are three applicants, ($\apa$, $\apb$, $\apc$), and three colleges, ($\ca, \cb$, $\cg$), with one seat each. Applicant preferences and college priority rankings are:
\begin{align*}
 \text{Applicant preferences:} &&    \apa:\ \cb\-\ca\-\cg && \apb:\ \ca\-\cb\-\cg && \apc:\ \ca\-\cb\-\cg\\
 \text{College priority ranking:}   &&    \ca:\ \apa \- \apc\- \apb && \cb:\ \apb \- \apc\- \apa && \cg: \text{arbitrary}
\end{align*}
An unstable matching ($\apa$--$\cb$, $\apb$--$\ca$, $\apc$--$\cg$) is a Nash equilibrium outcome under DA, with $\apa$ and $\apb$ adopting TT and $\apc$ listing only $\cg$ in her ROL. Applicant~$\apc$'s deviation to TT, her best possible deviation, is unprofitable, as it activates a chain of rejections that leaves $\apc$'s assignment unchanged. Specifically, $\apc$ knocks off $\apb$ from $\ca$, who knocks off $\apa$ from $\cb$, who knocks off $\apc$ from $\ca$ to $\cg$.  There is a failure of price-taking here. To see this, recall that colleges rank applicants by scores. Suppose that $\apa$'s scores at colleges~$\ca$ and $\cb$ are 1 and 0, respectively; $\apb$'s are 0 and 1; and $\apc$'s are 0.5 at both colleges. In the unstable equilibrium matching, the cutoffs of colleges~$\ca$ and $\cb$ are zero, determined by their matched applicant. When $\apc$ deviates to TT, the outcome will be the unique stable matching, ($\apa$--$\ca$, $\apb$--$\cb$, $\apc$--$\cg$), and the cutoffs of colleges~$\ca$ and $\cb$ both jump to one. Thus, price-taking is not satisfied.\footnote{Although in a different setting, this example may also help to illuminate Theorems~\ref{p1} and \ref{p2}. The stable matching ($\apa$--$\cb$, $\apb$--$\ca$, $\apc$--$\cg$) is an equilibrium outcome even when applicants do not adopt TT, as in Theorem~\ref{p1}, e.g., by ranking their matched colleges first. In contradiction to Theorem~\ref{p2}, the matching ($\apa$--$\cb$, $\apb$--$\ca$, $\apc$--$\cg$) is an equilibrium outcome but unstable. Both $\ca$ and $\cb$ have a zero cutoff, below $\apc$'s score (which is 0.5), and are therefore feasible to $\apc$. Applicant~$\apc$ is assigned her least preferred college, $\cg$, among her feasible colleges, violating stability. In Theorem~\ref{p2}, we find conditions that eliminate unstable equilibrium outcomes.}  Example~\ref{eg:1} in Section~\ref{sec:asymptotics} will show that merely increasing the size of the economy does not restore price-taking.

The key step to our results is to re-establish the price-taking behavior (Proposition~\ref{claim:uniform-convergence of P}), which presents a major challenge compared to the literature \citep[see, e.g.,][]{acy:15,azevedo/leshno:16,Agarwal-Somaini(2014),Fack-Grenet-He(2015),grigoryan2022convergence}. Those papers often study a fixed strategy, which allows them to impose conditions directly on the demand induced by that strategy. In our setting, such an approach would amount to imposing conditions on possible deviations, which is not justified. Instead, we impose conditions on the model primitives by assuming the full support of applicant types. To maintain full support when applicants adopt a strategy, we restrict ourselves to study \textit{regular} strategies that require TT being played with some arbitrarily low probability.\footnote{The full support assumption can be readily relaxed to allow for uni-dimensional applicant scores, as in Serial Dictatorship. Regularity can also be weakened to mean that there is a positive mass of applicants who report truthfully with some probability.} Intuitively, these two restrictions make it unlikely that a unilateral deviation triggers a massive rejection chain. Without assumptions on demand, we develop novel proof techniques that use the lattice structure of stable matching and the properties of DA (for both sides). We establish that ``demand curves'' are well-behaved in that an infinitesimal change in demand can only happen if there is an infinitesimal change in cutoffs. Lastly, we show that unilateral deviations do not significantly change demand or cutoffs faced by applicants in large economies.

The above arguments lead us to Proposition~\ref{claim:uniform-convergence of P}, stating that any (possibly non-robust-equilibrium) strategy profile must admit a subsequence of random cutoffs that converge almost surely to a vector of fixed cutoffs {\it uniformly} with respect to any possible unilateral deviation by applicants. This uniform convergence is of independent interest, as it generalizes the existing large economy convergence results. When applicants employ symmetric strategies---a special case of which is TT---, our result implies uniform almost-sure convergence of cutoffs to the cutoffs of the unique stable matching in the limit economy.

This proposition is the key to proving Theorems~\ref{p1} and \ref{p2}. As the economy grows, the uncertainty about cutoffs vanishes and the set of feasible colleges become apparent to applicants. For their ROL, applicants then know which colleges they can safely omit (hence Theorem~\ref{p1}) and which colleges they must include (hence Theorem~\ref{p2}). With the price-taking behavior restored, robust equilibrium behavior along {\it each} converging subsequence ensures that stability must hold asympotically, and all such outcomes must converge to the unique stable matching in the limit.

Our Theorems~\ref{p1} and \ref{p2} are reassuring news for the theoretical literature on DA. To the extent that outcomes are more important than applicant behavior, the existing results that rely on applicants' truthful behavior are largely robust to applicants' mistakes, at least in large economies.

Our results also yield important implications for empirical research. Strategy-proofness is sometimes taken literally in interpreting applicants' ROLs and leads to the assumption of weak truth-telling (WTT); see, for example, \citet[][]{Hallsten:2010} and \cite{Kirkeboen_2012}. WTT hypothesizes that an applicant ranks her most-preferred colleges truthfully, but may not rank all acceptable colleges. Theorem~\ref{p1} calls such an approach into question. When an applicant omits a more-preferred out-of-reach college, WTT infers that that college is less preferred than any college listed in the applicant's ROL, leading to biased estimates. At the same time, an alternative approach that assumes stability of the matching is justified in large enough economies by Theorem~\ref{p2}. This approach only makes inference about feasible colleges, but ``refuses'' to infer any preferences over infeasible ones. We illustrate the empirical implications of our theorems in Monte Carlo simulations. WTT estimates have low variance, but are substantially biased when applicants omit out-of-reach colleges.

Even though our results advise caution in relying on TT for preference estimation, they support using TT for the counterfactual analysis of policies. That is, our Theorem~\ref{p2} justifies the approach that uses estimated applicant preferences, say based on the stability hypothesis, but simply assumes TT in simulating the outcome, as long as preference estimates are consistent.  Despite the fact that applicants may make mistakes, the counterfactual outcome is well approximated by the outcome with TT applicants. 

One may be tempted to use an alternative approach to counterfactual analysis in which one skips the estimation step and assumes that an applicant submits the same ROL in both regimes, given that DA is strategy-proof.\GAo{A seemingly reasonable alternative approach in counterfactual analysis is to skip the estimation step and assume that an applicant submits the same ROL in both regimes, given that DA is strategy-proof.} If applicants play robust equilibrium, this assumption is not theoretically justified: if previously out-of-reach colleges become within-reach for an applicant under the counterfactual, we should not expect her to submit the same ROL. This possibility is not just of academic interest but of significant policy importance, as it arises under many reforms aiming at expanding access by disadvantaged students to high-quality schools. Counterfactual analyses using observed ROLs or using WTT-based estimates are likely to underestimate the impact of such policy by mis-inferring their preferences for high-quality schools that are out of reach under the pre-reform regime. Our Monte Carlo simulations illustrate this point: assuming the same ROLs across two regimes can mis-predict the assignment of 40\% of the applicants, and the WTT-based estimates mis-predict 25\%. In contrast, the mis-prediction rate is merely 4.5\% when we use the stability-based estimates to simulate counterfactual outcomes. 

\paragraph{Other Related Literature.} 

Our paper is the first to provide a theoretical foundation for stable matching in the presence of mistaken agents and hence lends strong support for using stability in theoretical and empirical studies. There is a long line of theoretical research recognizing that agents may not report truthfully even in a strategically straightforward environment (e.g., \citealp{li:17,Rabin_Reference-Dependent-DA:2019,Fack-Grenet-He(2015),Meisner2022lossAversion}).\footnote{Our results do not apply to strategically complex environments such as those using the Immediate Acceptance (IA) mechanism. IA is not strategyproof, which implies that TT is not necessarily the optimal strategy. How an applicant's sophistication in IA affects her outcome has been studied theoretically \citep[see, e.g.,][]{Miralles2008:WP,pathak/sonmez:08,acy:11} and empirically \citep[see, e.g.,][]{he:2017,Agarwal-Somaini(2014),Calsamiglia-Fu-Guell(2014)}.} Each of these papers offers a specific explanation for such behaviors, often maintaining the assumption that agents are rational in a certain sense. In contrast, we only postulate that the higher the payoff consequences of a mistake are, the rarer the mistake is in equilibrium.\footnote{This is consistent with the evidence reported in Table~\ref{tab:otherStudies} that shows that payoff-relevant mistakes are a small fraction of all identified mistakes. Furthermore, \citet{shorrer/sovag:2017} find that mistakes are more common when their expected utility cost is lower.} Our approach to accommodating mistakes is consonant with the previous literature on deviations from optimal behavior, e.g., rational inattention \citep{sims:03,matejka/mckay:15} and quantal response equilibria \citep{mckelvey/palfrey:95}. While based on similar ideas, our solution concept is designed for a different goal. We are interested in the implications of mistakes and therefore are agnostic about why agents make mistakes. Compared to these existing concepts, robust equilibrium imposes less structure, and is thus more permissive, on the types of mistakes allowed. At the same time, it is more tractable for our large economy analysis and admits a sharp prediction.\footnote{The rational inattention model and quantal response equilibria, as formalized in the papers cited above, are generally intractable for a rich choice environment like matching where a choice takes the form of a rank-order list.}

Among the papers cited in the above paragraph, only \cite{Fack-Grenet-He(2015)} (FGH, hereafter) study how non-truthful behaviors affect the stability of DA outcome. They assume fully rational agents and introduce application costs in DA. Deviations from TT occur when the probability of admission to a college is so low that it is not worth paying the application cost. FGH therefore cannot accommodate mistakes that have real payoff consequences. Even though our model is more general than theirs and thus requires new proof techniques, our prediction with respect to equilibrium outcome is sharper: Theorem~\ref{p2} shows that every regular robust equilibrium leads to asymptotic stability; in contrast, FGH show that there exists one such sequence of equilibria.

As we are interested in studying the implications of strategic mistakes for stable matching, our motivation and results are similar to \cite{kalai:04} and \cite{deb/kalai:15}. They also study approximate Bayesian equilibrium and show that it implies ``hindsight-stability.'' Critically, they assume that the effect any participant can unilaterally have on an opponent's payoff is uniformly bounded and decreases with the number of participants in the game. This assumption is tantamount to assuming ``price-taking'' behavior and does not hold in our setting even in an arbitrarily large economy (Example~\ref{eg:1}). Instead, we derive the result endogenously through elaborate asymptotics of large random economies.

Our setting of random economies is similar to Section IV.B of \citet{azevedo/leshno:16} (AL, hereafter). They assume that colleges are overdemanded (i.e., the total college capacity is less than the total number of applicants) and that the gradient of demand is invertible. These assumptions may not hold in our setting and our results do not rely on them.\footnote{{When studying convergence for purposes different from ours, some other papers also relax these conditions. For example, \cite{Agarwal-Somaini(2014)} do not require overdemandness while maintaining some restrictions on demand; \cite{grigoryan2022convergence} relaxes both conditions and studies the asymptotics of DA when there may be multiple stable matchings in the limit.}}  Further, AL perform price-theory analysis of stable matchings without a game-theoretic framework. Yet, when applicants are allowed to make mistakes, in accord with the evidence, they may not be price-takers even in a large economy.
A richer game-theoretical setup makes some of the key results in AL inapplicable. For instance, because applicants can adopt asymmetric strategies and make unilateral deviations, the induced submitted ROLs will not be i.i.d., while AL require i.i.d.\ draws of ordinal preferences.  We also allow mixed strategies, so the measure of submitted ROLs, which needs to be well-defined in AL and here depends on both strategies and the measure of types, requires a law of large numbers on the limit economy to be well-defined. That has usual conceptual difficulties \citep[see, e.g.,][]{Judd_continuum_iid:1985}.  We thus use a novel technique, exploiting the lattice structure of stability and the properties of DA. {As such, we are able to study the effects of any unilateral deviation by applicants, which is an innovative and necessary ingredient in our analysis.}

\vskip0.5cm

The rest of the paper is organized as follows. We first describe the model primitives in Section~\ref{sec:primitives}. Section~\ref{sec:theorems} presents the analysis of applicant behavior and outcome under our solution concept. In Section~\ref{sec:asymptotics}, we provide a sketch of the proofs and highlight the asymptotics of cutoffs with unilateral deviations. The implications of our results for market design are discussed in Section~\ref{sec:implications}. We conclude in Section~\ref{sec:conclusion}.

\def\p{\bm{p}}
\section{Model Primitives}\label{sec:primitives}

Consider an economy, $F^k$, in which $k$ applicants compete for admissions to a finite set of colleges, $\C=\{c_1, \dots, c_C\}$, $C \geq 2$, under the applicant-proposing Deferred Acceptance algorithm \citep{gale/shapley:62}.  Throughout, we refer to this algorithm simply as DA. A formal definition of DA can be found in Appendix~\ref{sec:DA-def}.  
    
Each applicant has a type $\t=(\bu, \s) \in \T=[\ul, \uh]^C \times [0,1]^C$, with $\ul < \uh$ and $\uh > 0$. $\bu = (u_1, \dots, u_C)$ is a vector of von-Neumann Morgenstern utilities of attending colleges, and $\s=(s_1, \dots, s_C)$ is a vector of scores representing the colleges' priorities, such that an applicant with a higher score has a higher priority at a college. We assume that being unassigned, or taking an outside option, gives an applicant a zero utility. Note that $\ul$ can be positive or negative. If $\ul <0$, an applicant can be assigned to a college with a negative utility and thus incur {some loss relative to her outside option.}
{A vector $\bu$ induces ordinal preferences over colleges, denoted by a rank-order list (ROL)  $\rho(\t)$, of colleges with positive utilities of length up to $C$.}

Colleges rank applicants by their scores; college capacities are a $C$-vector $k\cdot \S^k=[k\cdot \S]$, where $\S = (S_1,\ldots,S_C)$, $0<S_c<1$ for all $c$, is a fixed vector and $[\x]$ is the vector of integers nearest to $\x$ (rounded down in case of a tie). 

The economy $F^k$ is random in that applicant types are drawn identically and independently according to a full-support probability measure $\eta$ over $\Theta$;\footnote{Technically, we can weaken this condition. First, we only need positive density on $\t \in \left[\max\{\underline{u},0\},\overline{u}\right]^C \times [0,1]^C \subset {\T} $. That is, any truthful ROL of length $C$ can be realized. Second, we allow for an important special case where colleges' scores are uni-dimensional, i.e., $s_1 = \dots = s_C$, as in the Serial Dictatorship. In that case, the full-support assumption holds with a reduced dimensionality of support; applicants’ scores are one-dimensional numbers in $[0, 1]$.} the resulting empirical measure is denoted $\eta^k$.  

In this matching game, applicant types are private information, while $\eta$ and all other information about the economy is common knowledge. Such a specification corresponds to the matching games summarized in Table~\ref{tab:otherStudies} as well as many others in which admissions are based on scores \cite[for more examples, see Table 1 in][]{Fack-Grenet-He(2015)}.\footnote{In these settings, applicants know their scores but do not know the scores or preferences of other applicants. Applicants may form a belief about the ``typical'' distribution of scores and preferences (captured by $\eta$), but are also aware that the particular distribution they face, $\eta^k$, may differ from the typical distribution. Note that our model does not apply to the setting where priorities are induced by lotteries, such as school choice in New York City \citep{abdulkadiroglu/pathak/roth:09,abdulkadiroglu/agarwal/pathak:15,che/tercieux:15b}.} 

We are interested in applicant behaviors and outcomes in ``sufficiently large'' economies and thus study the asymptotics of behaviors and outcomes in a sequence of random economies $\{F^k\}_{k \in \mathbb{N}}$. As $k\to \infty$, the number of applicants and college capacities increase proportionally, while the number of colleges is fixed. The sequence of economies $\{F^k\}$ converges in the sense that $\eta^k$ converges in probability to $\eta$ and that $\S^k$ converges to $\S$.  It is, therefore, convenient, but not crucial, to view $(\eta,\S)$ as the description of the continuum economy that approximates the large finite economies.

Throughout, we assume that colleges are passive and rank applicants according to their scores. By contrast, we allow applicants not to rank colleges truthfully. In each random economy, an applicant's action is to choose an ROL from the set of possible ROLs, $\mathcal{R}$.  Applicant $i$'s \textit{strategy} is a measurable function $\sigma_i: \Theta \mapsto \Delta(\mathcal{R})$. One example is truth-telling, or TT, $\sigma_i(\theta) = \rho(\theta)$, which is a dominant strategy under DA \citep{dubins/freedman:81,roth:82}. A \textit{strategy profile} for $\{F^k\}_{k \in \mathbb{N}}$, denoted by $\bsigma$, is an infinite vector of individual strategies $\bsigma = (\sigma_1,\sigma_2,\dots)$, with the interpretation that an agent $i$ participates in all economies $k \geq i$, with a fixed strategy $\sigma_i$. That is, applicant~$i$'s ``identity'' is determined by her strategy, $\sigma_i$. Such a strategy profile enables us to keep track of a given applicant as the economy grows. By letting each applicant choose a different strategy, we allow for the possibility of an asymmetric strategy profile. We say $\bsigma$ is \textit{regular} if there exists $\gamma > 0$ such that for each $i$ and each $\t\in \T$, $\sigma_i(\t)$ assigns probability of at least $\gamma$ to playing $\rho(\t)$.\footnote{A regular strategy need not mean that every applicant reports truthfully with some probability. We can ``purify'' it by defining a richer type space, with a ``truthful'' type who always adopts TT.} We denote the truncation of a strategy profile for the economy $F^k$, which omits the strategies of applicants not in $F^k$, by $\bsigma^k = (\sigma_1,\dots,\sigma_k)$.

 DA uses applicants' submitted ROLs, their scores, and college capacities to calculate an outcome. 
An \textit{outcome}, or a \textit{matching}, is defined as a mapping $\m: C\cup \T\to 2^{\T} \cup (C\cup \T)$ satisfying the usual two-sidedness and consistency requirements. A \textit{stable matching} is also defined in the usual way to satisfy individual rationality and no-blocking.\footnote{Individual rationality requires that no participant (an applicant or a college) receives an unacceptable match. No blocking means that no applicant-college pair exists such that the applicant prefers the college over her match and the college has either a vacant position or admits another applicant whom the college ranks below that applicant.} When all applicants are TT (i.e., submitting $\rho(\theta)$) under DA, the resulting matching is stable \citep{gale/shapley:62}.     
    
Given an outcome $\mu$, we define a cutoff vector, $\p = (p_c)_{c \in C}$, such that college $c$'s cutoff $p_c$ is the lowest score among $c$'s matched applicants, $\mu(c)$, if its capacity is reached, and zero otherwise. When an applicant's score at college $c$, $s_c$, satisfies $s_c\geq p_c$,  the college is \textit{feasible} to her. An outcome is stable if everyone is matched with her most-preferred feasible college. DA ensures stability with respect to submitted ROLs as well as market clearing in the sense that no college admits more applicants than its capacity. When we consider a random economy $F^k$ operated by DA, the cutoffs, which depend on applicants' realized types via $\bsigma^k$, are random. We denote random cutoffs in $F^k$ by $\P^k = (P_c^k)_{c \in \C}$.\footnote{Our analysis will also  consider any arbitrary, non-random cutoff vector, $\p$, that need not clear the market.  We then  let applicants {\it demand} their highest-ranked feasible colleges given such $\p$ in their ROLs. }

\section{Analysis of Robust Equilibria}\label{sec:theorems}

    To accommodate the types of dominated strategies documented in empirical studies, we introduce the following solution concept:\footnote{A number of authors adopted a similar $\epsilon$-based solution concept to analyze approximate equilibrium behavior (see \cite{kalai:04}, \cite{deb/kalai:15}, \cite{azevedo/budish:15}, and \cite{che/tercieux:15b}, for instance).}

    \begin{defn} \label{def:robeq}
        A strategy  $\bsigma$ forms a \textbf{robust equilibrium} if, for any $\e>0$, there exists $K\in \mathbb{N}$ such that, for each $k>K$, $\bsigma^k$ is an {interim $\e$-Bayes Nash equilibrium} of a $k$-random economy $F^k$---namely, $\bsigma$ gives each applicant within $\e$ of the highest possible (supremum) payoff she can receive from any strategy when all the others employ $\bsigma$.
    \end{defn}
    
    Note that robust equilibrium relaxes the exact Bayesian Nash solution concept by allowing for mistakes that are payoff insignificant in a large economy.\footnote{In this sense, our concept of robustness differs from another notion of ``robustness,'' or ``incentives in the large'' (see  \cite{che/kojima:10}, \cite{liu/pycia:11},  \cite{azevedo/budish:15}, \cite{che/tercieux:15b}, and \cite{pycia:19}, for example).  This latter concept refers to the property of a mechanism (rather than a solution concept) which provides asymptotic incentives for agents to report truthfully, even though truth-telling may not be an exact equilibrium behavior in a finite economy.  By contrast, the current notion permits possible deviations from truth-telling even when it is a dominant strategy.} Such relaxation is necessary to accommodate mistakes in a finite economy. If cutoffs were known with certainty, a non-TT strategy, such as ranking only the most preferred feasible college with respect to the known cutoffs, may do just as well as TT in the continuum economy. However, such a strategy may not be optimal in a finite random economy because cutoffs are random, and a non-TT strategy may result in a payoff loss with a positive probability.\footnote{The distinction between fixed cutoffs and the cutoffs of a large but finite economy matters. Indeed, suppose that an applicant's score at her best feasible college~$c$ is precisely this college's fixed cutoff $p_c$. Then the applicant can submit an ROL that contains only $c$ and still suffer no payoff loss. Yet, no matter how large the economy is, submitting only $c$ would entail a loss because $c$'s cutoff is random and can be above her score with positive probability.} Hence, we instead require the equilibrium strategies to entail insignificant payoff loss in any sufficiently large but finite economies.
    
    Below we investigate the implications of this relaxation. In particular, we ask:  Does the robustness concept imply that most applicants report their preferences truthfully? Our first result shows that this is not the case. In fact, a robust equilibrium need not satisfy an even weaker notion of TT, weak truth-telling (WTT), which allows applicants to truncate their truthful ROL $\rho(\theta)$ from below. 
    To show that WTT may not hold, we construct a robust equilibrium in which all but a vanishing fraction of applicants adopt \textit{non}-WTT strategies.
    
    To begin, we define a \textit{stable-response strategy} (SRS) against an arbitrary, non-random cutoff vector ${\p}$ as \textit{any} strategy whereby an applicant demands the most preferred feasible college given ${\p}$ (i.e., she ranks that college ahead of all other feasible colleges). The set of SRSs is typically large. She could skip infeasible colleges, rank them ahead of feasible ones, or flip their order relative to her true preferences. 
    For a specific example, suppose that $\C=\{1,2,3,4\}$, an applicant's true preference order is $1\-2\-3\-4$, and 2, 3 and 4 are feasible for her. Then, out of the 65 ROLs she can choose from, 21 are SRS, including ROLs 2-4-3-1, 2-4-1-3, 2-1-4-3, 1-2-4-3, and 2-4-3 which do not even respect the true preference order among the ranked colleges. For each type $\t=(\bu,\s)$, there exists at least one SRS that violates WTT.\footnote{This can be shown as follows. If an applicant's most preferred college is infeasible (i.e., its cutoff at ${\p}$ is above the applicant's score at that college), then she can simply drop that college and rank order the remaining colleges truthfully. The resulting strategy is SRS but fails WTT. If an applicant's most preferred college is feasible, then she can rank that college at the top of her ROL but rank the remaining colleges untruthfully (in relative rankings). Again, the resulting strategy is an SRS but violates WTT.\label{fn:WTT}}

    Next, we consider a set of applicant types:
    $$\T^{\delta}(\p):=\left\{(\bu,\s)\in \T \mid  \exists j \in \C \mbox{ s.t. } |s_j - {p}_j|\le \delta\right\}.$$
   Although we define this set for arbitrary cutoffs $\p$, we are interested in $\T^\delta(\overline\p)$, where  $\overline\p$ are the cutoffs in the limit economy when all applicants report truthfully; we formally define $\overline\p$ in Appendix~\ref{app:uniform-convergence}. $\T^\delta(\overline\p)$ collects all applicants whose score is close to $\overline\p$  for at least one college. These applicants are required to be TT, but all other applicants---an arbitrarily large fraction of applicants for small $\delta$---adopt non-WTT SRSs against $\overline\p$ with an arbitrarily high probability. Theorem~\ref{p1} below shows that these strategies constitute a robust equilibrium.  When the cutoffs in the sequence of random economies where applicants play such strategies converge to $\overline{\p}$, the loss from adopting the SRSs vanishes, hence Theorem~\ref{p1}. It turns out that proving cutoff convergence is non-trivial, which we will revisit in Section~\ref{sec:asymptotics}. 
    
    \begin{thm} \label{p1} There exists ${\p} \in [0,1]^C$ such that, for any arbitrarily small $(\delta, \gamma)\in (0,1)^2$, the following strategy forms a robust equilibrium: in each $k$-random economy,
    	\begin{itemize}
    		\item all applicants with types $\t\in \T^{\delta}(\p)$ play TT and
    		\item all applicants with types $\t\not\in \T^{\delta}(\p)$ randomize between TT (with probability $\gamma$) and an SRS strategy against ${\p}$ that violates both WTT and TT (with probability $1-\gamma$). 
    	\end{itemize}
    \end{thm}

    Since  $(\delta, \gamma)$ is arbitrary, the following striking conclusion emerges.

    \begin{cor} \label{cor1}
        There exists a robust equilibrium in which every applicant plays a non-WTT strategy (hence, a non-TT strategy) with probability arbitrarily close to one.
    \end{cor}

    To the extent that a robust equilibrium is a reasonable solution concept, Theorem \ref{p1} implies that we should not be surprised to observe a non-negligible fraction of participants making ``mistakes''---more precisely, playing dominated strategies---even in a strategy-proof environment. Importantly, even among the colleges that an applicant includes in her ROL, the order may not respect her true preferences. This result raises some concerns about the empirical methods relying on WTT---any particular strategy relying on ROL data for that matter---as an identifying restriction.

    If strategic mistakes undermine the prediction of applicant behavior, do they also undermine the stability of the outcome?  This is an important question on two accounts. First, if mistakes jeopardize stability in a significant way, the rationale for using DA---to ensure a stable matching---should be called into question.  
    Second, stability is widely used as an empirical identification assumption \citep[see][for instance]{Fox(2009)Palgrave,Agarwal2015, Fox_Bajari:2013,Chiappori-Salanie(2015)JEL,Fack-Grenet-He(2015), HSS}. 
    Our second theorem shows that mistakes captured by robust equilibrium leave the stability property of DA largely unscathed. We begin by defining a notion of approximate stability in large economies.

    \begin{defn} 
    	 A strategy $\bsigma$ is \textbf{asymptotically stable} if the fraction of applicants matched with their most preferred feasible colleges (given the realized cutoffs) in economy $F^k$ under $\bsigma^k$ converges in probability to one as $k\to \infty$.\footnote{More formally, we require that for any $\e>0$ there exists $K\in \mathbb{N}$ such that in any $k$-random economy with $k>K$, with probability of at least $1-\e$, at least  a fraction $1-\e$ of all applicants are matched with their most preferred feasible colleges given the equilibrium cutoffs $\P^k$.}
    \end{defn}

    We now state the main theorem:

    \begin{thm}\label{p2} 
        Any regular robust equilibrium is asymptotically stable.
    \end{thm}

    While Theorem \ref{p2} already provides some justification for stability as an identification assumption for a sufficiently large economy, a question arises as to whether the concept of robust equilibrium would predict the same outcome as would emerge had all applicants reported their preferences truthfully.  Our answer is in the affirmative:\footnote{This result is reminiscent of the upper hemicontinuity of Nash equilibrium correspondence \cite[see][for instance]{fudenberg/tirole:91}. The current result is slightly stronger, however, since it implies that a sequence of $\epsilon$-BNE (which is weaker than BNE) converges to an exact BNE as the economy grows large.}

    \begin{cor} \label{cor2} 
        For a sequence of economies $\{F^k\}_k$, consider two sequences of outcomes: $\{\mu^k_{\bsigma}\}_k$, generated by any regular robust equilibrium strategy $\bsigma$, and $\{\mu^k_{{TT}}\}_k$, generated by TT. The fraction of applicants who receive their TT outcome while adopting $\bsigma$ (i.e., $\mu^k_{\bsigma}(\theta) = \mu^k_{{TT}}(\theta)$) converges in probability to one.
    \end{cor}

    Although Theorem \ref{p1} questions TT as a \emph{behavioral prediction}, Corollary \ref{cor2} supports TT as a means for predicting an \emph{outcome}. In this sense, the corollary validates the vast theoretical literature on DA that assume truth-telling.  This result also suggests that when one evaluates the \emph{outcome} of a counterfactual scenario involving DA, one can simply assume that applicants report their preferences truthfully in that scenario, as we will do in Section~\ref{sec:implications}.

    Taken together, our two theorems provide very different implications for the behavior and outcome under DA.  On the one hand, the behavioral prediction exhibits multiplicity and possibly a drastic departure from truth-telling. On the other, the prediction in terms of outcome is virtually unique, and the outcome is virtually the same as if all applicants reported their preferences truthfully.   This latter finding should ultimately be reassuring about the performance of DA. 
    
    The next section describes proof sketches of Theorems~\ref{p1} and \ref{p2} and highlights the challenges in proving asymptotic stability. We also present novel results on cutoff convergence that significantly extends the existing results in \cite{azevedo/leshno:16} and may be of interest in their own right. A reader who is more interested in an in-depth discussion of the implications for empirical studies may skip to Section~\ref{sec:implications}.

\section{Proof Sketches: Cutoff Convergence and Asymptotic Stability}\label{sec:asymptotics}

    To prove Theorems~\ref{p1} and \ref{p2}, we need to establish that the cutoffs in random economies, with applicants being able to unilaterally deviate, are still concentrated around those arising from TT. A crucial step towards this goal is to re-establish applicants' price-taking in large economies. Before sketching out our proof, we present an example to illustrate that such price-taking cannot be taken for granted, even in arbitrarily large economies. 
        
    \begin{example}\label{eg:1}
\normalfont

We return to the example discussed in the introduction. Recall that there are three applicants, ($\apa, \apb, \apc$), and three colleges, ($\ca, \cb, \cg$), with one seat each and that applicant preferences and college priority rankings are:
\begin{align*}
\text{Applicant preferences:} &&    \apa:\ \cb\-\ca\-\cg && \apb:\ \ca\-\cb\-\cg && \apc:\ \ca\-\cb\-\cg\\
\text{College priority ranking:}   &&    \ca:\ \apa \- \apc\- \apb && \cb:\ \apb \- \apc\- \apa && \cg: \text{arbitrary}
\end{align*}
Applicant $\apa$'s score are 0 at $\ca$ and 1 at $\cb$; $\apb$'s are 1 and 0, respectively; and $\apc$'s are 0.5 at both. To recap, we have shown that there is a Nash equilibrium where $\apa$ and $\apb$ adopt TT and $\apc$ submits $\cg$ only. The DA outcome is ($\apa\-\cb, \apb\-\ca, \apc\-\cg$), which is unstable. Yet, applicant~$\apc$'s deviation to TT activates a rejection chain that leads to no change in $\apc$'s assignment but a change in the cutoffs of $\ca$ and $\cb$, hence there is no price-taking.

One may hope that applicants become price-takers in a sufficiently large economy. It is not necessarily the case.  To see this, we consider ($\apa$, $\apb$, $\apc$) as applicant types and replicate the economy $k$-fold so that we have $k$ applicants of each type and $k$ seats at each college. Consistent with the baseline economy, at each college, the top-ranked applicant types have scores in $[2/3,1]$, the middle-ranked types are in $[1/3,2/3)$, and the bottom-ranked types are in $[0,1/3)$. Within each applicant type, the scores are drawn independently.  Note that this construction violates the full support condition assumed for our random $k$-economy $F^k$, a point we will return to later.  

Consider as before a candidate equilibrium in which applicants with types $\apa$ and $\apb$ adopt TT and applicants with type $\apc$ list only $\cg$ in their ROLs.  The outcome is again unstable: this is seen by the fact that the cutoffs for colleges  $\apa$ and $\apb$ lie below $1/3$, and converge to $1/3$ as $k\to\infty$, and thus are well below the scores type-$\apc$ applicants have for these colleges.  Yet, the unstable outcome is supported as---exact and hence robust---equilibrium no matter how large $k$ is.  To see this, suppose a type-$\apc$ applicant deviates to TT. It will knock off the type-$\apb$ applicant with the lowest score at $\ca$ from $\ca$, who will then knock off the type-$\apa$ applicant with the lowest score at $\cb$ from $\cb$.  The latter in turn knocks off a type-$\apb$ applicant with the second-lowest score at $\ca$ from $\ca$, who then knocks off the second-lowest score type-$\apa$ applicant, and so on.  The rejection chain continues until {\it all} applicants with types $\apa$ and $\apb$ are knocked off from their top choices, and the deviating type-$\apc$ applicant ends up with $\cg$. The process illustrates the failure of price-taking in a spectacular manner, as a single applicant's deviation triggers discontinuous jumps of cutoffs from below $1/3$ to values above $2/3$.  \qed
    \end{example}

One expects that this failure of price-taking behavior can be avoided with a sufficient ``smoothness'' in the economy.  This is indeed the case with our economy where  $\eta$ has full support and the types in $\eta^k$ are i.i.d.\ from $\eta$.  Given this, any regular strategy profile leads to the necessary smoothness in cutoff responses when an applicant deviates to TT unilaterally.  Although this seems intuitive, it is not trivial to establish the price-taking behavior formally.  

To this end, we consider any arbitrary regular strategy profile $\bsigma$. For any random $k$-economy $F^k$, the truncated profile $\bsigma^k$ induces random cutoffs denoted by $\P^k_{(0)}(\bsigma) \in [0,1]^{C}$. Now suppose each applicant $i\in \mathbb{N}$ unilaterally deviates to TT.  The corresponding truncated profile for $F^k$, denoted $\bsigma^k_{(i)}$, induces another set of random cutoffs $\P^k_{(i)}(\bsigma) \in [0,1]^{C}$.\footnote{Note that if $i$ is not in the $k$-economy $F^k$, i.e., $i>k$, then $\bsigma^k_{(i)} = \bsigma^k$ and $\P^k_{(i)}(\bsigma) = \P^k_{(0)}(\bsigma)$.} 

To establish the price-taking behavior in large economies, one needs to show that the cutoff profiles $\P^k_{(i)}(\bsigma)$ for each $i\in \mathbb{N}$ become  arbitrarily close to $\P^k_{(0)}(\bsigma)$ uniformly with high probability as $k\to \infty$.  Since $\bsigma$ could be asymmetric across all applicants, each profile  $\bsigma^k_{(i)}$ is potentially distinct, making the resulting cutoff profiles  $\P^k_{(i)}$ distinct across all $i\in \mathbb{N}\cup\{0\}$. Hence, price-taking behavior in a sequence of economies would hold if the infinite family of cutoffs $(\P^k_{(i)})_{i\in \mathbb{N}\cup\{0\}}$ converges uniformly almost surely to some cutoff vector $\overline {\p}$. It turns out that such convergence may not hold for an arbitrary asymmetric $\bsigma$:  the example in Appendix~\ref{app:ex-nonconvergence} shows that the cutoffs cycle across  distinct values (instead of converging). Nevertheless, we can show that there is always a \textit{subsequence} of economies such that the cutoffs induced by $\bsigma$ in that subsequence converge to some deterministic vector.  This turns out to be sufficient for our purposes. On the other hand, if $\bsigma$ is symmetric, all our results can be stated for the whole sequence of economies.

\def\bbsigma{;\bsigma}

	\begin{prop}\label{claim:uniform-convergence of P} 
		Let $\bsigma$ be any $\gamma$-regular strategy profile. Then there exists a subsequence $\left\{F^{k_{\ell}}\right\}_{\ell}$ such that
		$$
			\sup_{i\in \mathbb{N}\cup\{0\}} \lVert \P^{k_{\ell}}_{(i)}(\bsigma) - \overline\p(\bsigma) \rVert \overset{a.s.}{\longrightarrow }0 \mbox{ as }\ell \to \infty,
		$$
		where $\lVert\cdot\rVert$ denotes the sup norm; i.e., for any $\bm{x},\bm{x}'\in [0,1]^{|C|}$, $\lVert \bm{x}-\bm{x}'\rVert := \sup_c |x_c - x_c'|$.
		If $\bsigma$ is symmetric across all applicants, then the uniform almost-sure convergence holds  for the entire sequence of economies $\left\{F^{k}\right\}_{k\in \mathbb{N}}$ .
	\end{prop}

Proposition~\ref{claim:uniform-convergence of P} is interesting in its own right as it generalizes AL's result on cutoff convergence with truth-telling applicants (part 2 of their Proposition~3). First, while the existing literature shows convergence of cutoffs when applicants adopt TT,  the second part of Proposition~\ref{claim:uniform-convergence of P} establishes convergence for {\it any} regular symmetric strategies.   We also allow for unilateral deviations and show that convergence of cutoffs is uniform over an infinite family of cutoffs resulting from such deviations. This will prove useful for our analysis. In fact, the first part establishes the same uniform convergence albeit on a subsequence of economies.  Second, we do not require that $\sum_{c=1}^C S_c < 1$ (over-demanded systems); dropping this requirement may be practically important, as many matching markets, such as school choice, are not overdemanded. We also do not need that $\partial \D(\overline{\p}(\bsigma))$ is invertible, as was required by AL, where $\D(\overline{\p}(\bsigma))$ is the vector of demand for colleges at cutoffs~$\overline{\p}(\bsigma)$, which is to be formally defined below.  It is not clear whether this property holds under an arbitrary regular symmetric strategy $\bsigma$.

Theorem~\ref{p1} builds on the second part of Proposition~\ref{claim:uniform-convergence of P}.  Recall that the strategy profile we construct for the theorem is regular and symmetric across all applicants. Hence, the cutoffs resulting from the constructed strategies as well as those resulting from the most profitable unilateral deviation (namely, to TT) all converge to some deterministic cutoff vector $\overline{\p}$.
Further,  the constructed strategies guarantee that all applicants adopt SRS against $\overline{\p}$ with high probability as the economy grows large. This means that the constructed strategies must form a robust equilibrium, and in that equilibrium,  with high probability, all applicants must obtain their stable matches. Since our limit economy admits a unique stable matching, this means that  $\overline{\p}=\overline{\p}(\bm{\rho})$,  the cutoffs that would emerge in the limit if all applicants employed TT. In other words,  the cutoffs from the constructed strategy profile converge to $\overline{\p}(\bm{\rho})$, even though almost no one employs TT. These observations lead to Theorem~\ref{p1}.

Meanwhile, Theorem \ref{p2} crucially uses the first part of Proposition~\ref{claim:uniform-convergence of P}, namely the uniform convergence on a subsequence of economies. 
Fix any regular robust equilibrium  $\bsigma$.   Suppose by way of contradiction that $\bsigma$ is not asymptotically stable.  Then, there must be a subsequence of economies such that, with non-vanishing probability, a non-vanishing proportion of applicants do not get their favorite feasible colleges given the prevailing cutoffs along that subsequence. Proposition~\ref{claim:uniform-convergence of P} then ensures that there is a cutoff vector $\overline{\p}(\bsigma)$ and a further subsequence (of the subsequence) of economies such that the  cutoffs induced by $\bsigma$ converge to $\overline{\p}(\bsigma)$ along that sub-subsequence. Given the asymptotic instability, we can then easily identify a  set of applicants who would suffer discrete payoff losses from their matches given $\overline{\p}(\bsigma)$.  Recall further from uniform convergence that if any such applicant were to deviate to TT, it will not alter the cutoffs much.  This in turn implies that the applicant would enjoy a discrete payoff gain from the deviation. Then, $\bsigma$ could  not have been a robust equilibrium, delivering  a desired contradiction.

    We close this section by sketching the proof of Proposition~\ref{claim:uniform-convergence of P}.  Recall the cutoffs are defined to clear markets.  Hence, to study how such cutoffs behave in  large economies, we must first study how the demand system behaves in large economies.  To this end, we consider the ``empirical'' demand induced by $\bsigma$ for each college $c$ at any fixed cutoffs $\p$:
    \begin{align*}
  		D^{k}_{c}(\p\bbsigma) := \frac{1}{k} \sum_{j = 1}^{k} \I \left\{c\in \arg\max_{\text{w.r.t. } R_{j}}\left\{c'\in C: s_{j,c'} \geq p_{c'}\right\} \right\},
	\end{align*}
    where $\arg\max_{\text{w.r.t. } R_{j}}$ picks the highest-ranked college in  $R_j$ from the set of feasible colleges $\left\{c'\in C:s_{j,c'}\geq p_{c'}\right\}$ and $\I\{\cdot\}$ is an indicator function.  In words, $D^{k}_{c}(\p)$ is the fraction of applicants in economy $F^k$ for whom $c$ is the best feasible college, given a fixed strategy $\bsigma$, which we suppress in the notation below, and fixed cutoffs $\p$. These cutoffs are not necessarily market-clearing. The demands for all colleges form the vector $\D^{k}(\p)$, or $\D^{k}_{(0)}(\p)$. We are interested in an infinite family of demand systems $\{\D^{k}_{(i)}(\p)\}_{i\in \mathbb{N}\cup \{0\}}$ which also include the demand vectors that result from a unilateral deviation of applicant $i$ to TT. 
    
    Note that, the demand system thus defined is random, but, for each $\p$,  as the economy grows large, $\D^{k}_{(i)}(\p)$ converges pointwise to its expectation $\overline\D^{k}_{(i)}(\p)$ \citep{mcdiarmid:89}.    
    Meanwhile, the (deterministic)  functions $\overline\D^{k}_{(i)}(\p)$ are Lipschitz-continuous and, by the Arzela-Ascoli theorem, there is a subsequence $\overline\D^{k_\ell}_{(i)}(\p)$ that converges to some Lipschitz-continuous function $\overline{\D}(\p)$.  Then, using an argument in the spirit of the Glivenko-Cantelli theorem, we show that the random demands $\D^{k_{\ell}}_{(i)}(\cdot)$ converge uniformly (with respect to both its argument and $i$) to $\overline\D(\cdot)$ almost surely. Note that if $\bsigma$ is symmetric, the above convergence results apply to the whole sequence.
    
    Having established the convergence of random demand functions, we then show by induction on the steps of DA that the random cutoffs $\P^{k_\ell}_{(i)}$, which clear random demands $\D^{k_\ell}_{(i)}(\cdot)$, converge to $\overline{\p}$, which clears $\overline{D}(\cdot)$.
    To this end, we view $\P^{k_\ell}_{(i)}$ and $\overline{\p}$ respectively as the limiting outcomes of the monotonic cutoff adjustment processes that occur in the DA algorithms of random-$k$ and the continuum economies.  Specifically,  each step~$m$ of the DA proposal/acceptance adjusts cutoff $P^{k_\ell,m}_{(i),c}$ or $p^m_{c}$ to clear the market tentatively. 
    We are interested in the cutoffs arising in each step of the adjustment process because (i) they converge respectively to cutoffs $\P^{k_\ell}_{(i)}$ and $\overline{\p}$, the two key objects in Proposition~\ref{claim:uniform-convergence of P} and (ii) 
    we can bound the difference between the cutoffs from each step. We obtain the upper and lower bounds by defining this adjustment process for both the applicant- and college-proposing versions of DA and using the lattice structure of cutoffs and uniqueness of stable matching in the limit economy \citep[by Theorem~1 in][]{azevedo/leshno:16}.

    We show that, for large enough $k_\ell$, the difference between $\P^{k_\ell,m}_{(i)}$ and $\p^{m}$ is arbitrarily small in each step of DA using an induction argument. For the step $m=1$, the argument relies on the full-support assumption and the regularity of $\bsigma$. For $m>1$, the argument uses the convergence of random demands to limit demand and the Lipschitz continuity of limit demand in cutoffs.

    Summarizing the arguments in the last two paragraphs, we have established that the difference between $\P^{k_\ell,m}_{(i)}$ and $\p^{m}$ is small for all $m$, and that the former converges to $\P^{k_\ell}_{(i)}$, while the latter converges to $\overline{\p}$. Taken together, the difference between $\P^{k_\ell}_{(i)}$ and $\overline{\p}$ is small, delivering the proposition.  
    
    The argument tracking the monotonic tatonnement process, which allows us to prove the uniform convergence without imposing restrictive assumptions, is new in the large market asymptotic analysis and will be useful beyond the current context.

\section{Implications for Market Design Research}\label{sec:implications}

Our theoretical results, along with existing the field evidence, suggest that non-truthful behavior may be widespread but it rarely leads to significant payoff consequences. This observation has implications for empirical market design. In simulated data, we illustrate these implications for two exercises that are common in the literature: {(a) estimation of applicant preferences and (b) analysis of a counterfactual policy.}

\subsection{Estimating Applicant Preferences \label{sec:estimate}}

There are two typical identifying assumptions in the literature for the estimation of applicant preferences.

The first is WTT \citep{Hallsten:2010,Kirkeboen_2012}. Recall that applicant $i$ is WTT if her submitted ROL ranks her most-preferred colleges according to her true preferences, while every unranked college is less desirable to $i$ than any ranked college. Let $\succ_i$ denote the inferred preference relation of $i$. As an example, consider an applicant whose submitted ROL is $c_3$-$c_1$, while there are four colleges available, $\{c_1,c_2,c_3,c_4\}$.  WTT infers that $c_3 \succ_i c_1 {\succ_i c_2,c_4}$.

The second assumption is stability \citep{Akyol-Krishna(2017):EER,Bucarey(2018),Fack-Grenet-He(2015),Combe_Tercieux_Terrier_2022}. An outcome is stable if every applicant is matched with her most-preferred college among the feasible ones. Suppose that the aforementioned applicant has $c_2$ and $c_3$ feasible and is matched with $c_3$. Stability infers $c_3 \succ_i c_2$. In contrast to WTT, stability does not make any inference about infeasible colleges, $c_1$ and $c_4$.

According to our Theorem~\ref{p1}, in a robust equilibrium, applicant~$i$ may not be WTT, in which case preference inference would be incorrect. For example, she may rank more desirable but infeasible $c_4$ arbitrarily. Stability makes inference only about feasible colleges and, according to our Theorem~\ref{p2}, is satisfied asymptotically. Moving from WTT to stability, we gain robustness to untruthfulness but utilize less information. Therefore, the estimation based on stability will be less efficient than WTT, yet less likely to be biased because it uses fewer possibly incorrectly inferred preference relations.
\subsubsection{Monte Carlo Simulations \label{sec:mc_sum}}

We evaluate the performance of WTT and stability in simulated data that resembles the typical college admissions studied in the previous sections. 
There are 12 colleges and 1800 applicants. This matching market is operated by a Serial Dictatorship, a special case of DA, in which colleges rank applicants by an ex-ante known score. 
Applicant preferences follow a conditional logit model. Specifically, applicant~$i$'s utility from being matched with college~$c$ is:\footnote{We keep our Monte Carlo setting as simple as possible to illustrate our main points. It, therefore, ignores some potential challenges in practice. For example, in a real-life setting, factors affecting an applicant’s priority scores may directly enter their preferences. This is true if priority scores depend on an applicant's distance to colleges or academic ability. This may cause issues in the non-parametric identification of applicant preferences when one assumes stability; see \cite{Fack-Grenet-He(2015)} for a detailed discussion on sufficient conditions for identification.}
\begin{align}
u_{i,c} &= \beta_1 \cdot c + \beta_2 \cdot d_{i,c} + \beta_3 \cdot T_{i} \cdot A_{c} + \beta_4 \cdot Small_c + \epsilon_{i,c}, \forall i \text{ and } c, \label{eq:MC_utility}
\end{align}
where $\beta_1 \cdot c$ is college~$c$'s baseline quality; $d_{i,c}$ is the distance from applicant~$i$'s location to college~$c$; {$T_{i}\in \{0,1\}$} is applicant~$i$'s type (e.g., disadvantaged or not); $A_{c}\in \{0,1\}$ is college~$c$'s type (e.g., known for resources for disadvantaged applicants); $Small_c=1$ if college $c$ has a small capacity, 0 otherwise; and $\epsilon_{i,c}$ is a type-I extreme value and i.i.d.\ across $i$ and $c$. 

In the simulations, $T_i$ is equal to one for two-thirds of the applicants whose score is below the median, and we thus call them disadvantaged. 

We consider three data generating processes (DGPs). The first is the \textit{Truth-Telling} (TT): every applicant truthfully ranks \textit{all} colleges. The other two DGPs rely on a simulated cutoff distribution that we calculate from 1000 simulation samples with truth-telling applicants. Specifically, the second DGP is \textit{Payoff Irrelevant Mistakes} (PIM): a fraction of applicants skip colleges with which they would never be matched according to the simulated cutoff distribution. Those never-matched colleges for an applicant are likely to be almost always out of reach to her. Hence, PIM approximates the documented behavior that applicants choose not to apply to colleges at which they have a close-to-zero chance. We expect that stability is satisfied in both TT and PIM. 
The last DGP is \textit{Payoff Relevant Mistakes} (PRM): in addition to skipping those never-matched colleges, applicants may skip some colleges with which they have a low match probability according to the simulated cutoff distribution, leading to some payoff-relevant mistakes or violations of stability. According to our theory, such payoff-relevant mistakes, although rare, can happen in a robust equilibrium of a finite economy. The three DGPs, each of which has 150 simulation samples, are summarized in Table~\ref{tab:mc_setup}.

\begin{table}[h!]
  \centering \footnotesize
  \caption{ROLs and Mistakes in Monte Carlo Simulations}   \label{tab:mc_setup}
    \scalebox{0.875}{\begin{tabular}{rccccc}
    \toprule
          & \multicolumn{5}{c}{Data Generating Processes (DGPs) with Different Applicant Strategies} \\
          \cline{2-6}
          & \multicolumn{1}{c}{Truth-Telling} &       & \multicolumn{1}{c}{Payoff Irrelevant} &       & \multicolumn{1}{c}{Payoff Relevant} \\
          & \multicolumn{1}{c}{(TT)} &       & \multicolumn{1}{c}{Mistakes (PIM)} &       & \multicolumn{1}{c}{Mistakes (PRM)} \\
    \midrule
    \multicolumn{1}{l}{Average length of submitted ROLs} & 12   &       & 7.34    &       & 6.58  \\
    [0.25em]
    \multicolumn{1}{l}{WTT: \textit{Weak Truth-Telling} (\%)$^a$}  & 100     &       & 50    &       & 44 \\
    [0.25em]
    \multicolumn{1}{l}{Matched w/ favorite feasible college (\%)$^b$} & 100   &       & 100  &       & 97  \\
    \bottomrule
    \end{tabular}}
    \begin{tabnotes}
Each entry in the table is an average over the 150 simulation samples for a given DGP. In each sample, there are $1800$ applicants and $12$ colleges with a total of $1500$ seats. 
$^a$An applicant is WTT if she truthfully ranks her top $K_i$ ($1 \leq K_i\leq 12 $) preferred colleges, where $K_i$ is the observed number of colleges ranked by $i$. Omitted colleges are always less preferred than any ranked college. 
$^b$A college is feasible to an applicant, if the applicant's score is above the college's ex-post admission cutoff.
\end{tabnotes}
\end{table}

With the simulated data, we estimate the four unknown parameters, $(\beta_1, \dots, \beta_4)$, in equation~\eqref{eq:MC_utility}. We apply a rank-ordered logit model when assuming WTT and a conditional logit model when assuming stability to estimate four parameters. Appendix~\ref{app:mc} provides more details.

\paragraph{Bias-variance tradeoff.} 
Figure~\ref{Fig:Mis}  illustrates several patterns in the estimation for one of the parameters, $\beta_1=0.3$. When applicants report truthfully, WTT and stability are both consistent but WTT is more efficient (panel~a). However, WTT leads to a biased estimator whenever some applicants are not truthful, i.e., {under the PIM and PRM DGPs} (panels~b--c). In contrast, stability performs well when there are no, or just a few, payoff-relevant mistakes. {These results illustrate a bias-variance tradeoff: from WTT to stability, the variance of the estimator increases  while the bias decreases whenever it exists.}

\begin{mfignotesin}{\label{Fig:Mis} Distribution of Estimates based on Weak Truth-Telling or Stability ($\beta_1=0.3$)}
    {
            \begin{subfigure}{0.3\textwidth}
        \centering
        \caption{\footnotesize DGP: TT}
        \graphique{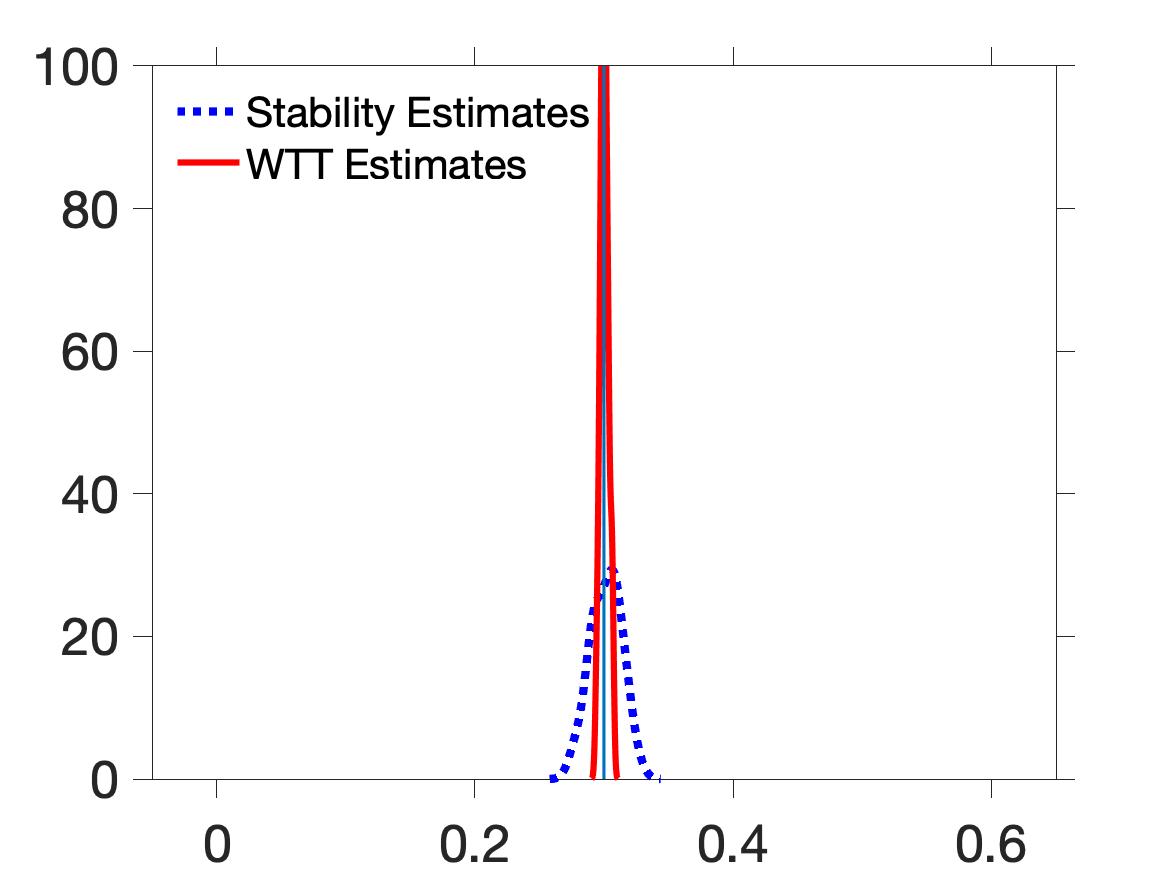}%
        \end{subfigure}%
        \begin{subfigure}{0.3\textwidth}
        \centering
        \caption{\footnotesize DGP: PIM}
        \graphique{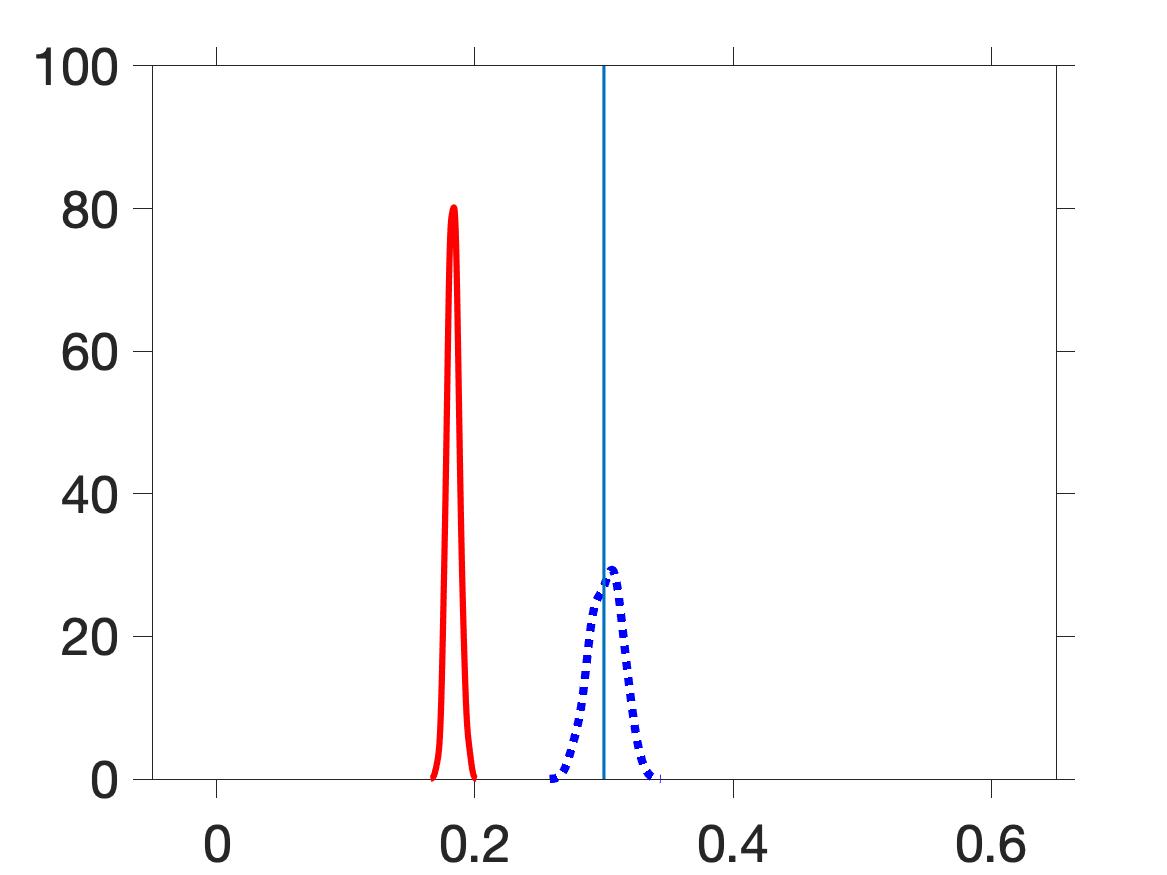}
        \end{subfigure}%
        \begin{subfigure}{0.3\textwidth}
        \centering
        \caption{\footnotesize DGP: PRM}
        \graphique{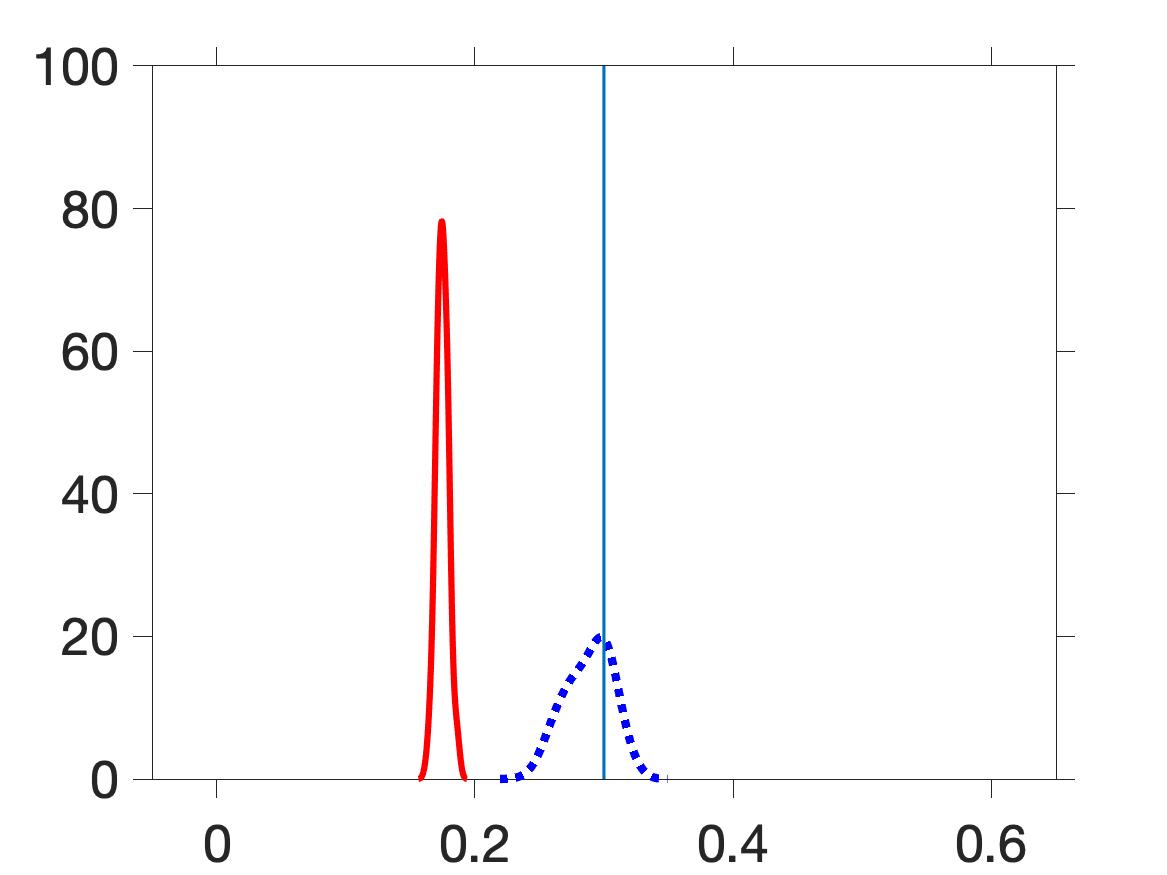}%
        \end{subfigure}%
}
    {The figures focus on the estimates of one parameter ($\beta_1=0.3$) from two approaches, weakly truth-telling (WTT, the solid line) and stability (the dotted line). Each panel uses the $150$ simulation samples given a DGP and reports an estimated density of the estimates based on a normal kernel function. See Table~\ref{tab:mc_setup} for more details on the three DGPs.}
\end{mfignotesin}

\paragraph{Mis-Estimated Preferences.} 
A direct consequence of an inconsistent estimator is the mis-estimation of applicant preferences. As an example, let us consider colleges~10 and 11. For a disadvantaged applicant ($T_i=1$) with an equal distance to these two colleges, the true probability that she prefers college~11 to college~10 is $0.91$ (the straight, dashed line in Figure~\ref{fig:reversal}).  Using the logit formula, we calculate the same probability based on the two sets of estimates, and Figure~\ref{fig:reversal} presents the average across the 150 samples given an estimation approach and a DGP. Clearly, WTT produces significant biases in PIM and PRM, while stability only leads to a small bias in PRM. 

\begin{mfignotesin}{\label{fig:reversal} True and Estimated Probabilities That an Applicant Prefers College~11 to College~10}
    {
        \begin{subfigure}{0.45\textwidth}
        \centering
        \graphique{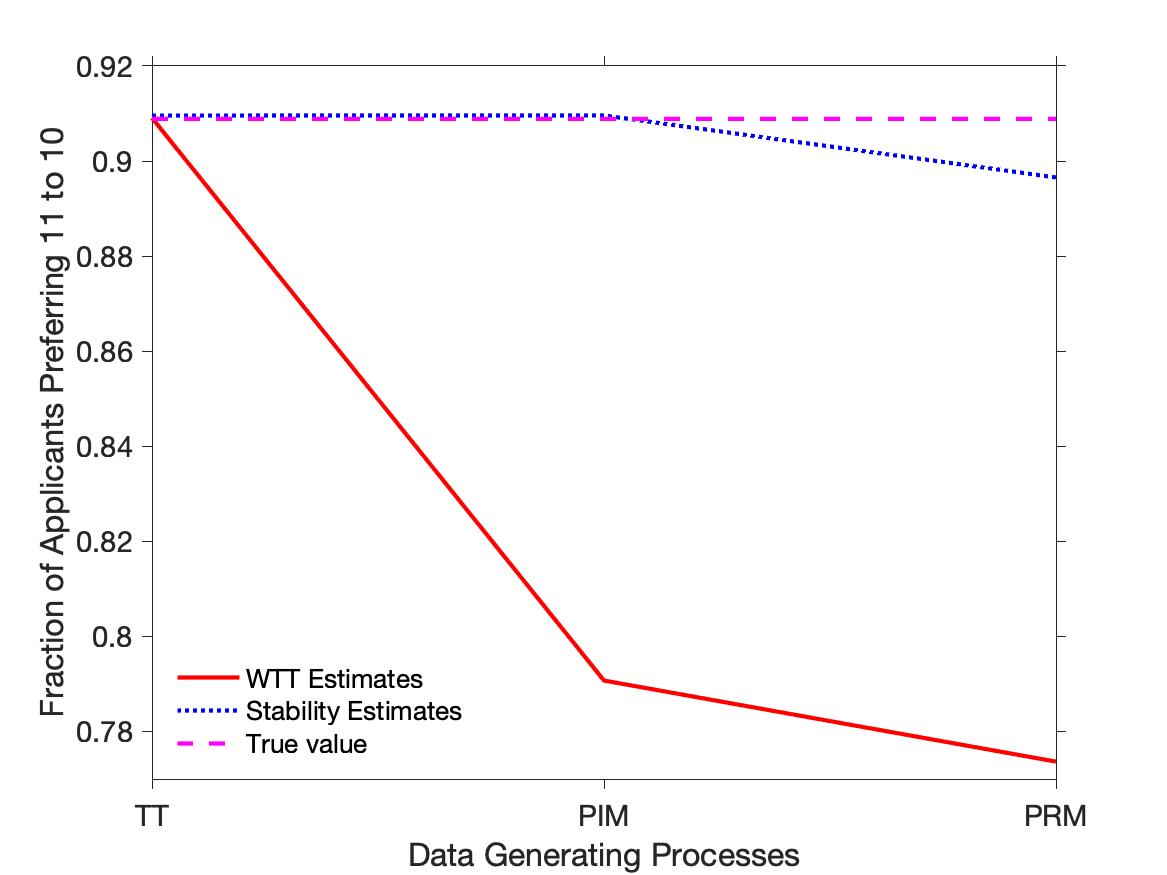}%
        \end{subfigure}%
    }
    {The figure presents the probability that a disadvantaged applicant ($T_i=1$), with an equal distance to both colleges, prefers college~11 to college 10. The true value is $0.91$ (the straight, dashed line). With the logit formula, we calculate the probability based on the WTT-based estimates, and the solid line presents the average over the 150 simulation samples in each DGP. Similarly, the dotted line describes those from the stability-based estimates.}
\end{mfignotesin}

\subsection{Counterfactual Analysis}
Making policy recommendations based on counterfactual analysis is one of the main objectives of market design research. Our theoretical results have some implications for this objective too.

The literature has two types of approaches to counterfactual analysis. The first is based on \textit{submitted ROLs}. See, for example, the analysis of the National Resident Matching Program by \cite{roth/peranson:99} and kindergarten allocation in Estonia by \cite{Veski/etal:2016}. It is assumed that submitted ROLs under the existing policy are true ordinal preferences and that an applicant will submit the same ROL under the counterfactual policy. Our Theorem~\ref{p1} implies that this assumption need not hold in a robust equilibrium. 

In the second type of approaches, the researcher uses estimated preferences and let every applicant submit a truthful ROL under the counterfactual policy. Assuming truth-telling in the counterfactual is justified by Corollary~\ref{cor2}, as any regular robust equilibrium leads to an asymptotically stable matching that is well approximated by the stable matching from truthful reporting. However, this approach crucially relies on the preference estimates being unbiased, because biased estimates will only lead to a misleading prediction about the counterfactual. Section~\ref{sec:estimate} has presented the two possible assumptions for preference estimation, WTT and stability. It is therefore important to choose the appropriate one.

As an illustration, we use the Monte Carlo simulations in Section~\ref{sec:estimate} and consider a counterfactual policy in which applicants with $T_i=1$ are given priorities over those with $T_i=0$, while applicants of the same type are still ranked according to their scores. The mechanism is still DA in which everyone can rank all colleges.

\subsubsection{Performance in Monte Carlo Simulations}
Recall that we have 150 samples in the simulations for each DGP (TT, PIM, or PRM). Additionally, for each DGP, we generate the true outcome under the counterfactual as a benchmark.  That is, we assume applicants potentially make mistakes under the counterfactual policy as they do under the current policy. 

We focus on the three approaches to counterfactual analysis: submitted ROLs, the WTT-based estimation, and the stability-based estimation. We calculate how each approach perform in terms of predicting the new policy's effects on outcomes and on welfare.

\paragraph{Mis-predicted Cutoffs.}
An informative statistic of an outcome is college cutoffs. Figure~\ref{Fig:CF_cutoff} shows, given each DGP, how the three approaches mis-predict cutoffs under the counterfactual policy. For each college, indexed from 1 to 12, we calculate the average difference between the predicted cutoffs and the true cutoffs across the 150 simulation samples. 

\begin{mfignotesin}{\label{Fig:CF_cutoff} Comparison of the Three Approaches: Biases in Predicted Cutoffs}
    {
        \begin{subfigure}{0.32\textwidth}
        \centering
        \caption{\footnotesize DGP: TT}
        \graphique{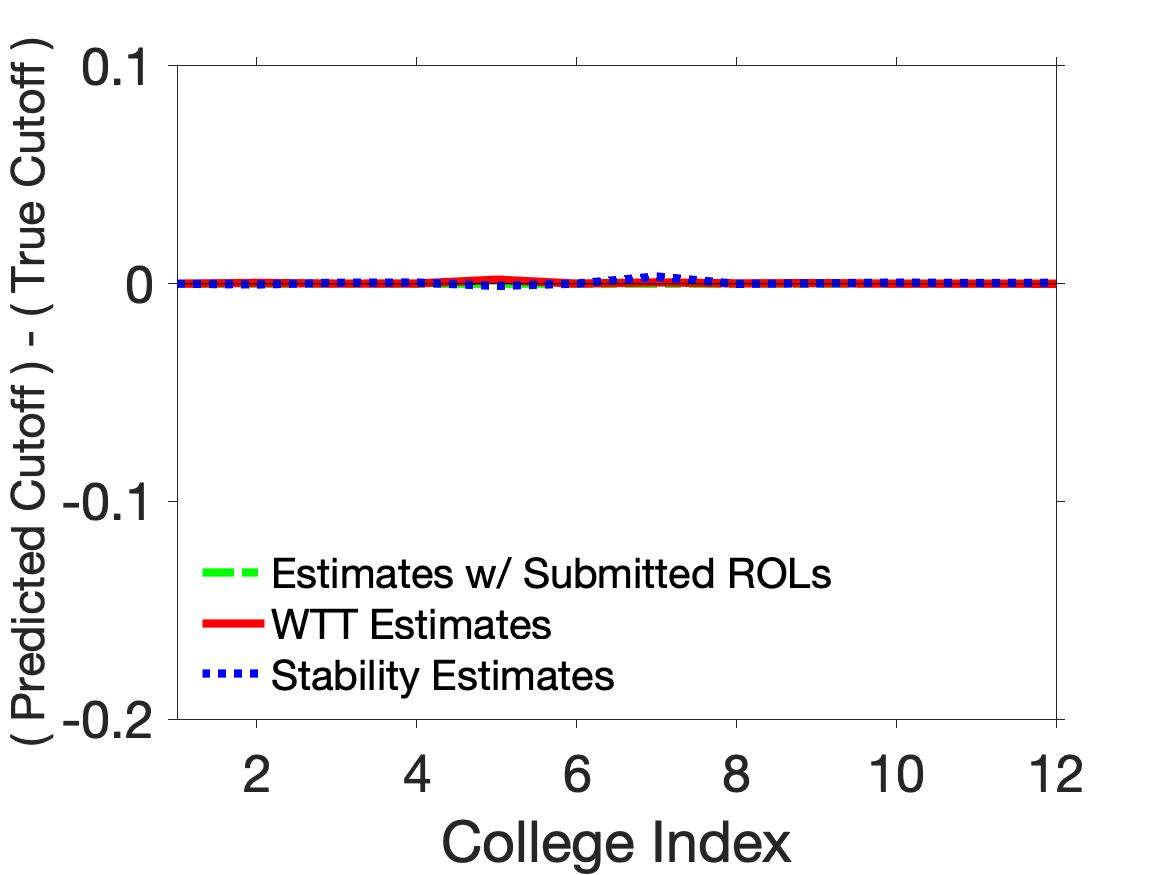}%
        \end{subfigure}%
        \begin{subfigure}{0.32\textwidth}
        \centering
        \caption{\footnotesize DGP: PIM}
        \graphique{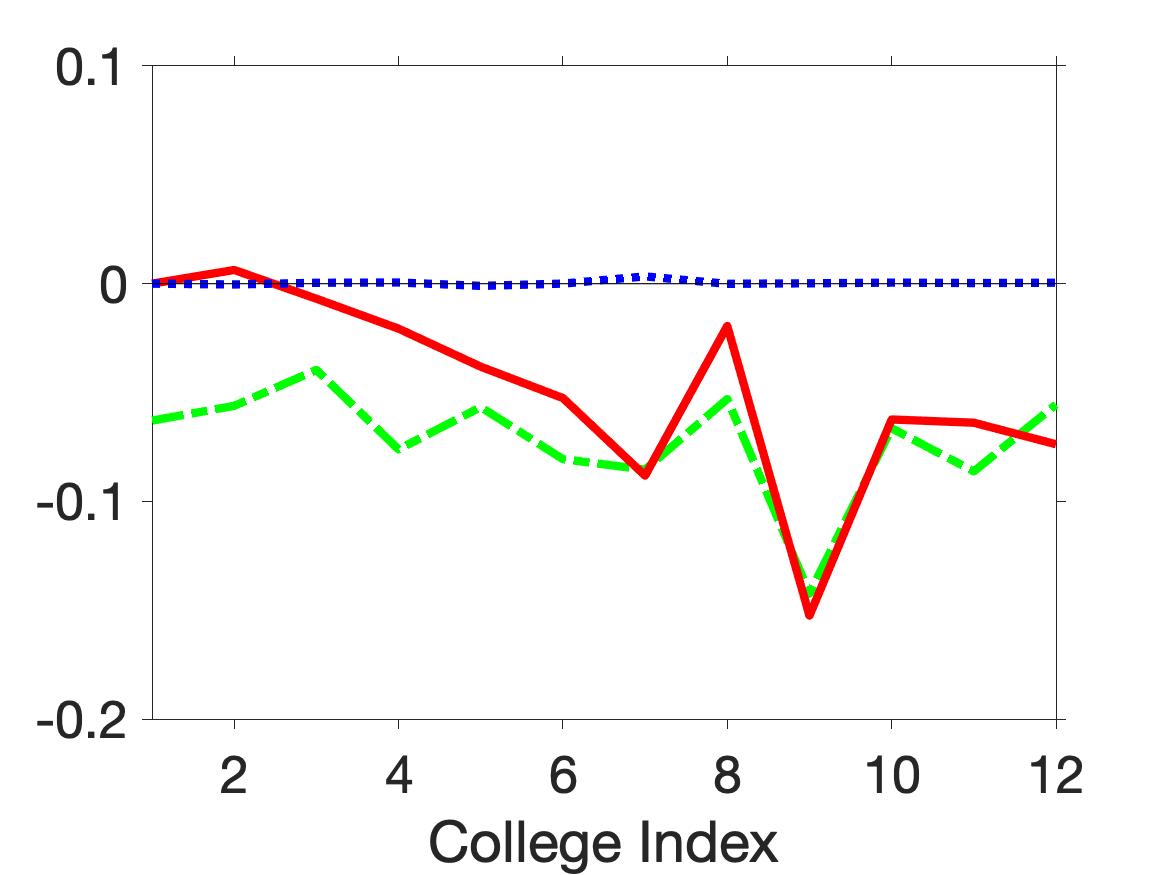}
        \end{subfigure}%
        \begin{subfigure}{0.32\textwidth}
        \centering
        \caption{\footnotesize DGP: PRM}
        \graphique{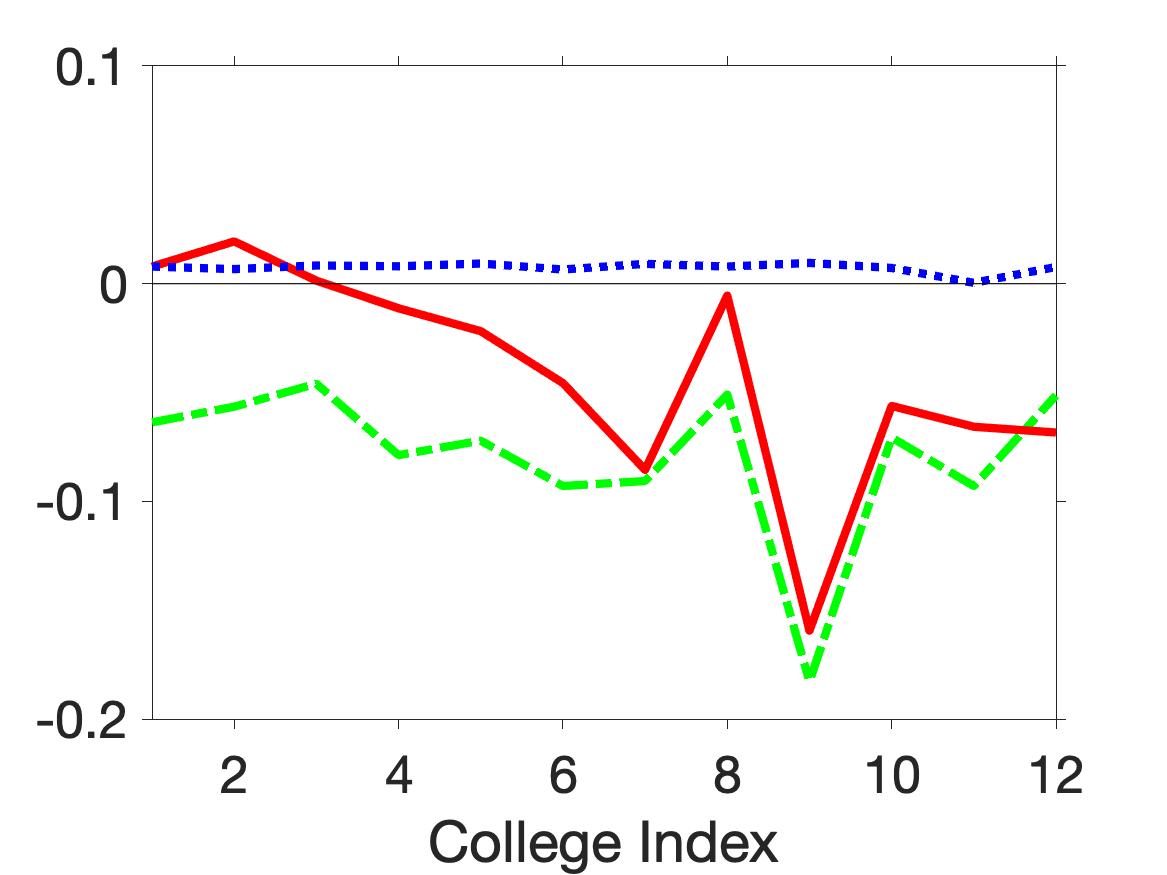}%
        \end{subfigure}%
}
    {In a given DGP, each panel presents how the predicted cutoffs from each approach differ from the true ones that are simulated based on the actual behavior (i.e., the true preferences with possible mistakes). Given a DGP, we simulate the colleges' cutoffs following each approach and calculate the mean deviation from the true ones.}  
\end{mfignotesin}

In panel~(a), the DGP is TT, and thus the submitted ROLs coincide with true ordinal preferences. Consequently, the predicted cutoffs from the submitted-ROLs approach are the true ones. The other two approaches also lead to almost the same cutoffs.

In panel~(b), which corresponds to DGP PIM, only the stability-based estimation is consistent, and indeed it has the smallest mis-predictions relative to the other two. As applicants tend to omit popular colleges, which have higher indices in our setting, from their submitted ROLs in this DGP, the approaches based on WTT and submitted ROLs systematically underestimate the demand for these colleges and thus their cutoffs. 

When the DGP contain payoff-relevant mistakes (PRM, in panel c), none of the approaches is unbiased. However, the stability-based estimates seem to have a negligible mis-prediction compared to the other two.

\begin{mfignotesin}{\label{Fig:CF_AA} Three Approaches to Counterfactual Analysis: Disadvantaged Applicants $T_i=1$}
    {
        \begin{subfigure}{0.475\textwidth}
        \centering
        \caption{\footnotesize Mis-predicted Match}
        \graphique{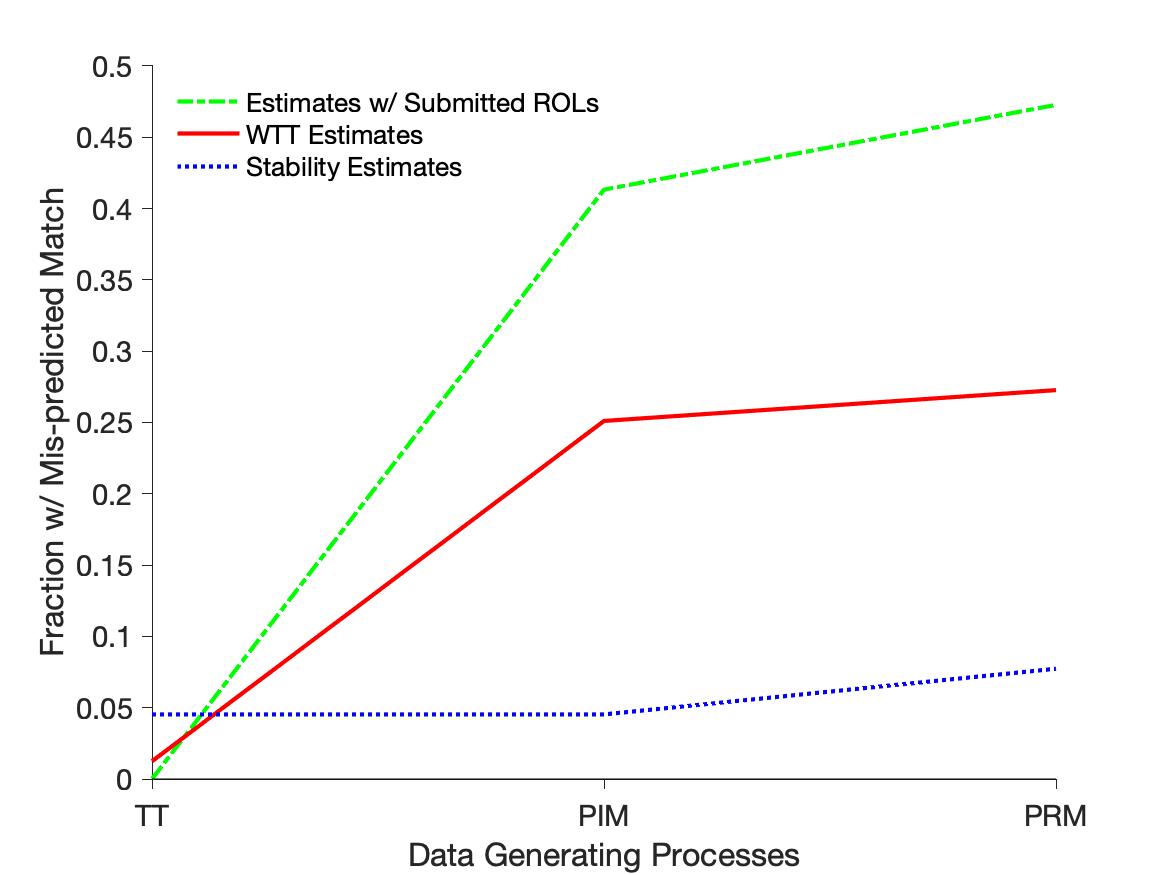}%
        \end{subfigure}%
        \begin{subfigure}{0.475\textwidth}
        \centering
        \caption{\footnotesize Predicted Welfare Effects}
        \graphique{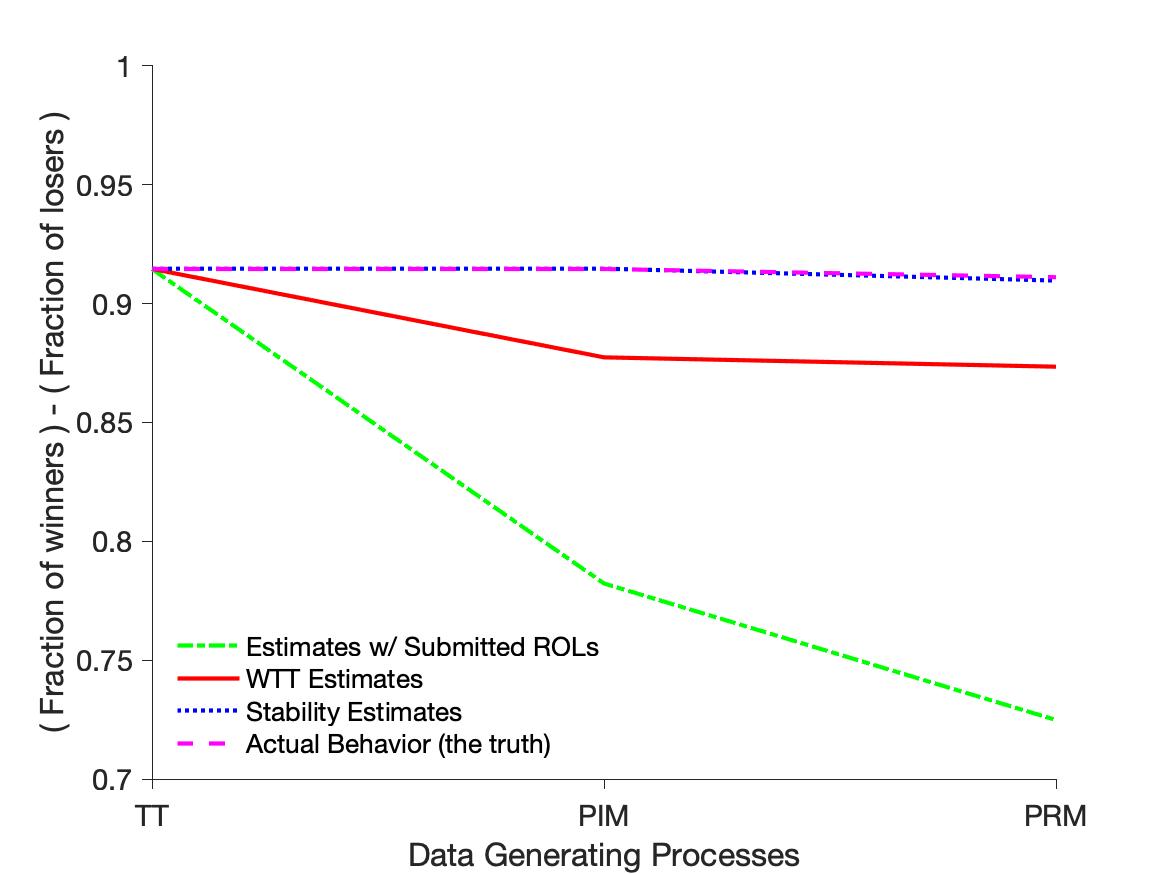}
        \end{subfigure}%
    }
    {The figure shows the averages among $T_i=1$ applicants across the 150 samples in each DGP. On average, there are 599 such applicants in a sample. Given a DGP, we simulate an outcome under the counterfactual policy and compare it to the truth from the actual behavior (i.e., the true preferences with possible mistakes).  Panel (a) shows the average mis-prediction rates. Panel~(b) shows the predicted welfare effects by each approach. It is measured by the difference between the fractions of winners and losers.  See Table~\ref{tab:cf_all} in Appendix~\ref{app:mc} for more details.}
\end{mfignotesin}

\paragraph{Mis-predicted Matches.} 

Panel (a) of Figure~\ref{Fig:CF_AA} further shows {the extent to which} each of the three approaches mis-predicts individual outcomes for applicants with $T_i=1$. Recall that the counterfactual policy is intended to help those applicants. 
The stability-based approach incorrectly predicts the match for $4.5$ percent of applicants on average in DGPs TT and PIM. Even in PRM, its mis-prediction rate is merely 7.7 percent. The WTT-based approach has a lower mis-prediction rate in DGP TT but under-performs relative to stability in the other two DGPs. The submitted-ROLs approach has the highest mis-prediction rates in all DGPs except TT. Among the applicants with $T_i=0$, Figure~\ref{Fig:CF_nAA} in Appendix~\ref{app:mc} shows that the comparison of the approaches follows the same pattern.

\paragraph{Mis-predicted Welfare Effects.} 

We now investigate the welfare effects on the $T_i=1$ applicants of the counterfactual policy. Given a simulation sample and a DGP, we compare the outcomes of each applicant under the two policies. If the applicant is matched with a ``more-preferred'' college according to the true/estimated preferences, she is a \textit{winner}; she is a \textit{loser} if she is matched with a ``less-preferred'' one.\footnote{Because each approach to counterfactual analysis estimates applicant preferences in a unique way, an applicant's utility associated with a college can differ across the approaches. Therefore, the measured welfare effects of the counterfactual policy may differ even when an applicant is matched with the same college.}

Panel (b) of Figure~\ref{Fig:CF_AA} shows the difference between the fractions of winners and losers, averaged across the 150 samples.\footnote{The outcome does not change for 9--28 percent of applicants. 
See Table~\ref{tab:cf_all} for more detailed summary statistics.} Among the $T_i=1$ applicants, the stability-based estimates are almost identical to the truth, even in the DGP with payoff-relevant mistakes (PRM). In contrast, the other two approaches' predictions are close to the true value only in DGP TT; both tend to be biased toward a zero effect when applicants make mistakes (DGPs PIM and PRM). The reason for the bias is clear.  Under WTT, the preferences for popular colleges are underestimated. Meanwhile, the submitted-ROLs approach ignores the likely changes in ROLs under the new policy.  In particular, disadvantaged applicants find previously out-of-reach colleges now within reach, so they may include these colleges in their ROLs. 

Despite being shown in simulations, these findings may provide important implications for policymaking, especially in public education. For example, many recent policy initiatives are designed to increase access to high-quality colleges and schools by traditionally disadvantaged students. Such an affirmative-action policy precisely changes popular schools from out-of-reach to within-reach for disadvantaged students. To predict the effects of such a policy, only the stability-based estimates can perform well if students may have chosen not to apply to some out-of-reach schools in the regime without affirmative action.

\section{Conclusion} \label{sec:conclusion}

Motivated by field evidence on non-truthful behavior in strategy-proof environments, we theoretically argue, using a robust equilibrium concept, that an \emph{outcome} of DA can be reliably predicted, but not participants' \emph{behavior}. Moreover, in a sufficiently large economy, the outcome approximates well the one that would have emerged if every participant plays the dominant strategy. While this result justifies the vast theoretical literature that assumes truthful reporting behavior to analyze outcomes of DA, it calls into question empirical methods that take truthful reporting as a literal behavioral prediction. Our theory suggests that {the alternative approach}  focusing on the stability property of the outcome may be robust to applicant mistakes. These implications are relevant to the estimation of participant preferences and counterfactual analysis.

Our paper focuses on environments where applicants know their scores according to which colleges rank them. However, the general insights can be extended to other settings, for example, where applicants are ranked by colleges according to a post-application lottery. \cite{CHH}, an ongoing project, provide such an extension.

\newpage
\begin{appendix}


\setcounter{equation}{0}
\renewcommand{\theequation}{\Alph{section}.\arabic{equation}}
\setcounter{table}{0}
\renewcommand{\thetable}{\Alph{section}.\arabic{table}}
\setcounter{figure}{0}
\renewcommand{\thefigure}{\Alph{section}.\arabic{figure}}
\setcounter{prop}{0}
\renewcommand{\theprop}{\Alph{section}.\arabic{prop}}

\begin{center}
{\Large Appendix to}\\
[0.7cm]
{\LARGE Stable Matching with Mistaken Agents}\\
[0.7cm]
{\large Georgy Artemov \hspace{1cm} Yeon-Koo Che \hspace{1cm} YingHua He} \\
[0.5cm]
{\large \today}\\
[5cm]
\end{center}

\listofappendices

\clearpage
\newpage

\def\ca{a}
\def\cb{b}
\def\cg{c_3}
\def\apj{\theta_j}
\def\apa{\theta_1}
\def\apb{\theta_2}
\def\apc{\theta_3}
\def\ep{\epsilon}

\section{Definition of the Deferred-Acceptance Mechanism}\label{sec:DA-def}
\appcaption{Appendix \ref{sec:DA-def}: Definition of the Deferred-Acceptance Mechanism}

The applicant-proposing Deferred-Acceptance (DA) mechanism uses each college's capacity and ranking over applicants as well as applicants' submitted ROLs to calculate a matching. It proceeds as follows:

\begin{itemize}
    \item[] {\bf Round 1.} Every applicant applies to her first choice. Each college holds the highest-ranked applicants up to its capacity and rejects the rest, if any.
    \item[] Generally, in 
    \item[] {\bf Round $m>1$.} Every applicant who is rejected in Round $(m-1)$ applies to the next choice college on her ROL if there is one. Each college pools together new applicants and those held from Round $(m-1)$; it holds the highest-ranked applicants up to its capacity and rejects the rest, if any.
\end{itemize}
The process terminates after any Round $m$ when no rejections are issued. Each college is then matched with the applicants it is currently holding.

\section{An Example of Non-convergent Cutoffs}\label{app:ex-nonconvergence}
\appcaption{Appendix \ref{app:ex-nonconvergence}: An Example of Non-convergent Cutoffs}

In this appendix, we construct a sequence of economies that satisfies the conditions of Proposition~\ref{claim:uniform-convergence of P}, in particular, full support of types, yet the sequence of cutoffs induced by a regular strategy does not converge. Note that this regular strategy is not a robust equilibrium.

In the example, we will use  $\lceil\cdot\rceil$ and $\lfloor\cdot\rfloor$ to denote ceiling and floor functions. Consider a market with $k$ applicants and two colleges, $\ca$ and $\cb$. Each college has capacity $\lfloor k/4 \rfloor$. Applicants draw their types independently and uniformly from $\{\ca\-\cb,\cb\-\ca\} \times [0,1]^2$. 

We will consider two strategies: $\rho$, which is TT, and $\hat\sigma$ which prescribes submitting an empty ROL with probability $1-\gamma$ and TT with probability $\gamma$, where $\gamma$ is close to zero. Let $k_{2m} = \lceil 1000/\gamma^{2m} \rceil$, for $m \in \mathbb{N}$ (note that $\gamma^{2m}$ is $\gamma$ to the power of $2m$). The strategy profile $\bsigma$ is constructed as follows. Applicants from 1 to $k_1$ play $\hat\sigma$. 
For any $m$, applicants from $k_{2m-1}+1$ to $k_{2m}$ play $\rho$ and applicants from $k_{2m} + 1$ to $k_{2m+1}$ play $\hat\sigma$.

We consider two subsequences of economies: $\{F^{k_{2m}}\}_{m \in \mathbb{N}}$ and $\{F^{k_{2m+1}}\}_{m \in \mathbb{N}}$.

For any economy $F^{k_{2m}}$ from the first subsequence, applicants with indices between $k_{2m-1}+1$ and $k_{2m}$ play TT. Their total number is $\lceil 1000/\gamma^{2m} \rceil - \lceil 1000/\gamma^{2m-1} \rceil \geq \lfloor 1000(1-\gamma)/\gamma^{2m}\rfloor$. As there are other applicants with lower indices who play TT, the fraction of applicant who are TT is more than $(1-\gamma)$. 
Given the type distribution and capacities, when the economy size grows, in this subsequence, the cutoff at each college tends to be no less than $(1-\gamma)/2$ with probability close to one.

For any economy $F^{k_{2m+1}}$ in the second subsequence, applicants with indices between $k_{2m} + 1$ and $k_{2m+1}$ constitute $(1-\gamma)$ fraction of all applicants. They submit an empty ROL with probability $1-\gamma$. Thus, there are fewer than $\lfloor k/4 \rfloor$ applicants at each college, with probability close to 1. Thus, with probability close to 1, $P_{\ca}^{k_{2m+1}} = P_{\cb}^{k_{2m+1}} = 0$.

Given that, for these two subsequences, the cutoffs are either above $(1-\gamma)/2$ or equal to 0 with probability close to 1, the sequence of cutoffs $\P^k$ does not converge in probability. Note that there is a convergent subsequence, in line with Proposition~\ref{claim:uniform-convergence of P}. This example also illustrates why AL's result cannot be applied. In particular, this setting does not have a symmetric strategy or an overall excess demand given the strategy profile.
 
\section{Preliminary Theoretical Results} \label{app:prel}
\appcaption{Appendix \ref{app:prel}: Preliminary Theoretical Results}

    Consider the continuum economy $E=[\eta, \S]$ with the full support assumption $\frac{1}{C}(\overline{u}-\max\{\underline{u},0\})^{C} \eta(\t) > \xi$ for all $\t \in \left[\max\{0,\underline{u}\},\overline{u}\right]^C \times [0,1]^C \subset {\T} $ and for some $\xi>0$.

    \def\bbsigma{;\bsigma}
    We now reiterate and give the formal definitions of demands and cutoffs introduced in Section~\ref{sec:asymptotics}. Fix $\bsigma$. The strategy profile $\bsigma^k$ induces a random ROL, $R_j$, for each applicant $j\in \{1,\dots, k\}$. For any $\p\in [0,1]^{C}$, we define a per capita profile of (random) demands for colleges---henceforth, simply called demand---$\D^{k} (\p\bbsigma) = \left(D^{k}_{c}(\p\bbsigma{})\right)_{c\in C}$;  the demand for college $c$ is given by
    \begin{align*}
 		D^{k}_{c}(\p\bbsigma) := \frac{1}{k} \sum_{j = 1}^{k} \I \left\{c\in \arg\max_{\text{w.r.t. } R_{j}}\left\{c'\in C: s_{j,c'} \geq p_{c'}\right\} \right\},
 	\end{align*}
    where $\arg\max_{\text{w.r.t. } R_{j}}$ picks the highest-ranked college in  $R_j$ from the set of feasible colleges $\left\{c'\in C:s_{j,c'}\geq p_{c'}\right\}$ and $\I\{\cdot\}$ is an indicator function. Similarly, we define a demand profile $\D^{k}_{(i)}(\p\bbsigma)=(D^{k}_{(i),c}(\p\bbsigma))_{c \in C}$ that arises when applicant $i$ employs truthful reporting $\rho$ and all other applicants $j \ne i$ continue to use $\sigma_j$. For notational convenience, we use $\D^{k}_{(0)}(\p\bbsigma) = \D^k(\p\bbsigma)$ to denote the demand arising from the original strategy $\bsigma^k$. Let $\overline{\D}^{k}_{(i)}(\p\bbsigma) := \mathbb{E}\left[\D^{k}_{(i)}(\p\bbsigma)\right]$, where the expectation is taken over the random draws of applicants' types and the random ROLs arising from  $\bsigma_{(i)}^k$ being (possibly) mixed.
    
    Random cutoffs $\P^k_{(i)}(\bsigma) \in [0,1]^{C}$, defined by DA with ROLs prescribed by $\bsigma$, clear the (random) demand system $\D^{k}_{(i)}(\p\bbsigma)$. Cutoffs $\overline{\p}(\bsigma)$ clear the (non-random) demand system $\overline{\D}(\p\bbsigma)$. Appendix~\ref{app:uniform-convergence} provides an argument that $\overline{\p}(\bsigma)$ exists for any regular strategy. When there is no ambiguity, we use $\overline\p$ instead of $\overline\p(\bsigma)$. 
    We omit $\bsigma$ from the expression of the demands in the following.  By construction, $D_{(i),c}^k(\p)$ are non-increasing in $p_c$ and non-decreasing in $p_{-c}$ for any $c \in C$ and $0 \leq i \leq k$.

    We now formally describe the outcome of an applicant-proposing deferred acceptance algorithm (DA) in the $k$-random economy by defining the \textit{DA cutoffs} of $k$-random economy,  $\P^k:= \lim_{m\to \infty} (  P_1^{k,m},...,  P_{C}^{k,m})$, where $\P^{k,0}=(0,...,0)$ and for $m\ge 1$, 
    $$
	    P^{k,m}_c=\sup\left\{p \in [0,1]: D_c^k(p, \P_{-c}^{k, m-1}) = S_c^k \right\}, \forall c\in C,
    $$ 
    if the set is nonempty and $P^{k,m}_c=0$ otherwise. Note that the iterative steps of defining the cutoffs correspond to the iterative steps of DA. Initially, the applicants who prefer college $c$ most apply to $c$ and $c$ tentatively accepts applicants from among them in the descending order of score $s_c$ up to its capacity. That is, for $c\in \C$,  $D_c^k(0,...,0)$ is the measure of applicants to $c$ and $S_c^k$ is the capacity of $c$, so $P_c^{k,1}$ becomes the cutoff for $c$ in step 1. More generally, in step $m$, a measure $ D_c^k(P_c^{k,m-1},\P_{-c}^{k,m-1})$ of applicants apply to $c$, and the same process determines the cutoff $P_c^{k,m}$ for college $c$.\footnote{The measure $D^k_c(P_c^{k,m-1},\P_{-c}^{k,m-1})$ includes applicants retained from the previous round. The description we provide is a slight modification to the usual DA: applicants who have never been rejected by a college and have a score below $P_c^{k,m-1}$ do not apply to college~$c$ in round~$m$. These applicants would have been rejected if they applied. Like the standard DA, the algorithm converges in at most $C k$ steps.} Due to the property of $\D^k(\p)$ observed above, $\P^{k,m} = (P_c^{k,m})_c$ is monotone non-decreasing, and the limit $\P^k$ is well defined. Importantly, the cutoffs at each step, and thus $\P^k$, are random since $\D^k$ is random.
	
    Even though we are interested in the outcome of DA (i.e., the applicant-proposing deferred acceptance), it is useful to define the cutoffs that arise from  CPDA (college-proposing deferred acceptance).
    Let the \textit{CPDA cutoffs} be defined by  $\Q^k:= \lim_{m \to \infty} (  Q_1^{k,m},\dots, Q_{C}^{k,m})$, where $\Q^{k,0}=(1,\dots,1)$ and for $m \ge 1$, 
    $$
	    Q^{k,m}_c=\sup\left\{p \in [0,1]:  D_c(p, \Q_{-c}^{k, m-1}) = S_c^k \right\}, \forall c\in C,
    $$ 
    if the set is nonempty and $Q^{k,m}_c=0$ otherwise. Similarly to before, we observe that $\Q^{k,m}:=(Q^{k,m}_c)_c$ are monotone non-increasing in $m$, so $\Q^k$ is well defined.  

    Finally, the standard lattice property of stable matchings and the extremality of DA and CPDA matchings imply that $\P^k\le \Q^k$.\footnote{A useful perspective is to view $\P^k$ and $\Q^k$ as the smallest and largest fixed points of a self map $\Phi: [0,1]^{C}\to [0,1]^{C}$ defined by $\Phi_c(\p):=\sup\{p_c \in [0,1]:  D_c(p_c, p_{-c}) = S_c^k \}$ if the set is nonempty and otherwise $\Phi_c(\p):=0$. The monotonicity of $\Phi$ means that by Tarski's fixed point theorem, the fixed points of $\Phi$ form a complete lattice, admitting extremal points. Such extremal fixed points are obtained via the iterative steps we have defined.}
		
    Next, suppose $i$ unilaterally deviates to TT ($\rho$). We can define the resulting DA and CPDA cutoffs analogously, and denote them respectively by $\P^k_{(i)}$ and $\Q^k_{(i)}$ and observe $\P^k_{(i)}\le \Q^k_{(i)}$.  It is notationally convenient to define the cutoffs when no one deviates from $\bsigma^k$ by $\P^k_{(0)} := \P^k$ and $\Q^k_{(0)} := \Q^k$.

    Our goal (Proposition~\ref{claim:uniform-convergence of P}) is to establish a desirable limit behavior of $(\P^k_{(i)})_{i\in \mathbb{N}_0}$ as $k\to \infty$, where $\mathbb{N}_0 := \mathbb{N} \cup 0$. We accomplish this goal in Section \ref{app:uniform-convergence}. To this end, however, we need to establish a few preliminary results on demands. We will first establish almost sure convergence of random demands to their expectation (Lemma~\ref{mcdiarmid}). We then establish that the expectation is Lipschitz-continuous (Lemma~\ref{equi}) and converges to a Lipschitz-continuous function $\overline{\D}$ (Lemma~\ref{aa}). These two results help us establish the key result that the family of random demands converges almost surely to non-random $\overline{\D}$ (Lemma~\ref{claim}). For an asymmetric strategy, we only establish it for a subsequence, because the whole sequence may not converge at all. The appropriate smoothness of random demands is what will help us establish the convergence of cutoffs.

    Our first step is to establish a probabilistic bound for the distance between $\D^k_{(i)}$ and its expectation.  For $i\in \mathbb{N}_0 $ and $\p\in [0,1]^{C}$, recall $\overline{\D}^{k}_{(i)}(\p) := \mathbb{E}\left[  \D^{k}_{(i)}(\p)\right]$, where the expectation is taken over the random draws of applicants' types (when $F^k$ is constructed) and the randomness in the ROLs arising from $\bsigma_{(i)}^k$ being (possibly) mixed. Because $\bsigma$ may be asymmetric, some lemmas below require selecting a subsequence of economies $\{F^{k_\ell}\}_\ell$ to deal with asymmetric strategies; all these lemmas can be stated for the whole sequence of economies $F^k$ if strategies are symmetric. 
    Recall that, throughout, we use $\lVert\cdot\rVert$ to denote the sup norm; i.e., for any $\bm{x},\bm{x}'\in [0,1]^{|C|}$, $\lVert \bm{x}-\bm{x}'\rVert := \sup_c |x_c - x_c'|$.  	
		
    \begin{lem} \label{mcdiarmid}
    	Fix any strategy $\bsigma$, any $\p\in [0,1]^{C}$, and any $i \in \mathbb{N}_0$. Then, for any $\alpha>0$,
    	\begin{align*}
    		&\Pr\left[\left\lVert \D^{k}_{(i)}(\p) - \overline{\D}^{k}_{(i)}(\p) \right\rVert > \alpha\right] \leq\left\vert C\right\vert \cdot e^{-2k\alpha^{2}}.
    	\end{align*}
    \end{lem} 

    \begin{proof} 
        By McDiarmid's inequality \citep[][]{mcdiarmid:89}, for each $c\in C$,
    	$$
    		\Pr\left\{ \left|D^k_{(i),c}(\p)-\overline D^k_{(i),c}(\p)\right| > \alpha\right\} \le e^{-2k\alpha^2},
    	$$
    	since for each $c\in C$,
    	$|D^k_{(i),c}(p; R_1,\dots,R_{k}) - D^k_{(i),c}(p; R_1',\dots,R_{k}')|\le 1/k$ whenever ROLs $(R_1,\dots,R_{k})$ and $(R_1',\dots,R_{k}')$ differ only in one component (recall that demands depend on ROLs, although ROLs are usually suppressed in the notation).
    		
    	It then follows that
    	\begin{align*}
    		&\Pr\Big[\left\Vert \D^{k}_{(i)}(\p) -\overline{\D}^{k}_{(i)}(\p)\right\Vert > \alpha\Big]
    		\\
    		= & \Pr\Big[\exists\ c\in \C\text{ s.t.}\left\vert D_{(i),c}^{k}(\p) - \overline{D}_{(i),c}^{k}(\p) \right\vert > \alpha\Big]
    		\\
    		\leq & \sum_{c \in \C}\Pr\Big[\left\vert D_{(i),c}^{k}(\p)  -\overline{D}_{(i),c}^{k}(\p) \right\vert > \alpha\Big] 
    		\leq C \cdot e^{-2k\alpha^{2}}.
    	\end{align*}	
    \end{proof}	

    Lemma~\ref{mcdiarmid} implies almost sure convergence via the first Borel-Cantelli lemma. In this sense, it can be thought of as an extension of a strong law of large numbers to a special case of non-i.i.d.\ random variables. When strategy $\bsigma$ is symmetric and $i=0$, the almost sure convergence  can be readily obtained from the strong law of large numbers because demands are just a sample average of bounded random variables. The next two lemmas establish Lipschitz-continuity of expected demands.

    \begin{lem} \label{equi} 
        For any strategy $\bsigma$ and each $(k,i) \in \mathbb{N} \times \mathbb{N}_0$, the function $\overline{\D}^{k}_{(i)}(\p)$ is Lipschitz continuous with a constant $L$ that is independent  of $(k,i)$.     
    \end{lem}
    \begin{proof}  
        Let $\p$ and $\p'$ be two arbitrary cutoff vectors in $[0,1]^{C}$.  
	    Define 
	    \[
		    \Theta_{\p,\p'}:=\left\{  \left(\bu, \s\right) \in\Theta: \exists\ c\in \C \text{ such that }   p_{c} < s_c < p_{c}' \text{  or  } p_{c}' < s_c < p_{c}\right\}.
	    \]
	    Since $\eta$ is absolutely continuous with respect to Lebesgue measure, we have $\eta (\Theta_{\p,\p'}) \leq L \lVert \p' - \p\rVert$, where $L$ is an upper bound for the density for all $\theta\in \T$. Then,
	    \begin{align*}
		    &\left\lVert \overline{\D}^{k}_{(i)}(\p') - \overline{\D}^{k}_{(i)}(\p) \right\rVert 
		    \\
	        =\, & \sup_{c\in \C}\left\vert\mathbb{E}_{\theta,\sigma}\left[D_{(i),c}^{k}(\p') - D_{(i), c}^{k}(\p)\right]\right\vert 
		    \\
		    =\, & \sup_{c\in \C}\left\vert \mathbb{E}_\theta\left[\frac{1}{k}\sum_{j=1}^{k}\sum_{R \in \R}\mathbb{P}(\sigma_j(\theta_j) = R)
		    \left(
		    \begin{array}[c]{rl}
    		    & \I\left\{c\in\argmax_{\text{w.r.t. }R}\left\{c' \in \C: s_{j,c'} \geq p_{c'}^{\prime}\right\}\right\}
    		    \\
    		    -& \I\left\{c\in\argmax_{\text{w.r.t. }R}\left\{c' \in \C: s_{j,c'} \geq p_{c'}\right\} \right\}
		    \end{array}
		    \right)
		    \right]\right\vert  
		    \\
		    \leq\, & \sup_{c\in \C} \frac{1}{k} \sum_{j=1}^{k}\mathbb{E}_\theta\left[\sum_{R \in \R} \mathbb{P}(\sigma_j(\theta_j) = R)\left\vert
		    \begin{array}[c]{rl}
    		    & \I\left\{c\in\argmax_{\text{w.r.t. }R}\left\{c' \in \C: s_{j,c'}\geq p_{c'}^{\prime}\right\}  \right\}
    		    \\
    		    - & \I\left\{c\in\argmax_{\text{w.r.t. }R}\left\{c' \in \C: s_{j,c'} \geq p_{c'}\right\} \right\}
		    \end{array}
		    \right\vert\right]    
		    \\
		    \leq\, &\frac{1}{k} \sum_{j=1}^{k}\mathbb{E}_\theta\left[\I\left\{\theta_{j}\in\Theta_{\p,\p'}\right\}\right]
		    \\
		    =\, &  \mathbb{E}_\theta \left[\I\left\{\theta_{j}\in\Theta_{\p,\p'}\right\}\right] 
		    = \eta\left(\Theta_{\p,\p'}\right) \leq L \lVert \p' - \p \rVert,
        \end{align*}
	    where the expectation $\mathbb{E}_{\theta,\sigma}$ is over applicant types and mixed strategies; $\mathbb{E}_{\theta}$ is an expectation over applicant types; and $\mathbb{P}(\sigma_j(\theta_j)=R)$ is the probability that applicant $j$ of type $\theta_j$ submits $R$ as prescribed by mixed strategy $\sigma_j(\theta_j)$ (with an abuse of notation, we denote $i$'s strategy by $\sigma_i$ even though $i$ deviates to truth-telling).	
	    The first inequality follows from Jensen's inequality and the second inequality holds since the two sets, $\left\{c' \in \C: s_{i,c'} \geq p_{c'}'\right\}$ and $\left\{c' \in \C: s_{i,c'} \geq p_{c'}\right\}$, are identical when $\theta_{i} \notin \Theta_{\p,\p'}$.	
    \end{proof}
	
	\begin{lem} \label{aa} 
	    There exists a subsequence of economies $F^{k_{\ell}}$ such that $\sup_{i,\p} \lVert \overline D^{k_{\ell}}_{(i)}(\p) - \overline D(\p) \rVert \to 0$ as $\ell \to \infty$. Function $\overline{D}(\p)$ is Lipschitz-continuous with the same constant $L$ as Lipschitz-continuous functions $\overline{D}^{k_{\ell}}_{(i)}(\p)$.
	\end{lem}

	\begin{proof} 
	    The sequence of functions $\{\overline{\D}^{k}(\p)\}_{k=1}^{\infty}$ defined on a compact set $\left[0,1\right]^{C}$ is uniformly bounded and uniformly equicontinuous (which follows from their Lipschitz property with a uniform constant $L$, as shown in Lemma \ref{equi}). By the Arzela-Ascoli theorem, we can find a subsequence $\{\overline{\D}^{k_{j}}(\p)\}_{j=1}^{\infty}$ which converges uniformly to Lipschitz-continuous function $\overline{\D}(\p)$ with the same constant $L$, since the Lipschitz property is preserved in the limit.
		
		Now consider any $i\neq 0$. For any $\p\in [0,1]^{C}$ and $i \in \mathbb{N}$,
		$$ 
		    \left\Vert \overline{\D}^{k_{j}}_{(i)}(\p) - \overline{\D}^{k_{j}}(\p)\right\Vert
		    = \left\Vert \mathbb{E}\left[\D^{k_{j}}_{(i)}(\p) -  \D^{k_{j}}(\p)\right ]\right\Vert 
		    \leq \mathbb{E}\left[\left\Vert \D^{k_{j}}_{(i)}(\p) -  \D^{k_{j}}(\p)\right \Vert \right]
		    \leq \frac{1}{k_j},
		$$
		since changing the strategy from $\bsigma^{k_j}$ to $\bsigma^{k_j}_{(i)}$ can change the demand for any college at most by $1/k_j$. Note that the upper bound of the difference, $\frac{1}{k_j}$, in the last inequality depends on neither $i$ nor $\p$, implying uniform convergence.
		
		Combining this result with the earlier observation, we conclude that there exists a subsequence in the sequence $\{\overline{\D}^{k}_{(i)}(\p)\}_{k=1}^{\infty}$ that converges uniformly to the same $\overline{\D}(\p)$ for all $i\in \mathbb{N}_0$ and $\p$. 
	\end{proof}

    Lemma~\ref{aa} is stated for a subsequence of economies $F^{k_\ell}$ because the convergence of demands is not guaranteed for the whole sequence when strategies are asymmetric; if we consider only symmetric strategies, the lemma could claim a uniform convergence of the whole sequence of expected demands, rather than a subsequence. Several of the results below are shown for a subsequence of economies described in Lemma~\ref{aa} and its associated strategy profile. It is useful to introduce a specific name for such subsequences. 
	\begin{defn}
	    A subsequence $\{F^{k_\ell},\bsigma^{k_\ell}\}_{\ell}$ which induces a subsequence of Lipschitz-continuous demands $\{\D^{k_{\ell}}_{(i)}(\p)\}_{k_\ell,0 \leq i \leq N}$ with the same constant $L$ is \emph{expected-demand convergent} if there exist Lipschitz-continuous demands $\overline{\D}(\p)$ with the same constant $L$ such that $\sup_{\p,i} \lVert \overline{\D}^{k_{\ell}}_{(i)}(\p) - \overline{\D}(\p) \rVert \to 0$ as $\ell \to \infty$.
	\end{defn}
	
	\begin{lem} \label{claim} 
	    Consider any expected-demand-convergent subsequence $\{F^{k_\ell},\bsigma^{k_\ell}\}_{\ell}$. Then, for any $\e>0$,
		\[
			\Pr\left\{\lim_{\ell\to \infty}\sup_{i,\p}\lVert \D^{k_{\ell}}_{(i)}(\p) - \overline{\D}(\p)\rVert > \e \right\} = 0.
		\]
	\end{lem}

	\begin{proof} 
		Since $\overline{\D}$ is Lipschitz-continuous, thus continuous, we can partition the space of $\p$'s into  finite intervals of the form $ Z({\bkappa}):=\prod_{c} [p_{\kappa_c}, p_{\kappa_c+1}]$, where $\bkappa = (\kappa_c)_c\in \{0,\dots,n\}^{C}$, for some $n\in \mathbb{N}$,\footnote{Naturally, we have $p_{k_{\mathbf{0}}}=\mathbf{0}$ and $p_{k_{\mathbf{n}}}=\mathbf{1}$.} 
		such that for each $c$, 
		\[
			\left|\overline D_c(\p) - \overline{D}_c(\p')\right| < \frac{\e}{2}
		\]
		for all $\p,\p'\in Z ({\bkappa})$, $\forall\bkappa=(\kappa_c)_c$. There are $n^{C}$ such intervals. Note that these intervals partition the whole space of $\p$'s and do not consider deviations of applicants; thus, they do not depend on specific $(i,\p)$.
	
		Consider any $\p$ and $c$. Let $\bkappa$ be the index of the interval such that $\p \in Z(\bkappa)$. Let 
		  \begin{align*}
		\p'_{\bkappa} := & (p_{\kappa_1 + 1}, \dots, p_{\kappa_{c-1} + 1}, p_{\kappa_c},p_{\kappa_{c+1} + 1}, \dots, p_{\kappa_{C} + 1}), \\
		\p''_{\bkappa} := & (p_{\kappa_1}, \dots,  p_{\kappa_{c-1}}, p_{\kappa_c + 1},p_{\kappa_{c+1}},\dots, p_{\kappa_{C}}). 
		\end{align*}
		The demand for $c$, $\overline{D}_c(\cdot)$, is the highest at $\p'_{\bkappa}$ and the lowest at $\p''_{\bkappa}$ among the prices in $Z(\bkappa)$. Consider a randomly drawn economy $F^{k_\ell}$ and the correspondent demand for $c$ $D_{(i),c}^{k_{\ell}}(\p)$. Then,\footnote{This inequality follows from an argument used for the proof of the Glivenko-Cantelli theorem extended to a multidimensional case.}
		\begin{equation}\label{eq:DemandDiff1}
			\left|D_{(i),c}^{k_{\ell}}(\p) - \overline{D}_c(\p)\right| 
			\leq \max\left\{|D^{k_{\ell}}_{(i),c}(\p'_{\bkappa}) - \overline D_c(\p'_{\bkappa})|, |D^{k_{\ell}}_{(i),c}(\p''_{\bkappa}) - \overline D_c(\p''_{\bkappa})|\right\} + \frac{\e}{2}.
		\end{equation}

	    Suppose the event $\{\lVert \D^{k_{\ell}}_{(i)}(\p) - \overline\D(\p)\rVert > \e\}$ occurs for some $\p$ and $i$. Then, since $\lVert \D^{k_{\ell}}_{(i)}(\p) - \overline\D(\p)\rVert 
	    = \sup_c |D^{k_{\ell}}_{(i),c}(\p) - \overline D_c(\p)|$, 
	    there must exist $c$ such that $|D_{{(i)},c}^{k_{\ell}}(\p) - \overline D_c(\p)|\ge \e.$ From (\ref{eq:DemandDiff1}), there is $\p^*_{\bkappa} \in \{\p'_{\bkappa},\p''_{\bkappa}\}$ is such that $|D^{k_{\ell}}_{(i),c}(\p^*_{\bkappa}) - \overline D_c(\p^*_{\bkappa})| \ge \frac{\e}{2}$. Since $\overline{D}^{k_{\ell}}_{(i),c}(\p)$ converges to $\overline D_c(\p)$ in sup norm by Lemma \ref{aa}, there exists $N'$ such that for all $\ell > N'$, $\sup_{i,\hat{\p}}|\overline{D}^{k_{\ell}}_{(i),c}(\hat\p)-\overline{D}_c(\hat\p)| < \frac{\e}{4}$. Consequently, for $\ell > N'$ and $\p^*_\kappa$, we must have
	    $$
		    \left|D^{k_{\ell}}_{(i),c}(\p^*_{\bkappa}) - \overline{D}^{k_{\ell}}_{(i),c}(\p^*_{\bkappa})\right| \geq \frac{\e}{4}.
	    $$
 		
	    Combining the arguments so far, we conclude:
	    \begin{align*}
		    & \Pr\left\{\sup_{i,\p} \lVert \D^{k_{\ell}}_{(i)}(\p) - \overline\D(\p)\rVert > \e\right\}
		    \\
		    = & \Pr\left\{\exists (\p,i) \mbox{ s.t. } \lVert \D^{k_{\ell}}_{(i)}(\p) - \overline\D(\p)\rVert > \e\right\} 
		    \\
		    \leq &\sum_{i=0}^{k_{\ell}} \Pr\left\{\exists c \mbox{ and } \p^*_{\bkappa} \mbox{ s.t. } |D^{k_{\ell}}_{(i),c}(\p^*_{\bkappa}) - \overline{D}^{k_{\ell}}_{(i),c}(\p^*_{\bkappa})| \geq \frac{\e}{4}\right\}
		    \\
	    	\leq & \sum_{i=0}^{k_{\ell}} \sum_{c=1}^{C} \sum_{\bkappa \in \{0,\dots,n\}^{C}} \Pr\left\{|D^{k_{\ell}}_{(i),c}(\p^*_{\bkappa}) - \overline{D}^{k_{\ell}}_{(i),c}(\p^*_{\bkappa})| \geq \frac{\e}{4}\right\}
		    \\
		    \leq & n^{C}C \sum_{i=0}^{k_{\ell}} e^{-{k_{\ell}}{\e}^2/8}
		    \\
		    = & n^{C}C (k_{\ell}+1)e^{- {k_{\ell}}{\e}^2/8 } \to 0 \mbox{ as } \ell \to \infty,
	    \end{align*}	
	    where the last inequality follows from McDiarmid inequality (see Lemma \ref{mcdiarmid}).

        Note that 
        \[\sum_{k_\ell} \Pr\left\{\sup_{i,\p} \lVert \D^{k_{\ell}}_{(i)}(\p) - \overline\D(\p)\rVert > \e\right\} \leq n^{C}C e^{\e^2/8} \sum_{k_\ell}  (k_{\ell}+1)e^{- {k_{\ell}}} < \infty.\]
        Hence, by the first Borel-Cantelli lemma, $\D^{k_{\ell}}_{(i)}(\p)$ almost surely converges to $\overline\D(\p)$ uniformly over $(i,\p)$.
	\end{proof}
		
\subsection{Asymptotics of Cutoffs for a Regular Strategy} \label{app:uniform-convergence}

    We are now ready to establish, for a regular strategy, the uniform almost-sure convergence of cutoffs $\P^{k_{\ell}}_{(i)}$ to some deterministic cutoffs as $\ell\to \infty$ along any expected-demand-convergent subsequence. In fact, we show that the limit cutoffs are the cutoffs defined by the limit demand system (i.e., $\overline{\D}(\p)$). 
    Specifically, let $\overline\p=(\overline p_1,\dots, \overline p_{C})$ be the DA cutoffs  defined by $\overline p_c := \lim_{m\to \infty} \overline p^m_c$, where $\overline\p^0 = (0,...,0)$ and for $m \ge 1$, and $\overline\p^m = (\overline p^m_c)_c$ is given by 
	$$
	    \overline p^m_c=\sup\left\{p \in [0,1]: \overline D_c(p, \overline\p_{-c}^{m-1}) = S_c\right\}, \forall c\in C,
	$$ 
	if the set is nonempty, and $\overline p^m_c=0$ otherwise.
	
	Similarly, let $\overline\q = (\overline q_1,..., \overline q_{C})$ be the CPDA (College-Proposing Deferred Acceptance) cutoffs defined by $\overline q_c := \lim_{m \to \infty} \overline q^m_c$ for each $c$, where $\overline\q^0 = (1,...,1)$ and for $m \geq 1$, and $\overline\q^m = (\overline q^m_c)_c$ is given by 
	$$
	    \overline q^m_c = \sup\left\{p \in [0,1]: \overline D_c(p, \overline \q_{-c}^{m-1}) = S_c\right\}, \forall c\in C,
	$$ 
	if the set is nonempty and $\overline q^m_c=0$ otherwise. The interpretation is the same as the applicant-proposing DA cutoffs.
	
	We note that $\overline D_c(p_c, \p_{-c})$ is non-increasing in $p_c$ and non-decreasing in $\p_{-c}$. This is because this property, which holds for each realization of the $k$-random economy, is preserved when one takes expectation  to obtain $\overline\D^k$ and takes a limit along a subsequence. It then follows that $\overline\p^m$ is a monotone non-decreasing sequence and $\overline\q^m$ is a monotone non-increasing sequence, and their limits are well defined.  Moreover, $\overline\q \ge \overline\p$.

	Since $\bsigma$ is $\gamma$-regular, ROL $\rho(\theta)$ is chosen by an applicant of type $\theta$ with probability at least $\gamma$. The full support of $E=[\eta,S]$ implies that there is a positive lower bound on the density of $\theta$ on the original limit economy. Thus, the resulting limit economy induced by $\bsigma$ is full support in terms of ordinal preferences and scores. Then, Theorem 1-(a) of \cite{azevedo/leshno:16} guarantees that the induced limit economy has a unique stable matching, which in turn implies that $\overline\p = \overline\q$.
	
	\begin{proof}[Proof of Proposition~\ref{claim:uniform-convergence of P}]  
	
	    Consider an arbitrary expected-demand-convergence subsequence $\{F^{k_\ell},\bsigma^{k_\ell}\}_\ell$. Recall that there exists at least one such sequence (Lemma~\ref{aa}). 
	
	    Fix $\e > 0$. We will show that for some $N \in \mathbb{N}$, for all $\ell>N$, 
		$$
			\Pr\left\{\sup_{i\in \mathbb{N}\cup\{0\}} \lVert\P^{k_{\ell}}_{(i)} - \overline\p \rVert < \epsilon\right\} = 1.
		$$
		
		To begin, let $M$ be such that for all $m \ge M$, $\max\{\lVert \overline\p - \overline\p^m \rVert, \lVert \overline\p - \overline\q^m \rVert\} < \epsilon/2$. Such an $M$ exists due to the convergence of $\overline\p^m$ and $\overline\q^m$ to $\overline\p$ (where the latter uses the fact that $\overline\q = \overline\p$).
	    We next observe that for each $c$, for any $p,p'\in [0,1]$
	    \begin{align}
		    \lvert\overline D_c(p', \p_{-c}) - \overline D_c(p, \p_{-c})\rvert \geq \xi \gamma \lvert p' - p\rvert, \forall \p_{-c} \in [0,1]^{C-1}. \label{eq:deltagamma}
	    \end{align}
	
	    To obtain (\ref{eq:deltagamma}), first note that $|\overline D_c(p', \p_{-c}) - \overline D_c(p, \p_{-c})| \geq |\overline D_c(p', \bm{0})- \overline D_c(p, \bm{0})|$, where $|\overline D_c(p', \bm{0})- \overline D_c(p, \bm{0})|$ is the mass of applicants for whom $c$ is top-ranked by $\bsigma$ and whose scores at $c$ are between $p$ and $p'$. Assuming, without loss of generality, that $p' > p$, we have
		\begin{align*}
			\left|\overline D_c(p', \bm{0}) - \overline D_c(p, \bm{0})\right| 
			\geq 
			& \gamma \int_{(u,s) \in \Theta: u_c > u_{c'} \forall c' \neq c, u_c > 0, s_c \in [p,p']} \eta(\theta) d\theta \\
			\geq &\gamma \frac{\xi C}{(\overline{u}-\max\{0,\underline{u}\})^{C}} (p'- p) \int_{u_c > u_{c'} \forall c' \neq c, u_c>0} 1 du_1\dots du_C
			\\
			\geq & \xi \gamma (p'- p).
		\end{align*}
		Recall that $\gamma$ is the lower bound on the probability of truth-telling and that $\xi$ determines the lower bound on density $\eta$.\footnote{In the case of Serial Dictatorship, in which the full-support assumption holds with a reduced dimensionality of support, the same inequality holds. As other elements of the proof do not invoke full support, all our results hold for this mechanism.}
		
	    Similarly, recall from Lemma \ref{equi} that one can find $L>0$ such that $|\overline D_c(p_c,\p_{-c}) - \overline D_c(p_c,\p_{-c}')| \leq L \lVert \p_{-c}'- \p_{-c}\rVert$ for all $c$. Let $\lambda := \max\left\{1, \frac{L}{\gamma\xi}, \frac{1}{L}\right\}$ and $\nu = \frac{1}{2M \lambda^M }\epsilon$.  
		
	    It follows from Lemma \ref{claim} and the convergence of $\S^{k_\ell} \to \S$ that, for any $\nu > 0$, there exists $N(\nu)$ such that for any $\ell>N(\nu)$, we have 
	    $$
		    \sup_{i,\p}\lVert \D^{k_{\ell}}_{(i)}(\p) - \overline \D(\p)\rVert + \lVert \S^{k_\ell} - \S \rVert < \nu
	    $$ 
	    with probability 1. Below, we fix any such $\ell>N(\nu)$ and condition on the event $\mathcal{E}:= \{\sup_{i,\p}\lVert \D^{k_{\ell}}_{(i)}(\p) - \overline\D(\p)\rVert + \lVert \S^{k_\ell} - \S \rVert < \nu\}.$ Hence, the probability of having event $\mathcal{E}$ is one.
 
	    Consider a random economy $F^{k_\ell}$. We argue inductively that, for each step of DA $m=1,\dots, M$,\footnote{Recall that $M$ is defined so that for all $m \ge M$, $\max\{\lVert \overline\p - \overline\p^m \rVert, \lVert \overline\p - \overline\q^m \rVert\} < \epsilon/2$.} $|P^{k_{\ell}, m}_{(i),c} - \overline p_c^m|\le m \lambda^m \nu$, for each college $c$. Fix any college $c$. Consider any $m$, assuming that the result holds true up to step $m-1$. There are two possibilities. Suppose first $P_{(i),c}^{k_{\ell},m}> \overline p_c^{m}\ge 0$. Then,
	    \begin{align*}
		    0&= D_c^{k_{\ell}}(P^{k_{\ell},m}_{(i),c}, \P^{k_{\ell},m-1}_{(i),-c}) - S^{k_\ell}_c 
		    \\
		    &\leq \overline D_c(P^{k_{\ell},m}_{(i),c}, \P^{k_{\ell},m-1}_{(i),-c}) - S_c + \nu  
		    \\
		    &\leq \overline D_c(P^{k_{\ell},m}_{(i),c}, \overline\p_{-c}^{m-1}) - S_c + \nu + L\left\lVert (P^{k_{\ell},m}_{(i),c}, \P^{k_{\ell},m-1}_{(i),-c}) - (P^{k_{\ell},m}_{(i),c}, \overline \p_{-c}^{m-1})\right\rVert
	    	\\
		    &\le \overline D_c(P^{k_{\ell},m}_{(i),c}, \overline\p_{-c}^{m-1}) - S_c +\nu +  L(m-1) \lambda^{m-1} \nu 
	    	\\
		    &\le \overline D_c(P^{k_{\ell},m}_{(i),c}, \overline\p_{-c}^{m-1}) - \overline D_c(\overline p_c^m, \overline\p_{-c}^{m-1}) +\nu +L(m-1) \lambda^{m-1} \nu
		    \\
		    &= \overline D_c(P^{k_{\ell},m}_{(i),c}, \overline\p_{-c}^{m-1}) - \overline D_c(\overline p_c^m, \overline\p_{-c}^{m-1}) + ( 1 +L(m-1) \lambda^{m-1}) \nu,
	    \end{align*}
		where the first equality follows from the definition of DA cutoff at step $m$ and upon noting that  $P^{k_{\ell}, m}_{(i),c}>0$ (meaning that the set over which $\sup$ is taken is well defined and the condition is an equality at $P^{k_{\ell}, m}_{(i),c}$); the first inequality follows as we are conditioning on event $\mathcal{E}$; the second inequality follows from the Lipschitz bound of $L$ for $\overline{\D}$; the third inequality follows from the induction hypothesis that $|P^{k_{\ell}, m-1}_{(i),c'} - \overline p_{c'}^{m-1}|\le  (m-1) \lambda^{m-1} \nu$ for any $c' \in C$;\footnote{Note that $P^{k_{\ell}, 0}_{(i),c'} = \overline p_{c'}^{0} = 0$, and hence the inequality holds for $m = 1$.} and the fourth inequality follows from the definition of cutoff for the limit economy at step $m$ (which implies $\overline D_c(\overline p_c^m, \overline\p_{-c}^{m-1}) \le S_c$).
	
	    Rewrite the string of inequalities and use (\ref{eq:deltagamma}) to obtain
	    $$
		    (1 +L(m-1) \lambda^{m-1}) \nu \geq \overline D(\overline p_c^m, \overline\p_{-c}^{m-1}) - \overline D(P^{k_{\ell},m}_{(i),c}, \overline\p^{m-1}_{-c}) \geq \gamma \xi (P^{k_{\ell}, m}_{(i),c}- \overline p_c^m),
	    $$
	    which in turn implies that 
	    \begin{align}
		    P^{k_{\ell}, m}_{(i),c} - \overline p_c^m \le \frac{(1 +L(m-1) \lambda^{m-1})}{\gamma \xi} \nu = \frac{(1 - L\lambda^{m-1} + Lm \lambda^{m-1})}{\gamma \xi} \nu \le m \lambda^m \nu. \label{eq:Lineq}
	    \end{align}
	    Recall that $\lambda = \max\left\{1, \frac{L}{\gamma\xi}, \frac{1}{L}\right\}$.  
		
	    Suppose next $\overline p_c^{m} > P_{(i),c}^{k_{\ell},m} \ge 0$.  Then, 
	    \begin{align*}
		    0 &\ge D_c^{k_{\ell}}(P^{k_{\ell},m}_{(i),c}, \P^{k_{\ell},m}_{(i),-c}) - S^k_c 
		    \\& 
		    \geq \overline D_c(P^{k_{\ell},m}_{(i),c}, \P^{k_{\ell},m}_{(i),-c}) - S_c - \nu  
		    \\& 
		    \geq \overline D_c(P^{k_{\ell},m}_{(i),c}, \overline\p_{-c}^m) - S_c - \nu - L(m-1) \lambda^{m-1} \nu  
		    \\& 
		    = \overline D_c(P^{k_{\ell},m}_{(i),c}, \overline\p_{-c}^m) - \overline D_c( \overline p_c^m, \overline\p_{-c}^m) - \nu - L(m-1) \lambda^{m-1} \nu \\& 
		    = \overline D_c(P^{k_{\ell},m}_{(i),c}, \overline\p_{-c}^m) - \overline D_c( \overline p_c^m, \overline\p_{-c}^m) - (1 + L(m-1) \lambda^{m-1}) \nu.
	    \end{align*}
	    which follows analogously to the earlier string of inequalities except that the first line is an inequality because we need to allow for $P_{(i),c}^{k_{\ell},m} = 0$ and the fourth line is an equality because $\overline p_c^{m} > 0$.  
	
	    As above, we use (\ref{eq:deltagamma}) to obtain
	    $$
		    (1 + L(m-1) \lambda^{m-1}) \nu \ge \overline D(P^{k_{\ell},m}_{(i),c}, \overline\p^{ m}_{-c}) - \overline D(\overline p_c^m, \overline\p_{-c}^m) \geq \gamma \xi (\overline p_c^m - P^{k_{\ell}, m}_{(i),c}),
	    $$
	    which in turn implies that 
	    \begin{align}
		    \overline p_c^m-P^{k_{\ell},m}_{(i),c} \le \frac{(1 + L(m-1) \lambda^{m-1}) \nu}{\gamma \xi} \le m \lambda^m \nu.  \label{eq:Rineq}
	    \end{align}
	    Combining (\ref{eq:Lineq}) and  (\ref{eq:Rineq}), we have 
		\begin{align}
		    |P^{k_{\ell}, m}_{(i),c} - \overline p_c^m|\le m \lambda^m \nu.  \label{eq:ineq}
	    \end{align}
        Since this result holds for all $m=1, \dots, M$, we now conclude that, for each $c$, 
	    \begin{align}
		    P^{k_{\ell} }_{(i),c} \ge P^{k_{\ell}, M }_{(i),c} 
		    \geq \overline p_c^M - M \lambda^M \nu 
		    \geq \overline p_c - M \lambda^M \nu - \frac{\epsilon}{2}, \label{eq:L-bd}
	    \end{align} 
	    where the first inequality follows from the fact that $P^{k_{\ell} }_{(i),c}$ is the limit of a monotone non-decreasing sequence $(P^{k_{\ell},m}_{(i),c})_{m}$ as $m\to\infty$,   the second inequality follows from (\ref{eq:ineq}), and the third follows from the definition of $M$.
   
        The exact same argument works for the CPDA process. Namely, for each   $c$, $|Q^{k_{\ell}, m}_{(i),c} - \overline q_c^m| \leq  m \lambda^m   \nu$ for $m = 1, \dots, M$ so that we have 
        \begin{align}
            Q^{k_{\ell}}_{(i),c} 
            \leq Q^{k_{\ell}, M}_{(i),c} 
            \leq \overline q_c^M + M \lambda^M \nu
            \leq \overline q_c + M \lambda^M \nu + \frac{\epsilon}{2}. \label{eq:R-bd}
        \end{align}  
  
        Combining (\ref{eq:L-bd}) and (\ref{eq:R-bd}), recalling $\overline p_c=\overline q_c$ and $Q^{k_{\ell} }_{(i),c}\ge P^{k_{\ell} }_{(i),c}$, $\forall c$, we obtain 
	    $$
		    \sup_{i} \lVert\P^{k_{\ell}}_{(i)} - \overline\p\rVert 
		    \leq M \lambda^M \nu + \frac{\epsilon}{2},
	    $$ 
	    for all $i\in \mathbb{N}\cup\{0\}$ and for an arbitrary random economy $F^{k_\ell}$.  
	    Recall that $\nu = \frac{1}{2M \lambda^M }\epsilon$, where $M$, $\lambda$ and $\e$ do not depend on either $k_\ell$ or the random economy $F^k_\ell$. Then, in the event $\mathcal{E}$, for all $\ell > N\left(\nu\right)$, where $N\left(\nu\right) \in \mathbb{N}$ is defined above, we have 
	    $$
		    \sup_{i\in \mathbb{N}\cup\{0\}} \lVert\P^{k_{\ell}}_{(i)} - \overline\p\rVert
		    \leq \epsilon.
	    $$
	    Recall that event $\mathcal{E}$ occurs with probability 1, which completes the first part of the proposition.
	    
	    The second part of the proposition follows immediately once we observe that $\bsigma = (\sigma,\sigma,\dots)$, hence the whole sequence $\{F^k,\bsigma^k\}_k$ is expected-demand-convergent.
	\end{proof}
		
\section{Proofs of Theorems}\label{app:thm}
\appcaption{Appendix \ref{app:thm}: Proofs of Theorems}

\subsection{Proof of Theorem \ref{p1}}
\begin{proof}[Proof of Theorem \ref{p1}]
Let $\p = \overline{\p}(\bm{\rho})$, where $\overline{\p}(\bm{\rho})$ is the unique market-clearing cutoff for the limit demand system induced by TT. In this proof, we refer to $ \overline{\p}(\bm{\rho})$ simply as  $\overline{\p}$.
    	
Recall that 
    	$$\T^{\delta}(\overline{\p}):=\left\{(\bu,\s)\in \T \mid \exists j \in \C \mbox{ s.t. } |s_j - \overline{p}_j|\le \delta\right\}$$ 
is the set of types whose score for some college is $\delta$-close to its market-clearing cutoff in the limit demand system.

For each type $\t=(\bu,\s)$, there exists at least one SRS strategy against $\overline{\p}$ that violates WTT (see footnote~\ref{fn:WTT}); denote this strategy by $\hat R(\t)$. The applicants with types $\t\in \T^{\delta}(\overline{\p})$ play $\rho(\t)$ and the applicants with types $\t\not\in \T^{\delta}(\p)$ randomize between $\rho(\t)$ (with probability $\gamma$) and $\hat R(\t)$ (with probability $1-\gamma$).

Fix any $\e>0$. Take any $\e'>0$ such that $\e' \uh < \e$. By Proposition~\ref{claim:uniform-convergence of P}, there exists $K\in \mathbb{N}$ such that for all $k>K$, $\Pr\{\lVert {\P}^k_{(i)} - \overline{\p}\rVert < \delta\}  \ge 1 - \e',$	where ${\P}^k_{(i)}$ is the vector of cutoffs associated with the  matching in $F^k$ under the prescribed strategy with at most one applicant deviating to TT. Let $\mathcal{E}^k$ denote the event where $\lVert {\P}^k_{(i)} - \overline{\p}\rVert < \delta$ holds. We now show that the prescribed strategy profile forms an interim $\e$-Bayesian Nash equilibrium for each $k$-random economy for $k>K$.
    		
First, for any type $\t\in \T^{\delta}(\overline{\p})$, the prescribed strategy, $\rho(\t)$, is trivially optimal given the strategy-proofness of DA. Consider an applicant with any type $\t\not\in \T^{\delta}(\overline{\p})$, and suppose that all other applicants employ the prescribed strategy. Now condition on event $\mathcal{E}^k$. Recall that the set of feasible colleges is the same for type $\t\not\in \T^{\delta}(\overline{\p})$ whether the cutoffs are $\hat \P^k_{(i)}$ or $\overline{\p}$, provided that  $\lVert \hat{\P}^k_{(i)} - \overline{\p}\rVert < \delta$. Hence, given event $\mathcal{E}^k$, strategy $\hat R(\t)$ is a best response and the prescribed mixed strategy attains the maximum payoff for type $\t\not\in \T^{\delta}(\overline{\p})$.
    		
Of course, the event $\mathcal{E}^k$ may not occur, but that probability is no greater than $\e'$ for $k > K$, and the maximum payoff loss in that case from failing to play her best response is $\uh$ (if an applicant becomes unmatched).  Hence, the payoff loss she incurs by playing the prescribed mixed strategy is at most
    	$$\e' \uh < \e.$$
This proves that the strategy profile forms a robust equilibrium.
    \end{proof}

\subsection{Proof of Theorem \ref{p2}} \label{app:t2-proof}

		\begin{proof}[Proof of Theorem \ref{p2}]
		Fix any $\gamma$-regular robust equilibrium strategy profile $\bsigma$, for any arbitrary $\gamma\in (0,1]$. Suppose to the contrary that $\bsigma$ is not asymptotically stable. Then, by definition, there exists $\varepsilon>0$ and a subsequence of finite economies $\left\{F^{k_{j}}\right\}_{j}$ such that, for all $k_j$,
 		\begin{align}
 			\Pr\left( \text{The fraction of applicants playing SRS against }  \P^{k_{j}}%
 		\text{ is at least } 1-\varepsilon\right) < 1-\varepsilon,
 		\label{*}
 		\end{align} 
 		where the applicants play $\bsigma^{k_{j}}$, a $k_{j}$-truncation of $\bsigma$.
		
		By Lemma \ref{claim}, there exists a subsubsequence $\{\D^{k_{j_{\ell}}}\}_{\ell}$ that converges to $\overline{\D}$ uniformly and almost surely. By Proposition \ref{claim:uniform-convergence of P},
		$\P^{k_{j_{\ell}}}_{(i)}$ converges to $\overline{\p}$ uniformly over $i$ almost surely, where $\overline{\p}$ is the deterministic cutoffs defined by the limit of demands.
		
		Define a set of applicants, for $0< \delta < \ul$,
		\[
		\hat{\Theta}:=
		\left\{\left(\bu,\s\right): \left\vert u_{c} - u_{c'}\right\vert > \delta\text{ for all }c\neq c^{\prime}\right\}
		\cap
		\left\{\left(\bu,\s\right): \left\vert s_{c} - \overline{p}_{c}\right\vert >\delta \text{ for all }c\right\}.
		\]
		These are the applicants whose payoffs from two distinct colleges (or from being matched and unmatched) differ by at least $\delta$ and whose score at each college $c$ differs from its limit economy cutoff $\overline p_c$ by at least $\delta$.
		
	    Take $\delta$ to be small enough s.t. $\eta (  \hat{\Theta}) > \left(1 - \varepsilon\right)^{1/3}$. This can be done since $\eta$ is absolutely continuous.
		
		By WLLN, we know that $\eta^{k_{j_{\ell}}} (  \hat{\Theta})$ converges to $\eta(\hat{\Theta})$ in probability, and therefore there exists $L_{1}$ such that for all $\ell > L_{1}$ we have
	    \begin{align}
		    \Pr\left(  \eta^{k_{j_{\ell}}} (  \hat{\Theta}) \geq \left(1 - \varepsilon\right)^{1/2}\right) \geq \left( 1 - \varepsilon\right)^{1/2}. \label{L1} 
	    \end{align}
 	
        Consider the event 
		\[
		    A^{k_{j_{\ell}}}:=\left\{\sup_{0\le i\le k_{j_{\ell}}} \lVert \P^{k_{j_{\ell}}}_{(i)} - \overline{\p}\rVert < \delta \right\}.
		\]
		
        Since $\P^{k_{j_{\ell}}}_{(i)}\overset{p}{ \rightarrow } \overline{\p}$ uniformly over $i\in \mathbb{N}_0$, there exists	$L_{2}$ such that, for all $\ell > L_{2}$, we have 
	    \begin{align}
	 	    \Pr\left(  A^{k_{j_{\ell}}}\right)  \geq\max\left\{\left(1-\varepsilon\right)^{1/6}, 1 - \left(1-\varepsilon\right)^{1/2} \left[\left(1 - \varepsilon\right)^{1/3} - \left(1 - \varepsilon\right)^{1/2}\right]\right\}. \label{**}
        \end{align}
		
        Since $\bsigma$ forms a robust equilibrium, there exists $L_{3}$ such that for all $\ell > L_{3}$, $\bsigma^{k_{j_{\ell}}}$
		is a $\delta\left[  \left(  1-\varepsilon\right)^{1/6}-\left(
		1-\varepsilon\right)  ^{1/3}\right]$-BNE for economy $F^{k_{j_{\ell}}}$.
		
		By WLLN, there exists $\hat{L}\in \mathbb{N}$ such that  $\hat{L}$ i.i.d. Bernoulli random
		variables with parameter $p=\left(  1-\varepsilon\right)  ^{1/3}$ have a sample mean
		greater than $\left(  1-\varepsilon\right)  ^{1/2}$ with probability no less
		than $\left(1 - \varepsilon\right)^{1/3}$.  Next, let $L_{4}$ be such that
		$\ell > L_{4}$ implies $\left(1 - \varepsilon\right)^{1/2}k_{j_{\ell}} > \hat{L}$.
		
		Now let's fix an arbitrary $\ell > \max\left\{  L_{1},L_{2},L_{3},L_{4}\right\}$. We wish to show that in economy $F^{k_{j_{\ell}}}$,
		\[
		    \Pr\left(\text{The fraction of applicants playing SRS against }\P^{k_{j_{\ell}}} \text{ is no less than } 1 - \varepsilon\right) \geq 1 - \varepsilon,
		\]
		which would contradict (\ref{*}) and complete the proof.
		
		We first prove that in economy $F^{k_{j_{\ell}}}$, an applicant with $\t \in
		\hat{\T}$ plays SRS against $\overline{\p}$ with probability no less than
		$\left(1 - \varepsilon\right)^{1/3}$. To see this, suppose to the  contrary that
		there exists some applicant $i$ and some type $\theta \in \hat{\Theta}$ such that 
		\[
		    \Pr\left(\sigma_i(\theta) \text{ plays SRS
			against } \overline{\p}\right) < \left(  1 - \varepsilon\right)^{1/3}.%
		\]

        Suppose now the applicant $i$ deviates to TT. By doing so, she will do weakly better in all circumstances (since TT is a dominant strategy) and strictly so  by at least $\delta$ (since  $\t\in\hat \T$) conditional on the deviation changing her match. Her match would change (at least) whenever she was not playing SRS against $\overline{\p}$ under $\sigma_i(\theta)$ \emph{and} event $A^{{k_j}_{\ell}}$ occurs. This is because in event $A^{{k_j}_{\ell}}$, the strategy $\sigma_i(\t)$ is SRS against $\P^{k_{j_{\ell}}}_{(0)}$ if and only if $\sigma_i(\t)$ is SRS against $\overline{\p}$ for type $\t \in \hat\T$, and deviating to truthful reporting would produce a stable match against $\overline{\p}$ for such a type. In sum, applicant $i$ with type $\theta \in \hat{\Theta}$ would gain from deviation by at least 
        \begin{align*}
		    & \delta\cdot\Pr\left(\sigma_i(\theta) \text{ is not SRS against } \overline{\p} \text{ \emph{and} event } A^{k_{j_{\ell}}} \text{ occurs}\right)  
		    \\
		    \geq & \delta \left[\Pr\left(  A^{k_{j_{\ell}}}\right) - \Pr\left(\sigma_i(  \theta) \text{ plays SRS against } \overline{\p}\right)\right]  
		    \\
		     \geq & \delta\left[\left(1 - \varepsilon\right)^{1/6} - \left(		1 - \varepsilon\right)^{1/3}\right],
		\end{align*}

        The above inequalities contradict the construction of $L_{3}$.\footnote{Recall that $L_{3}$ was defined so that  $\ell > L_3$ means that the strategy profile $\bsigma^{k_{j_{\ell}}}$ is a
		$\delta\left[\left(1 - \varepsilon\right)^{1/6} - \left(1 - \varepsilon\right)^{1/3}\right]$-BNE for the economy $F^{k_{j_{\ell}}}$.}
		Therefore, in economy $F^{k_{j_{\ell}}}$, for each applicant $i=1,\ldots,k_{j_{\ell}}$
		and each $\theta \in \hat{\Theta}$, we have
		\begin{align}
			& \Pr\left(\sigma_i(\theta) \text{ plays SRS against } \overline{\p} \mid \eta^{k_{j_{\ell}}} (\hat{\Theta})
			\geq \left(1 - \varepsilon\right)^{1/2}\right) \cr
			 = &  \Pr\left(\sigma_i(\theta) \text{ plays SRS	against }\overline{\p}\right)  \geq \left(1 - \varepsilon\right)^{1/3}, \label{***}
		\end{align}
		where the first equality holds because applicant $i$'s choice of a mixed strategy is independent of random draws of the applicants' types.\footnote{In other words, how a fixed type plays in equilibrium does not depend on how many of them are drawn.}  
		
		It then follows that 
		\begin{align}
    		& \Pr\left(  \left.
    		\begin{array}
    		[c]{c}%
        		\text{The fraction of applicants with } \theta \in \hat{\Theta}\\
        		\text{playing SRS against }\overline{\p} \text{ is no less than } \left(  1 - \varepsilon\right)^{1/2}%
    		\end{array}
    		\right\vert \eta^{k_{j_{\ell}}} (  \hat{\Theta}) \geq \left(1 - \varepsilon\right)^{1/2}\right) 
    		\cr
    		 \geq & \Pr\left(\left.
    		\begin{array}
    		[c]{c}%
        		\eta^{k_{j_{\ell}}} (\hat{\Theta})  \cdot k_{j_{\ell}}\text{ i.i.d.
        			Bernoulli random variables with }\\p = \left(1 - \varepsilon\right)^{1/3}
        		\text{have a sample mean no less than } \left(1 - \varepsilon\right)^{1/2}%
    		\end{array}
    		\right\vert \eta^{k_{j_{\ell}}} (  \hat{\Theta} )  \geq\left(
    		1-\varepsilon\right)  ^{1/2}\right)  
    		\cr
    		 \geq & \Pr\left(
    		\begin{array}
        	[c]{c}%
        		\hat{L}\text{ i.i.d. Bernoulli random variables with }p=\left(1 - \varepsilon
        		\right)^{1/3} \cr
        		\text{have a sample mean no less than }\left(  1-\varepsilon\right)^{1/2}%
    		\end{array}
    		\right)  
    		\cr
    		 \geq & \left(1 - \varepsilon\right)^{1/3},   \label{****}
		\end{align}
		where the first inequality follows from (\ref{***}) and the fact that $\sigma_i(\theta)$'s are independent across applicants, and the second inequality holds since $\ell > L_{4}$ and since, by the definition of $L_4$, for any such $\ell$, $\eta^{k_{j_{\ell}}} (\hat{\Theta}) \geq \left(1 - \varepsilon \right)^{1/2}$ implies $\eta^{k_{j_{\ell}}} (\hat{\Theta}) \cdot k_{j_{\ell}}>$ $\hat{L}$.
		
	    Comparing the finite economy random cutoffs $\P^{k_{j_{\ell}}}$ with the deterministic limit cutoffs $\overline{\p}$ yields: 
    	\begin{align}
    		& \Pr\left(\left.
    		\begin{array}[c]{c}%
    			\text{The fraction of applicants with } \theta \in \hat{\Theta}\\
    			\text{playing SRS against }\P^{k_{j_{\ell}}} \text{ is no less than } \left(1-\varepsilon\right)^{1/2}%
    		\end{array}
    		\right\vert \eta^{k_{j_{\ell}}} (\hat{\Theta}) \geq \left(1-\varepsilon\right)^{1/2}\right) 
    		\cr
    		 \geq & \Pr\left(\left.
    		\begin{array}
    			[c]{c}%
    			\text{The fraction of applicants with }\theta\in\hat{\Theta}\cr
    			\text{playing SRS against }\overline{\p} \text{ is no less than } \left(  1-\varepsilon\right)^{1/2}\\
    			\text{\emph{and} event }A^{k_{j_{\ell}}} \text{ occurs}
    		\end{array}
    		\right\vert \eta^{k_{j_{\ell}}} (  \hat{\Theta}) \geq \left(1 - \varepsilon\right)^{1/2}\right)  
    		\cr
    		 \geq & \Pr\left( \left.
    		\begin{array}[c]{c}%
    			\text{The fraction of applicants with } \theta \in \hat{\Theta} \cr
    			\text{playing SRS against }\overline{\p} \text{ is no less than } \left(1 - \varepsilon\right)^{1/2}%
    		\end{array}
    		\right\vert \eta^{k_{j_{\ell}}} (  \hat{\Theta}) \geq \left(1 - \varepsilon\right)^{1/2}\right)  
    		\cr
    		& \quad - \Pr\left(\left.  {A}^{k_{j_{\ell}}} \text{ does not occur}\right\vert \eta^{k_{j_{\ell}}} (
    		\hat{\Theta} )  \geq \left(1 - \varepsilon\right)^{1/2}\right)  
    		\cr
    		\geq & \left(1 - \varepsilon\right)^{1/3} - \frac{1 - \Pr\left({A}^{k_{j_{\ell}}}\right)} {\Pr\left(\eta^{k_{j_{\ell}}} (\hat{\Theta}) \geq \left(1 - \varepsilon\right)^{1/2}\right)}
    		\cr
    		\geq & \left(1 - \varepsilon\right)^{1/3} - \frac{\left(1-\varepsilon\right)^{1/2}\left[\left(1-\varepsilon\right)^{1/3}-\left(1-\varepsilon \right)^{1/2}\right]  }{\left(  1 - \varepsilon\right)^{1/2}}
    		\cr
    		 =& \left(1 - \varepsilon\right)^{1/2},  \label{last}
    	\end{align}
	    where the first inequality follows since in event ${A}^{k_{j_{\ell}}}$, the strategy  $\sigma_i(\t)$ is SRS against $\P^{k_{j_{\ell}}}$ if and only if $\sigma_i(\t)$ is SRS against $\overline{\p}$ for type $\t\in\hat \T$; the third inequality follows from (\ref{****}); and the fourth inequality follows from (\ref{**}).
		
    	We finally have in economy $F^{k_{j_{\ell}}}$%
	    \begin{align*}
		    & \Pr\left(\text{The fraction of applicants playing SRS against }\P^{k_{j_{\ell}}} \text{ is no less than } 1 - \varepsilon\right)  
		    \\
		    \geq &\Pr\left(
    		\begin{array}
    		[c]{c}%
        		\text{At least a fraction } \left(  1 - \varepsilon\right)^{1/2} \text{ of applicants with } 	\theta \in \hat{\Theta}
        		\text{ play SRS against }\P^{k_{j_{\ell}}}\\
        		\text{\it and }\eta^{k_{j_{\ell}}} (\hat{\Theta})  \geq \left(
        		1 - \varepsilon\right)^{1/2}%
    		\end{array}
		    \right)  \\
		    =&\Pr\left(\eta^{k_{j_{\ell}}} (  \hat{\Theta} ) \geq \left(1 - \varepsilon\right)^{1/2}\right)
		    \\
		    & \qquad \times \Pr\left(\left.
    		\begin{array}
    		[c]{c}%
        		\text{at least a fraction } \left(  1 - \varepsilon\right)^{1/2} \text{ of applicants }\\
        		\text{ with }	\theta \in \hat{\Theta} \text{ play SRS against } \P^{k_{j_{\ell}}}\\
    		\end{array}
    		\right\vert \eta^{k_{j_{\ell}}}\left(  \hat{\Theta}\right)  \geq\left(1-\varepsilon\right)^{1/2}\right)  \\
    		\geq&\left(1-\varepsilon\right)^{1/2}\cdot\left(  1-\varepsilon\right)^{1/2}\\
    		=&1-\varepsilon,
    	\end{align*}
	where the second inequality follows from the construction of $L_1$ (see \ref{L1}) and from (\ref{last}). Therefore, we have obtained a contradiction to (\ref{*}).
	\end{proof}

\section{Monte Carlo Simulations\label{app:mc}}
\appcaption{Appendix \ref{app:mc}: Monte Carlo Simulations}

Complementing Section~\ref{sec:implications} in the main text, this appendix provides additional details on the Monte Carlo simulations that we perform to assess the implications of our theoretical results. Section~\ref{sec:MC_Model_Specification} specifies the environment, Section~\ref{sec:MC_DGP} describes the data generating processes,  Section~\ref{sec:MC_Results} presents the estimation and the results, and, finally, Section~\ref{sec:MC_CF} presents some additional results on the counterfactual analysis.

\subsection{Simulated Environment \label{sec:MC_Model_Specification}}

We consider a finite economy in which $k=1,800$~applicants compete for admission to $C=12$~colleges. The vector of college capacities is specified as follows:
\begin{equation*}
\{S_{c}\}_{c=1}^{12} = \{150,75,150,150,75,150,150,75,150,150,75,150\}.
\end{equation*}
Setting the total capacity of colleges (1,500~seats) to be strictly smaller than the number of applicants (1,800) ensures that each college has a strictly positive cutoff in equilibrium.

The economy is located in an area within a circle of radius~1. The applicants are uniformly distributed across the area, and the colleges are evenly located on a circle of radius~$1/2$ around the centroid. The Cartesian distance between applicant~$i$ and college~$c$ is denoted by $d_{i,c}$.

Applicants are matched with colleges through a serial dictatorship, a special case of DA. Applicants are asked to submit an ROL of colleges, and there is no limit on the number of choices to be ranked. Without loss of generality, colleges have a priority structure such that all colleges rank applicant $i$ ahead of $i'$ if $i'<i$. One may consider the order being determined by certain test scores. 

To represent applicant preferences over colleges, we adopt a parsimonious random utility model without an outside option.  As is traditional and more convenient in empirical analysis, we now let the applicant utility functions take any value on the real line; we continue to use $u$ as a notation for utility functions. That is, applicant~$i$'s utility from being matched with college~$c$ is specified as follows:
\begin{align}
u_{i,c} &= \beta_1 \cdot c + \beta_2 \cdot d_{i,c} + \beta_3 \cdot T_{i} \cdot A_{c} + \beta_4 \cdot Small_c + \epsilon_{i,c}, \forall i \text{ and } c, \label{Eq:MC_model}
\end{align}
where $\beta_1 \cdot c$ is college~$c$'s baseline quality; $d_{i,c}$ is the distance from applicant~$i$'s location to college~$c$; $T_{i}=1$ or $0$ is applicant~$i$'s type (e.g., disadvantaged or not); $A_{c}=1$ or $0$ is college~$c$'s type (e.g., known for resources for disadvantaged applicants); $Small_c=1$ if college $c$ is small, 0 otherwise; and $\epsilon_{i,c}$ is a type-I extreme value and i.i.d.\ across $i$ and $c$.

The type of college $c$, $A_c$, is $1$ if $c$ is an odd number; otherwise, $A_c=0$. The type of applicant $i$, $T_{i}$, is $1$ with probability $2/3$ among the lower-ranked applicants ($i\leq 900$); $T_i=0$ for all $i>900$. This way, we may consider those with $T_i=1$ as the disadvantaged.

The coefficients of interest are $(\beta_1,\beta_2,\beta_3,\beta_4)$ which are fixed at $(0.3,-1,2,0)$ in the simulations. By this specification, colleges with larger indices are of higher quality, and $Small_c$ does not affect applicant preference. The purpose of estimation is to recover these coefficients and therefore the distribution of preferences.

\subsection{Data Generating Processes\label{sec:MC_DGP}}

Each simulation sample contains an independent preference profile obtained by randomly drawing $\{d_{i,c},\epsilon_{i,c}\}_c$ and $T_i$ for all $i$ from the distributions specified above. In all samples, applicant scores, college capacities, and college types ($A_c$) are kept constant.

We first simulate the joint distribution of the 12 colleges' cutoffs by letting every applicant submit an ROL ranking all colleges truthfully. After running the serial dictatorship, we calculate the cutoffs in each simulation sample. Figure~\ref{Fig:cutoff_dis} shows the marginal distribution of each college's cutoff from the $1000$ samples. Note that colleges with smaller capacities tend to have higher cutoffs. For example, college~11, with 75 seats, often has the highest cutoff, although college 12, with 150 seats, has the highest baseline quality.

\begin{mfignotesin}{\label{Fig:cutoff_dis} Simulated Distribution of Cutoffs when Everyone is Truth-telling}
    {
        \begin{subfigure}{0.5\textwidth}
        \centering
        \graphique{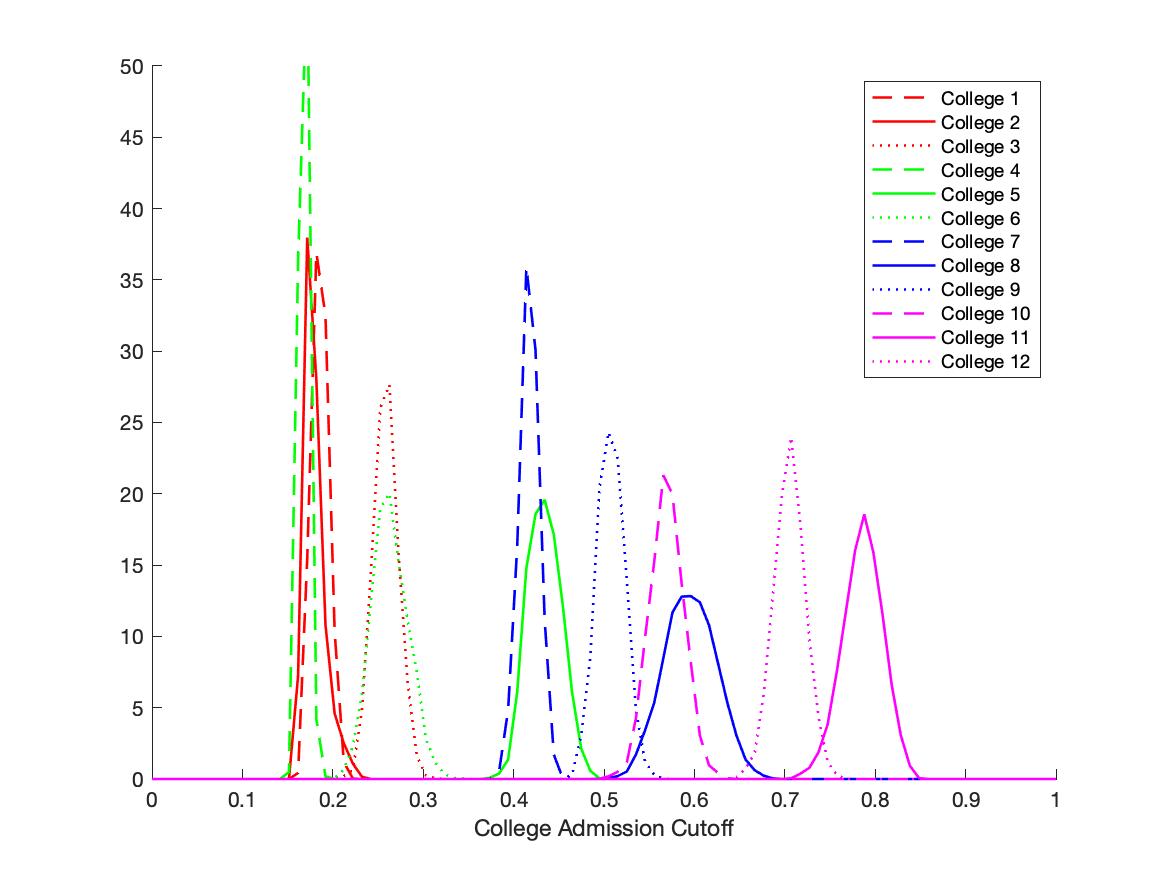}%
        \end{subfigure}%
    }
    {Assuming everyone is truth-telling, we calculate the cutoffs of all colleges in each simulation sample. The figure shows the marginal distribution of each college's cutoff, in terms of percentile rank (between 0 (lowest) and 1 (highest)). Each curve is an estimated density based on a normal kernel function. A solid line indicates a small college with 75, instead of 150, seats. The simulation samples for cutoffs use independent draws of $\{d_{i,c},\epsilon_{i,c}\}_c$ and $T_i$. }
\end{mfignotesin}

To generate data on applicant behaviors and outcomes, we simulate another 150 samples with new independent draws of $\{d_{i,c},\epsilon_{i,c}\}_c$ and $T_i$ for all $i$. These samples are used for the estimation and counterfactual analysis, and, in each of them, we consider three types of data generating processes (DGPs) with different applicant strategies.

\begin{enumerate}[(i)]
  \item \textbf{TT (Truth-Telling)}: Every applicant submits an ROL of 12 colleges according to her true preferences. Because everyone finds every college acceptable, this is TT as defined in our theoretical model.\footnote{This is equivalent to the definition of \textit{strict truth-telling} in \cite{Fack-Grenet-He(2015)} when there are no unacceptable colleges.}

  \item \textbf{PIM (Payoff Irrelevant Mistakes)}: A fraction of applicants omit from their ROLs some of the colleges with which they are never matched according to the simulated distribution of cutoffs. For a given applicant, an omitted college may have a high (expected) cutoff and thus be ``out of reach;'' alternatively, an omitted college may have a low cutoff, but the applicant is always accepted by one of her more-preferred colleges. There are 55 percent of applicants who omit at least one college. As applicants with $T_i=1$ have lower scores, they are more likely to omit than those with $T_i=0$: $61$ percent of $T_i=1$ drop at least one college, compared to $51$ percent of $T_i=0$. Among applicants who are never matched with any college, we randomly choose some colleges for them to include in their ROLs.

  \item \textbf{PRM (Payoff Relevant Mistakes)}: Taking the data generated under PIM, we let more applicants to omit never-matched colleges and also let some of them make payoff-relevant mistakes. That is, some applicants omit some of the colleges with which they have a chance of being matched lower than 30 percent according to the simulated distribution of cutoffs. Recall that the joint distribution of cutoffs is only simulated once under the assumption that everyone is truth-telling. On average, 60 percent of applicants drop at least one college. 
\end{enumerate}

To summarize, for each of the $150$ samples, we simulate the matching game $3$ times: TT (Truth-Telling), PIM (payoff-irrelevant mistakes), and PRM (payoff-relevant mistakes). See Table~\ref{tab:mc_setup} in the main text for summary statistics.

\subsection{Estimation and Results\label{sec:MC_Results}}

With the simulated data, the random utility model described by equation~\eqref{Eq:MC_model} is estimated under two different identifying assumptions.

We first re-write the random utility model (equation~\ref{Eq:MC_model}) as follows:
\begin{align*}
u_{i,c} & = \beta_1 \cdot c + \beta_2 \cdot d_{i,c} + \beta_3 \cdot T_{i} \cdot A_{c} + \beta_4 \cdot Small_c + \epsilon_{i,c}\\
        &\equiv V_{i,c} + \epsilon_{i,c}, \forall i =1,\cdots, k \text{ and } c=1,\dots,C;
\end{align*}
we also define $\mathbf{X}_{i} = (\{d_{i,c}, A_{c},Small_c\}_{c}, T_i)$ to denote the observable applicant characteristics and college attributes; and $\bm\beta$ is the vector of coefficients, $\bm\beta = (\beta_1,\beta_2,\beta_3,\beta_4)$. 

The key for each estimation approach is to characterize the choice probability of each ROL or each college, where the uncertainty originates from $\epsilon_{i,c}$, because the researcher does not observe its realization. In contrast, we do observe the realization of $\mathbf{X}_{i} $, submitted ROLs, and outcomes.

\subsubsection{Weak Truth-Telling (WTT)}
Naturally, one may start by a truth-telling assumption such as TT in which every applicant truthfully ranks every college in her ROL. The fact that applicants rarely rank all available colleges motivates a weaker version of truth-telling.  Weak truth-telling, or WTT, can be considered as a truncated version of TT, entails two assumptions: (a) the observed number of choices ranked in any ROL is exogenous to applicant preferences and (b) every applicant ranks her top preferred colleges according to her preferences, although she may not rank all colleges.  Although WTT is weaker than TT, it is still susceptible to untruthful ROLs from our robustness perspective: the robust equilibrium constructed in Theorem~\ref{p1} fails WTT.

The submitted ROLs specify a rank-ordered logit model that can be estimated by Maximum Likelihood Estimation (MLE). We define this as the WTT-based estimator.

The probability of applicant~$i$ submitting $R =r^{1}$-$r^{2}$-$\dotsc$-$r^{|R|} \in  \mathcal{R}$ is:
\begin{align*}
  &\Pr \left( \sigma_i(\bm{u}_i,\bm{s}_i) = R \mid \mathbf{X}_{i};{\bm\beta}\right)  \\
= &\Pr \left( u_{i,r^1} > \cdots > u_{i,r^{|R|}} > u_{i,c}, \forall c \notin \{r^{1},\dotsc,r^{|R|}\} \mid \mathbf{X}_{i};{\bm\beta}; |\sigma_i(\bm{u}_i,\bm{s}_i)| = |R| \right) \\
  & \times
   \Pr\left( |\sigma_i(\bm{u}_i,\bm{s}_i)| = |R|  \mid \mathbf{X}_{i};{\bm\beta}  \right).
\end{align*}

Under the assumptions that $|\sigma_i(u_i,s_i)|$ is orthogonal to $u_{i,c}$ for all~$c$ and that $\epsilon_{i,c}$ is a type-I extreme value, we can focus on the choice probability conditional on $|\sigma_i(u_i,s_i)|$ and obtain:
\begin{align*}
  &  \Pr \left( \sigma_i(\bm{u}_i,\bm{s}_i) = R \mid \mathbf{X}_{i};{\bm\beta}; |\sigma_i(\bm{u}_i,\bm{s}_i)|= |R|\right) \\
= & \Pr \left( u_{i,r^1} > \cdots > u_{i,r^{|R|}} > u_{i,c}, \forall c \notin \{r^{1},\dotsc,r^{|R|}\} \mid \mathbf{X}_{i};{\bm\beta}; |\sigma_i(\bm{u}_i,\bm{s}_i)| = |R| \right) \\
= & \prod_{c \in \{r^{1},\dotsc,r^{|R|}\}} \left(\frac{ \exp(V_{i,c})} {\sum_{c^{\prime} \nsucc_{R} c} \exp(V_{i,c^{\prime}})}\right)
\end{align*}
where $c^{\prime} \nsucc_{R} c$ indicates that $c^{\prime}$ is not ranked before $c$ in $R$, which includes $c$ itself and the colleges not ranked in $R$.

With a location normalization (e.g., $V_{i,1}=0$), the model can be estimated by MLE with the following log-likelihood function:
\begin{align*}
&\ln L_{WTT}\big({\bm\beta} \mid  \mathbf{X}, \{\vert \sigma_i(\bm{u}_i,\bm{s}_i)\vert\}_i \big)\\
= & \sum_{i=1}^{k}\sum_{c \text{ ranked in } \sigma_i(\bm{u}_i,\bm{s}_i)}
V_{i,c} - \sum_{i=1}^{k} \sum_{c \text{ ranked in } \sigma_i(\bm{u}_i,\bm{s}_i)} \ln\Big(\sum_{c^{\prime} \nsucc_{\sigma_i(\bm{u}_i,\bm{s}_i)} c} \exp(V_{i,c^{\prime}}) \Big).
\end{align*}
The WTT-based estimator, $\widehat{\bm\beta}^{WTT}$, is the solution to $\max_{\bm\beta} \ln L_{WTT}\big({\bm\beta} \mid  \mathbf{X}, \{\vert \sigma_i(\bm{u}_i,\bm{s}_i)\vert\}_i \big) $.

\subsubsection{Stability}
The assumption of stability implies that every applicant is matched with her favorite feasible college given the ex-post cutoffs. The random utility model can be estimated by MLE based on a conditional logit model where each applicant's choice set is restricted to the ex-post feasible colleges and where the matched college is the favorite among all her feasible colleges. If applicants play a regular robust equilibrium, stability is satisfied asymptotically according to Theorem \ref{p2}. We define this estimator as the stability-based estimator.

Suppose that the matching is $\mu$, which leads to a vector of cutoffs $\P$. With information on how colleges rank applicants, we can find a set of colleges that are ex-post feasible to $i$, $\mathcal{C}(s_i,\P)$.

The conditions specified by the stability of $\mu$ imply the likelihood of applicant $i$ matching with $c^*$ in $\mathcal{C}(s_i,\P)$:
$$\Pr \left(c^*=\mu(i) = \argmax_{c \in \mathcal{C}(s_i,\P)} u_{i,c}  \vert \mathbf{X}_i, \mathcal{C}(s_i,\P); \bm\beta \right).$$

Given the parametric assumptions on utility functions, the corresponding (conditional) log-likelihood function is:
\begin{align*}
\ln L_{ST}\left({\bm\beta} \mid \mathbf{X}, \mathcal{C}(s_i,\P) \right) =
\sum_{i=1}^{k} V_{i,\mu(i)} - \sum_{i=1}^{k} \ln\Big( \sum_{c^{\prime} \in \mathcal{C}(s_i,\P)} \exp(V_{i,c^{\prime}}) \Big). \label{MLE:stable}
\end{align*}
The stability-based estimator, $\widehat{\bm\beta}^{ST}$, is the solution to $\max_{\bm\beta} \ln L_{ST} \big({\bm\beta} \mid  \mathbf{X}, \mathcal{C}(s_i,\P )\big)$.

A key assumption of this approach is that the feasible set $\mathcal{C}(s_i,\P)$ is exogenous to $i$. It is satisfied when the mechanism is the serial dictatorship. 

\subsubsection{Estimation Results}

Table~\ref{tab:mc_est} provides summary statistics on the estimates from the WTT and stability approaches. 

\begin{table}[h]
  \centering \scriptsize
  \caption{Estimation Using Different Approaches: Monte Carlo Results}  \label{tab:mc_est}%
    \resizebox{\textwidth}{!}{\begin{tabular}{lrcccccccccccc}
    \toprule
    \multicolumn{1}{c}{\multirow{2}[0]{*}{DGPs}} & \multirow{1}[0]{*}{Identifying} &       & \multicolumn{2}{c}{Quality ($\beta_1=0.3$)} &       & \multicolumn{2}{c}{Distance ($\beta_2=-1$)} &       & \multicolumn{2}{c}{Interaction ($\beta_3=2$)} &       & \multicolumn{2}{c}{Small college ($\beta_4=0$)} \\
    \cline{4-5} \cline{7-8} \cline{10-11} \cline{13-14}
    \multicolumn{1}{c}{} &  Assumption     &       & mean  & s.d.  &       & mean  & s.d.  &       & mean  & s.d.  &  & mean  & s.d. \\

    \midrule
    & & \multicolumn{12}{l}{\textit{A. Both approaches are consistent.}} \\
        [0.25em]
    \multirow{2}[0]{*}{TT} & WTT   &       & 0.30  & 0.00  &       & 2.00  & 0.03  &       & -1.00 & 0.03  &       & 0.00  & 0.02 \\
          & Stability &       & 0.30  & 0.01  &       & 2.01  & 0.12  &       & -1.00 & 0.09  &       & 0.00  & 0.07 \\
          [0.5em]
    & & \multicolumn{12}{l}{\textit{B. Only stability is consistent.}}  \\
        [0.25em]
    \multirow{2}[0]{*}{PIM} & WTT   &       & 0.18  & 0.00  &       & 1.22  & 0.04  &       & -0.64 & 0.04  &       & -0.07 & 0.02 \\
          & Stability &       & 0.30  & 0.01  &       & 2.01  & 0.12  &       & -1.00 & 0.09  &       & 0.00  & 0.07 \\
                       [0.5em]
    \multirow{2}[0]{*}{PRM} & WTT   &       & 0.17  & 0.00  &       & 1.12  & 0.04  &       & -0.60 & 0.04  &       & -0.06 & 0.02 \\
          & Stability &       & 0.29  & 0.02  &       & 1.92  & 0.21  &       & -0.97 & 0.09  &       & -0.02 & 0.10 \\
          \bottomrule
    \end{tabular}}
    \begin{tabnotes}
This table presents estimates (mean and standard deviation across 150 samples) of the random utility model described in equation~(\ref{Eq:MC_model}). The true values are $(\beta_1,\beta_2,\beta_3,\beta_4)=(0.3,-1,2,0)$. It shows results in the three data generating processes (DGPs) with two identifying assumptions, WTT and stability.
\end{tabnotes}
\end{table}%

\subsection{Counterfactual Analysis}\label{sec:MC_CF}
We now provide some details on the counterfactual analysis. Recall that we consider the following counterfactual policy: applicants with $T_i=1$ are given priority over those with $T_i=0$, while within each type, applicants are still ranked according to their indices. That is, given $T_i=T_{i^{\prime}}$, $i$ is ranked higher than $i^{\prime}$ by all colleges if and only if $i>i^{\prime}$. The matching mechanism is still the serial dictatorship in which everyone can rank all colleges.

The effects of the counterfactual policy are evaluated by the following four approaches.
\begin{enumerate}[(i)]
  \item \textbf{Actual behavior (the truth)}: We use the true coefficients in utility functions to simulate counterfactual outcomes. They will be used as a benchmark against which alternative approaches will be evaluated. In keeping with our DGPs above, the ``actual behavior'' ranges from TT to untruthful reporting (see Section~\ref{sec:MC_DGP}). Specifically, DGP TT requires everyone to submit a truthful 12-college ROL; in DGP PIM, some applicants omit their never-matched colleges; and in DGP PRM, some applicants omit some colleges with which they have a low chance of being matched.
  \item \textbf{Submitted ROLs}: One assumes that the submitted ROLs under the existing policy are true ordinal preferences and that applicants will submit the same ROLs even when the existing policy is replaced by the counterfactual.
  \item \textbf{WTT}: One assumes that the submitted ROLs represent top preferred colleges in true preference order, and therefore applicant preferences can be estimated from the data with WTT as the identifying assumption. Under the counterfactual policy, we simulate applicant preferences based on the estimates and let applicants submit truthful 12-college ROLs.
  \item \textbf{Stability}: We estimate applicant preferences from the data with stability as the identifying assumption. Under the counterfactual policy, we simulate applicant preferences based on the estimates and let applicants submit truthful 12-college ROLs.
\end{enumerate}

Note that we assume truthful reporting in the counterfactual in the last two approaches. This is necessary because none of these approaches estimates how applicants choose ROLs, while we have to specify applicant behavior in counterfactual analysis.  This assumption of truthful reporting in the counterfactual analysis is justified by Corollary~\ref{cor2}.\footnote{Corollary \ref{cor2} rests on the uniqueness of stable matching in $E=[\eta, S]$, guaranteed  by the full support assumption on $\eta$.  While the current priority structure violates full support,  serial dictatorship produces a unique stable outcome, and thus validates the corollary for the current environment.}

When simulating counterfactual outcomes, we use the same $150$ simulated samples for estimation. In particular, we use the same simulated $\{\epsilon_{i,c}\}_c$ when constructing preference profiles after preference estimation. By holding constant $\{\epsilon_{i,c}\}_c$, we isolate the effects of different estimators.

To summarize, for each of the 150 simulation samples, we conduct $12$ different counterfactual analyses: $3$ (DGPs: TT, PIM, and PRM) $\times$ $4$ (actual behavior and 3 counterfactual approaches---submitted ROLs, WTT, and stability).

\subsubsection{Performance of the Approaches in Counterfactual Analysis}

Taking the counterfactual outcomes based on the actual behavior as our benchmark, we evaluate the three approaches from two perspectives: predicting the policy's effects on outcomes and on welfare.

\begin{mfignotesin}{\label{Fig:CF_nAA} Three Approaches to Counterfactual Analysis: Applicants $T_i=0$}
    {
        \begin{subfigure}{0.45\textwidth}
        \centering
        \caption{\footnotesize Mis-predicted Match}
        \graphique{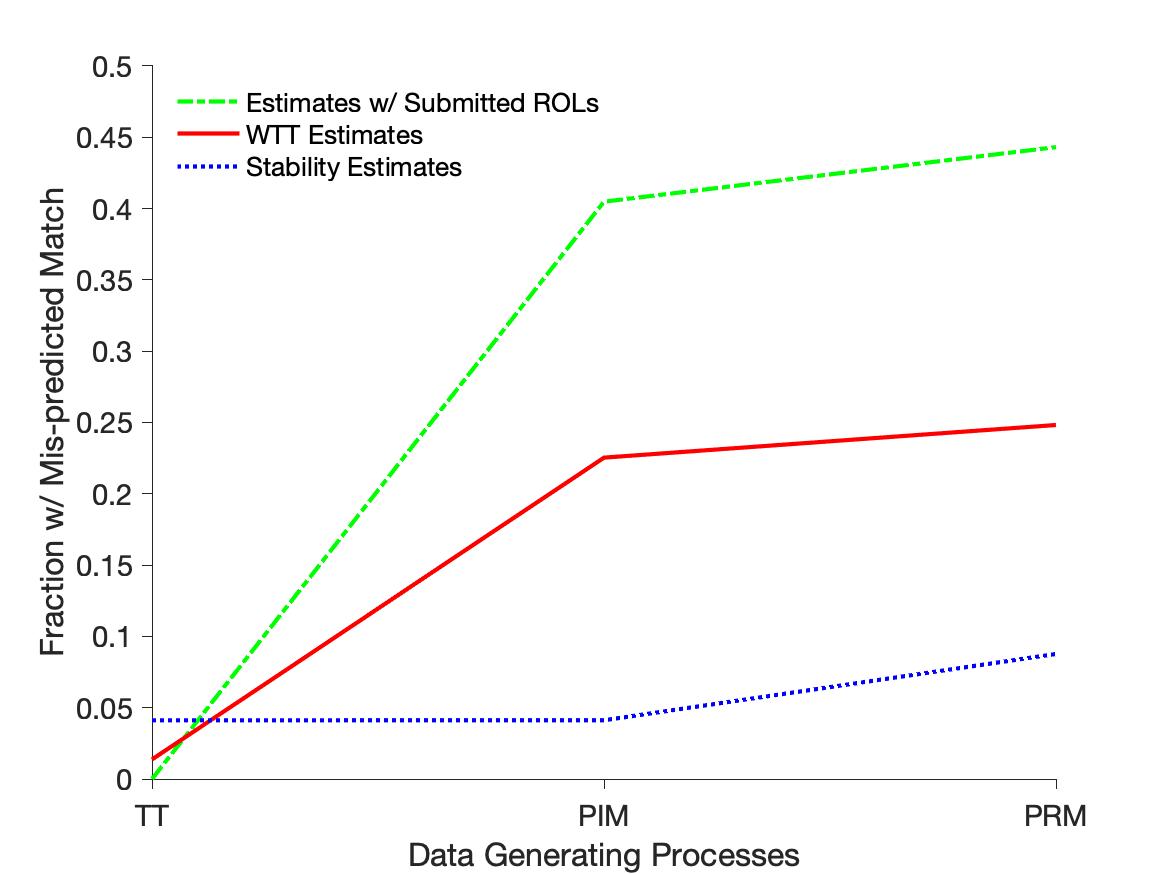}%
        \end{subfigure}%
        \begin{subfigure}{0.45\textwidth}
        \centering
        \caption{\footnotesize Predicted Welfare Effects}
        \graphique{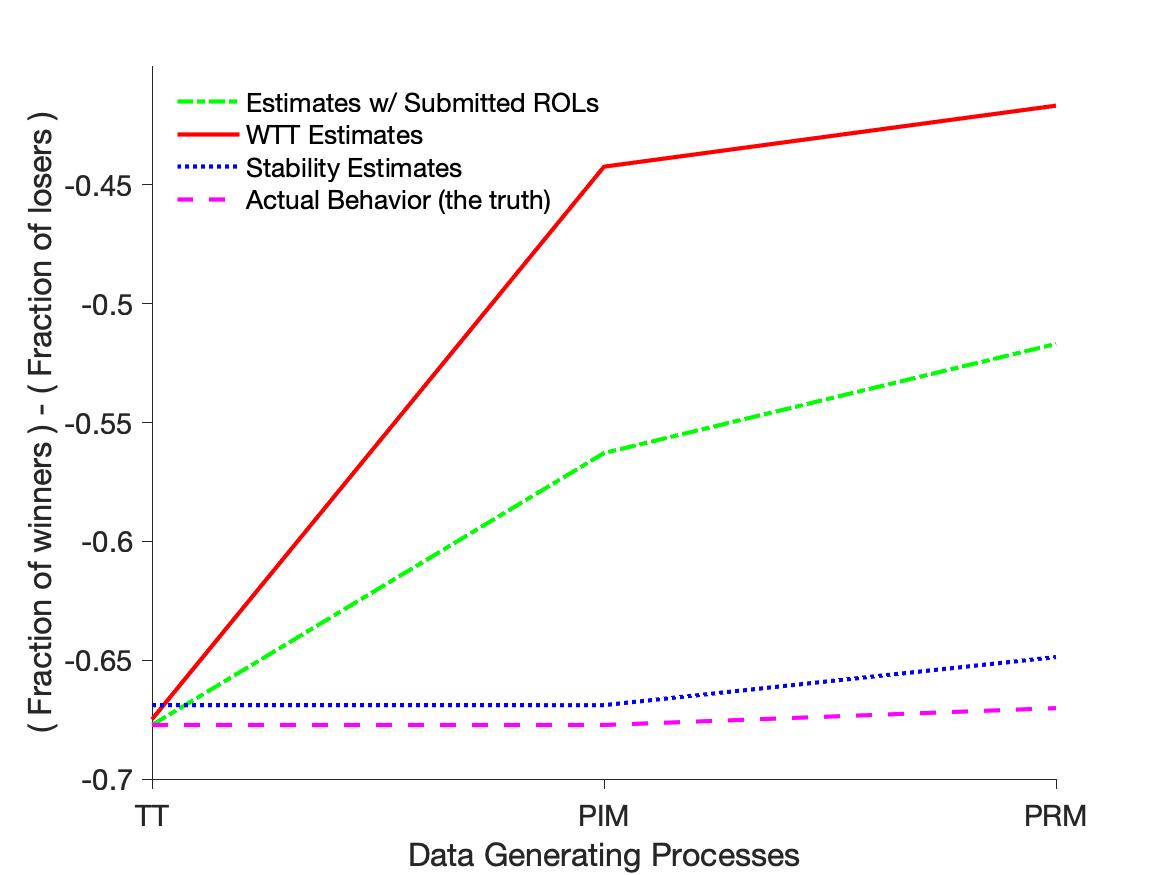}
        \end{subfigure}%
    }
    {This figure shows the averages among $T_i=0$ applicants across the 150 samples in each DGP. On average, there are 1201 such applicants in a sample. Given a DGP, we simulate an outcome under the counterfactual policy and compare it to the truth from the actual behavior (i.e., the true preferences with possible mistakes).  Panel (a) shows the average mis-prediction rates. Panel~(b) shows the predicted welfare effects by each approach. It is measured by the difference between the fractions of winners and losers.  See Table~\ref{tab:cf_all} for more details.}
\end{mfignotesin}

Complementing Figure~\ref{Fig:CF_AA} for applicants with $T_i=1$ in the main text, Figure~\ref{Fig:CF_nAA} shows the mis-predicted match and predicted welfare effects for applicants with $T_i=0$. The general patterns are the same as in Figure~\ref{Fig:CF_AA}: WTT and submitted ROLs produce biased predictions whenever some applicants are not truthful, while stability performs well in all DGPs.  

\begin{table}[h!]
	\centering \scriptsize
	\caption{Welfare Effects of the Counterfactual Policy (percentage points)}  \label{tab:cf_all}%
	\begin{tabular}{llcccccccc}
		\toprule
		&    Approaches to    & \multicolumn{2}{c}{Worse off} &       & \multicolumn{2}{c}{Better off} &       & \multicolumn{2}{c}{Indifferent} \\
		\cline{3-4} \cline{6-7} \cline{9-10}
		&   counterfactual    & mean  & s.d.  &       & mean  & s.d.  &       & mean  & s.d. \\
		\midrule
		&& \multicolumn{8}{l}{\it Panel A: Applicants with $T_i=1$}  \\
		[0.5em]
		\multirow{4}[0]{*}{DGP: TT} & Submitted ROLs & 0     & 0     &       & 91    & 1     &       & 9     & 1 \\
		& WTT   & 0     & 0     &       & 91    & 1     &       & 9     & 1 \\
		& Stability & 0     & 0     &       & 91    & 1     &       & 9     & 1 \\
		& Actual Behavior (the Truth)  & 0     & 0     &       & 91    & 1     &       & 9     & 1 \\
		&       &       &       &       &       &       &       &       &  \\
		\multirow{4}[0]{*}{DGP: PIM} & Submitted ROLs & 0     & 0     &       & 78    & 2     &       & 22    & 2 \\
		& WTT   & 0     & 0     &       & 88    & 1     &       & 12    & 1 \\
		& Stability & 0     & 0     &       & 91    & 1     &       & 9     & 1 \\
		& Actual Behavior (the Truth)  & 0     & 0     &       & 91    & 1     &       & 9     & 1 \\
		&       &       &       &       &       &       &       &       &  \\
		\multirow{4}[0]{*}{DGP: PRM} & Submitted ROLs & 0     & 0     &       & 72    & 2     &       & 28    & 2 \\
		& WTT   & 0     & 0     &       & 87    & 1     &       & 13    & 1 \\
		& Stability & 0     & 0     &       & 91    & 1     &       & 9     & 1 \\
		& Actual Behavior (the Truth)  & 0     & 0     &       & 91    & 1     &       & 9     & 1 \\
		\\
			&& 	\multicolumn{8}{l}{\it Panel B: Applicants with $T_i=0$}  \\
		[0.5em]
		    \multirow{4}[0]{*}{DGP: TT} & Submitted ROLs & 68    & 2     &       & 0     & 0     &       & 32    & 2 \\
		& WTT   & 68    & 2     &       & 0     & 0     &       & 32    & 2 \\
		& Stability & 67    & 2     &       & 1     & 0     &       & 32    & 2 \\
		& Actual Behavior (the Truth)  & 68    & 2     &       & 0     & 0     &       & 32    & 2 \\
		&       &       &       &       &       &       &       &       &  \\
		\multirow{4}[0]{*}{DGP: PIM} & Submitted ROLs & 56    & 2     &       & 0     & 0     &       & 44    & 2 \\
		& WTT   & 53    & 2     &       & 9     & 1     &       & 37    & 2 \\
		& Stability & 67    & 2     &       & 1     & 0     &       & 32    & 2 \\
		& Actual Behavior (the Truth)  & 68    & 2     &       & 0     & 0     &       & 32    & 2 \\
		&       &       &       &       &       &       &       &       &  \\
		\multirow{4}[0]{*}{DGP: PRM} & Submitted ROLs & 52    & 2     &       & 1     & 0     &       & 47    & 2 \\
		& WTT   & 52    & 2     &       & 10    & 1     &       & 38    & 2 \\
		& Stability & 66    & 3     &       & 1     & 1     &       & 32    & 2 \\
		& Actual Behavior (the Truth)  & 67    & 2     &       & 0     & 0     &       & 32    & 2 \\
		\bottomrule
	\end{tabular}
	\begin{tabnotes}
		This table presents the estimated effects of the counterfactual policy (giving $T_i=1$ applicants priority in admission) on applicants with $T_i=1$ (Panel A) and those with $T_i=0$ (Panel B).  On average, there are 599 applicants with  $T_i=1$ (standard deviation 14) and 1201 applicants with  $T_i=0$  (standard deviation 14)  in each simulation sample. The table shows results in the three data generating processes (DGPs) with four approaches. The one using submitted ROLs assumes that submitted ROLs represent applicant true ordinal preferences; WTT assumes that every applicant truthfully ranks her top $K_i$ ($1<K_i\leq 12 $) preferred colleges ($K_i$ is observed); and stability implies that every applicant is matched with her favorite feasible college, given the ex-post cutoffs. The truth is simulated with the possible mistakes in each DGP. The welfare change of each applicant is calculated in the following way: we first simulate the counterfactual match and investigate if a given applicant is better off, worse off, or indifferent by comparing the two matches according to estimated/assumed/true ordinal preferences. In each simulation sample, we calculate the percentage of different welfare change; the table then reports the mean and standard deviation of the percentages across the 150 simulation samples.
	\end{tabnotes}
\end{table}%

Moreover, Table~\ref{tab:cf_all} (Panel A for applicants with $T_i=1$ and Panel B for those with $T_i=0$) presents detailed statistics on the fractions of applicants being worse off, better off, and indifferent based on different approaches.

\end{appendix}

\clearpage
\setstretch{1.02}
\bibliographystyle{economet}
\bibliography{bibmatching}

\end{document}